\documentclass[12pt,twoside,english]{article}
\usepackage{a4wide}
\pdfoutput=1

\makeatletter
\usepackage[utf8]{inputenc}            
\usepackage[T1]{fontenc} 			
\usepackage{graphicx}
\usepackage{fancyhdr}
\usepackage{amsmath, amssymb, amsfonts, bm}
\usepackage{mathtools}
\usepackage{caption}
\usepackage[labelformat=simple]{subfig}
\usepackage{setspace}
\usepackage{verbatim} 
\usepackage[english]{babel}
\usepackage[novbox]{pdfsync} 
\usepackage{caption}
\usepackage{booktabs}
\usepackage{hyperref}
\usepackage{float}
\usepackage{relsize} 

\usepackage{siunitx}
\usepackage{xfrac}
\usepackage{pdfpages}
\usepackage{overpic}

\usepackage{color}

\includepdfset{turn=true,pagecommand={}}

\hypersetup{
	colorlinks = false,
	linktocpage = true,
	allbordercolors = {0.8 0.8 0.8}
}

\addto\captionsenglish{}
\addto\captionsenglish{}
\addto\extrasenglish{}
\addto\extrasenglish{}
\addto\extrasenglish{}
\addto\extrasenglish{}
\addto\extrasenglish{}
\addto\extrasenglish{}
\addto\extrasenglish{}

\usepackage{etex}
\usepackage{tikz}
\usepackage{pgfplots}
\usepackage{tikz-3dplot}
\usepackage{fp}

\pgfplotsset{compat=newest}

\ifdefined\bUseTikzExternalize
	\usetikzlibrary{external}
	\tikzsetexternalprefix{tmp/}
\else
	\usetikzlibrary{external}
	\tikzexternaldisable
\fi

\usetikzlibrary{calc,intersections}
\usetikzlibrary{shapes.geometric,shapes.arrows,decorations.pathmorphing}
\usetikzlibrary{matrix,chains,scopes,positioning,arrows,fit}
\usetikzlibrary{spy}
\usetikzlibrary{fadings}
\usetikzlibrary{shadings}
\usetikzlibrary{backgrounds}

\usepgfplotslibrary{groupplots}
\usepgfplotslibrary{patchplots}

\def\pgfplotfontsizetitle{\small}
\def\pgfplotfontsize{\small}

\pgfplotsset{
  mystyle/.style ={%
    grid = major,
    every tick label/.append style={font=\pgfplotfontsize},
    every axis label/.append style={font=\pgfplotfontsize},
    legend style={font=\scriptsize},
    label style={font=\pgfplotfontsize},
    title style={font=\pgfplotfontsizetitle},
    /pgf/number format/set thousands separator = {}, 
  }
}

\tikzset{
	>=stealth',
	axis/.style={<->},
	important line/.style={thick},
	connection/.style={thick, dotted},
	dot/.style = {
		draw,
		fill = white,
		circle,
		solid,
		thin,
		inner sep = 0pt,
		minimum size = 3pt
	},
	defP/.style = {
		inner sep = 0pt
	}	
}
\pgfarrowsdeclaredouble{doubled}{doubled}{stealth'}{stealth'}

\usepackage[outline]{contour}
\usepackage{standalone} 

\newcommand{\vek}[1]{\mathchoice{\displaystyle\boldsymbol#1}
	{\textstyle\boldsymbol#1}{\scriptstyle\boldsymbol#1}
	{\scriptscriptstyle\boldsymbol#1}}
\newcommand{\mat}[1]{\mathchoice{\displaystyle\mathbf#1}
	{\textstyle\mathbf#1}{\scriptstyle\mathbf#1}
	{\scriptscriptstyle\mathbf#1}}

\newcommand{\ten}[1]{ \ensuremath{\bm #1} }
\renewcommand{\d}{ \ensuremath{\mathrm{d} }}
\newcommand{\p}{ \ensuremath{\partial} }
\renewcommand{\t}[1]{ \ensuremath{\text{#1}} }
\newcommand{\T}{{\ensuremath{\mathrm{T}}}}

\newcommand{\divG}[1]{\ensuremath{\text{div}_{\Gamma} #1}}

\newcommand{\gradG}[1]{\ensuremath{\nabla_\Gamma #1}}
\newcommand{\gradGD}[1]{\ensuremath{\nabla_\Gamma^\text{dir} #1}}
\newcommand{\gradGC}[1]{\ensuremath{\nabla_\Gamma^\text{cov} #1}}
\newcommand{\nG}{\ensuremath{\vek{n}_\Gamma}}
\newcommand{\nCo}{\ensuremath{\vek{n}_{\p\Gamma}}}
\newcommand{\tB}{\ensuremath{\vek{t}_{\p\Gamma}}}

\newcommand{\vu}{\ensuremath{\vek{v}_u}}
\newcommand{\vw}{\ensuremath{\vek{v}_w}}

\definecolor{mygreen}{rgb}{0, 0.51, 0}
\definecolor{myred}{rgb}{0.46, 0., 0.05}

\newcommand{\lyxaddress}[1]{
	\par {\raggedright #1
		\vspace{1.4em}
		\noindent\par}
}

\numberwithin{equation}{section}

\makeatother

\graphicspath{{Pics/},{tmp/}}
\def\pathToBibFile{References}

\newcommand{\revStart}{\color{black}}
\newcommand{\revEnd}{\color{black}}

\begin{document}

\title{A Higher-order Trace Finite Element Method for Shells}

\author{D.~Sch{\"o}llhammer$^1$, T.P.~Fries$^2$}
\maketitle

\lyxaddress{\begin{center}
$^{1,2}$Institute of Structural Analysis\\
Graz University of Technology\\
Lessingstr. 25/II, 8010 Graz, Austria\\
\texttt{www.ifb.tugraz.at}\\
$^{1}$\texttt{schoellhammer@tugraz.at}, $^{2}$\texttt{fries@tugraz.at}
\end{center}}

\begin{abstract}

A higher-order fictitious domain method (FDM) for Reissner-Mindlin shells is proposed which uses a three-dimensional background mesh for the discretization. The midsurface of the shell is immersed into the higher-order background mesh and the geometry is implied by level-set functions. The mechanical model is based on the Tangential Differential Calculus (TDC) which extends the classical models based on curvilinear coordinates to implicit geometries. The shell model is described by PDEs on manifolds and the resulting FDM may typically be called Trace FEM. The three standard key aspects of FDMs have to be addressed in the Trace FEM as well to allow for a higher-order accurate method: (i) numerical integration in the cut background elements, (ii) stabilization of awkward cut situations  and elimination of linear dependencies, and (iii) enforcement of boundary conditions using Nitsche’s method. The numerical results confirm that higher-order accurate results are enabled by the proposed method provided that the solutions are sufficiently smooth.


\end{abstract} 
\newpage\tableofcontents\newpage
\section{Introduction}
\label{sec:intro}

A shell is a curved, thin-walled structure and due to their high bearing capacity, they occur in many engineering applications such as automotive, aerospace, biomedical- and civil-engineering \cite{Calladine_1983a,Farshad_1992a,Zingoni_2018a}. In the mechanical modelling of a shell, a dimensional reduction from the 3D shell body to its midsurface is a key step. The equilibrium equations result into partial differential equations (PDEs) on manifolds embedded in the physical space $\mathbb{R}^3$. In the classical approach of modelling shells, the midsurface is typically defined by a (piecewise) parametrization which may be called an explicit definition. Then, there exists a map from some two-dimensional parameter space $\hat{\Omega}$ to $\mathbb{R}^3$. In the context of the classical Surface FEM, the discrete midsurface is rather defined by an atlas of local element-wise mappings from the reference element to the physical elements. An overview of classical shell theory is given, e.g., in \cite{Chapelle_2000a,Simo_1989a,Simo_1989b,Bischoff_2017a} or in the text books \cite{Calladine_1983a,Blaauwendraad_2014a,Basar_1985a,Wempner_2002a,Zingoni_2018a}. Alternatively, the shell geometry may also be defined \emph{implicitly} following the level-set method \cite{Fries_2017a,Fries_2017b,Osher_2003a,Sethian_1999b}. The main advantage of implicit geometry descriptions is that the application of recent finite element techniques such as fictitious domain methods (FDM) for shells is enabled.

Herein, we propose a higher-order accurate fictitious domain method for \emph{implicitly} defined Reissner--Mindlin shells. The geometry of the shell is defined implicitly by means of level-set functions. The shell boundary value problem (BVP) is discretized with a higher-order accurate Trace FEM approach.\par

\begin{figure}[ht]
	\centering\def\mywidth{.32}
	\subfloat[implicit shell midsurface]{\includegraphics[width=\mywidth\textwidth]{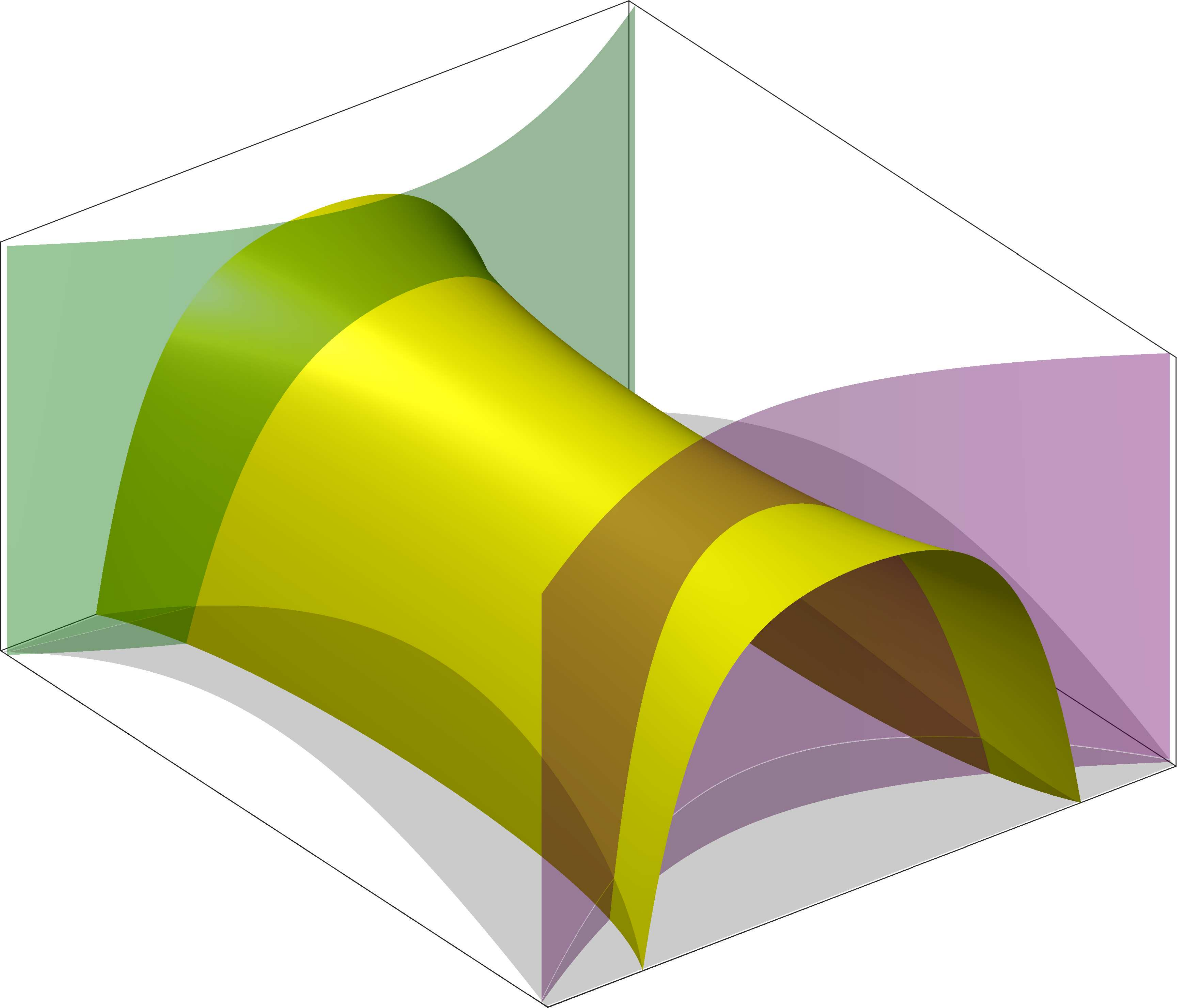}\label{fig:introtraceFEM1}}
	\hfil
	\subfloat[cut elements]{\includegraphics[width=\mywidth\textwidth]{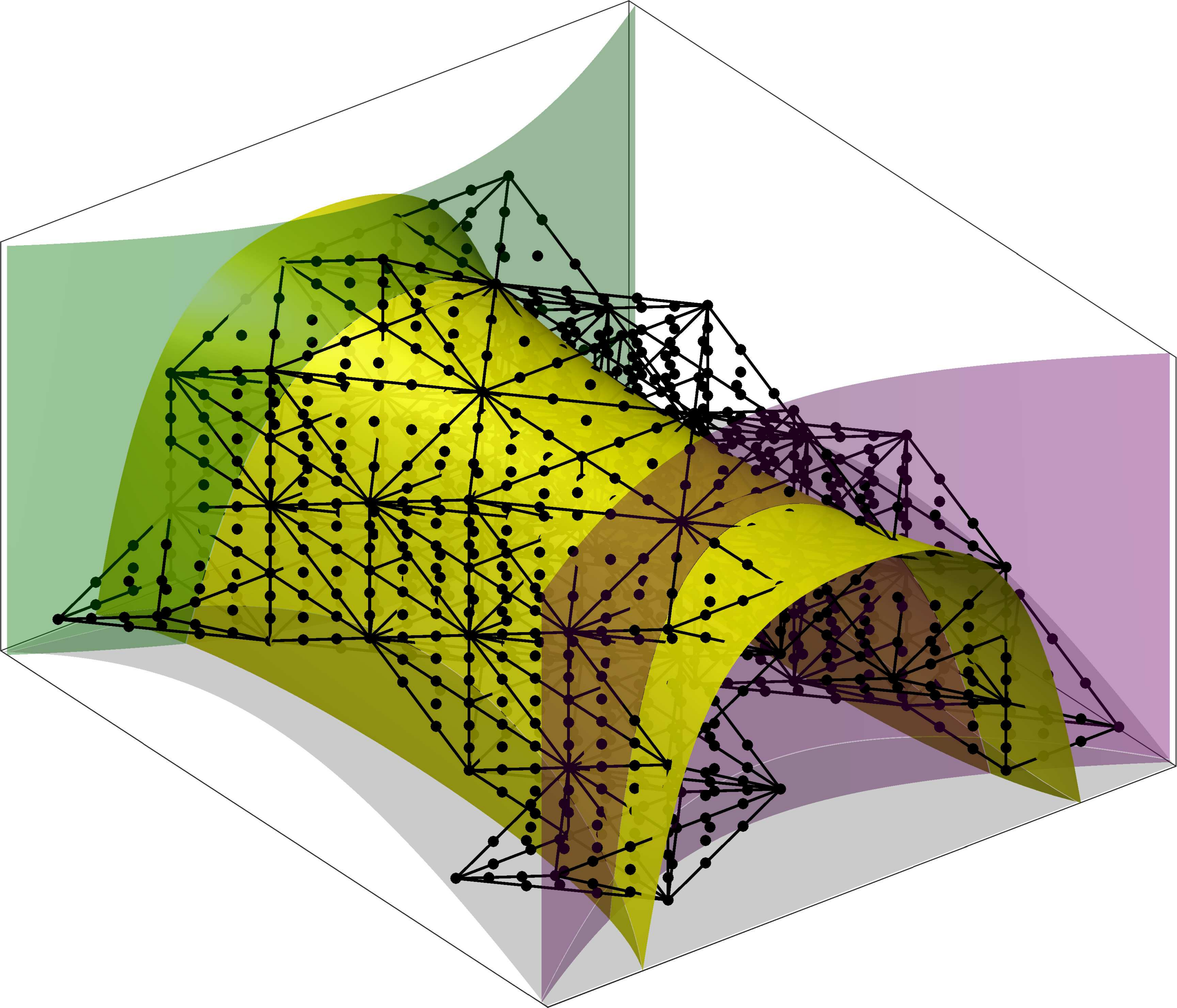}\label{fig:introtraceFEM2}}
	\hfil
	\subfloat[integration points on the zero-isosurface]{\includegraphics[width=\mywidth\textwidth]{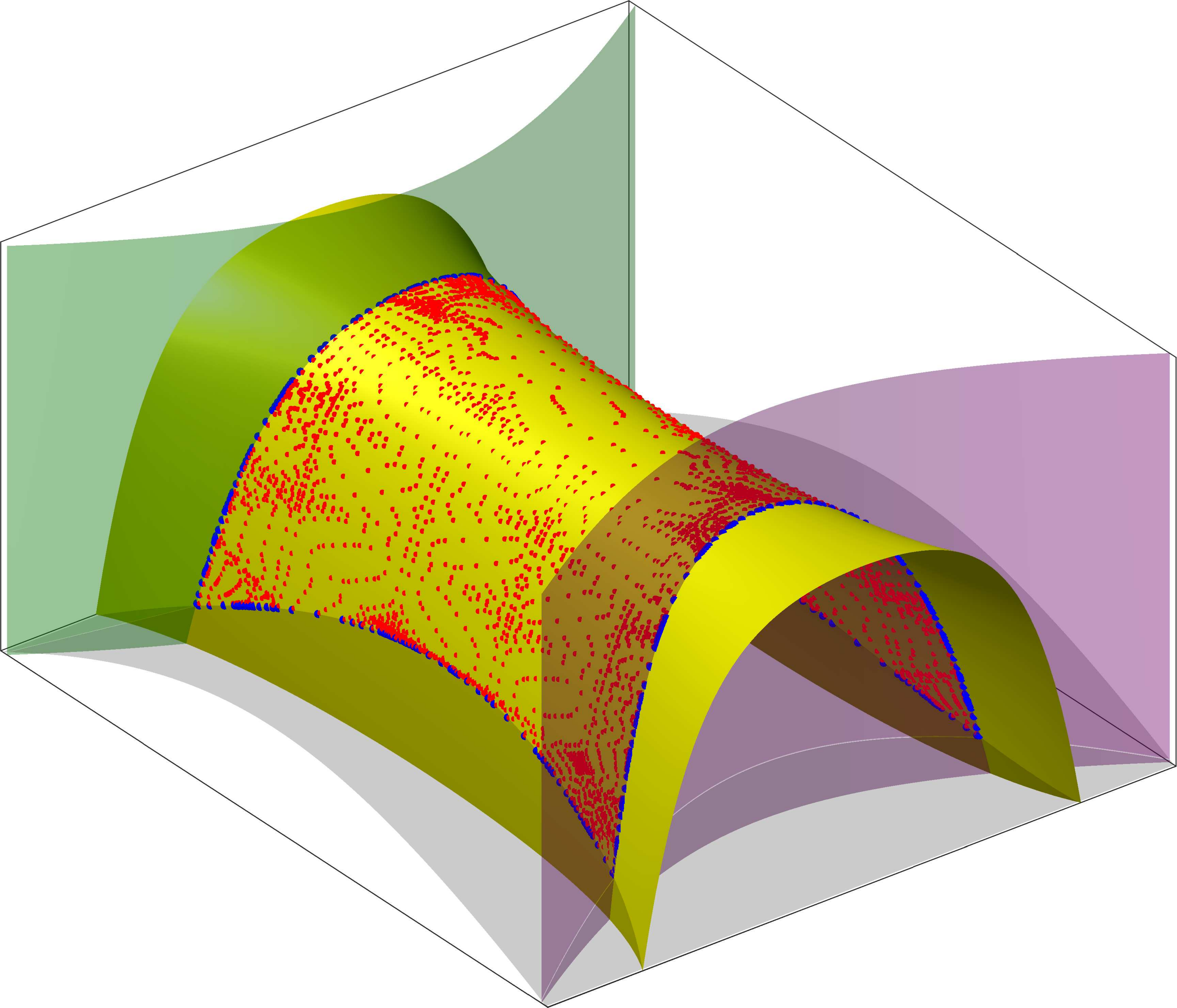}\label{fig:introtraceFEM3}}	
	\caption{Overview of the Trace FEM: (a) Implicit definition of the shell geometry by means of level-set functions, (b) the set of cut three-dimensional background elements are labelled as active mesh and their nodes imply the DOFs for the numerical simulation, (c) integration points on the zero-isosurface plotted in red for the domain and in blue on the boundary.\label{fig:introtraceFEM}}
\end{figure}

\begin{figure}[ht]
	\centering\def\mywidth{.32}
	\subfloat[explicit shell midsurface]{\includegraphics[width=\mywidth\textwidth]{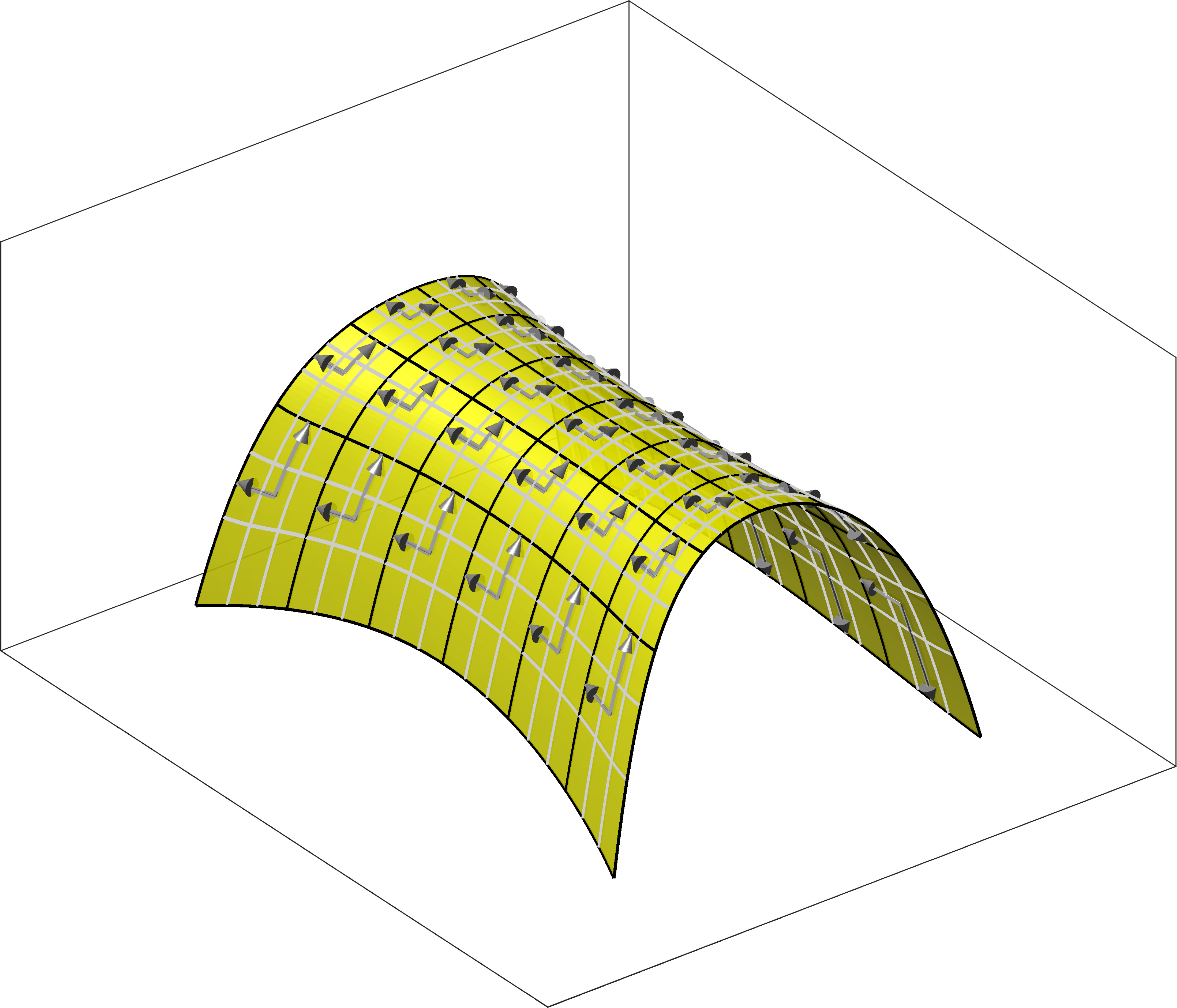}\label{fig:introtraceFEM4}}
	\hfil
	\subfloat[surface mesh]{\includegraphics[width=\mywidth\textwidth]{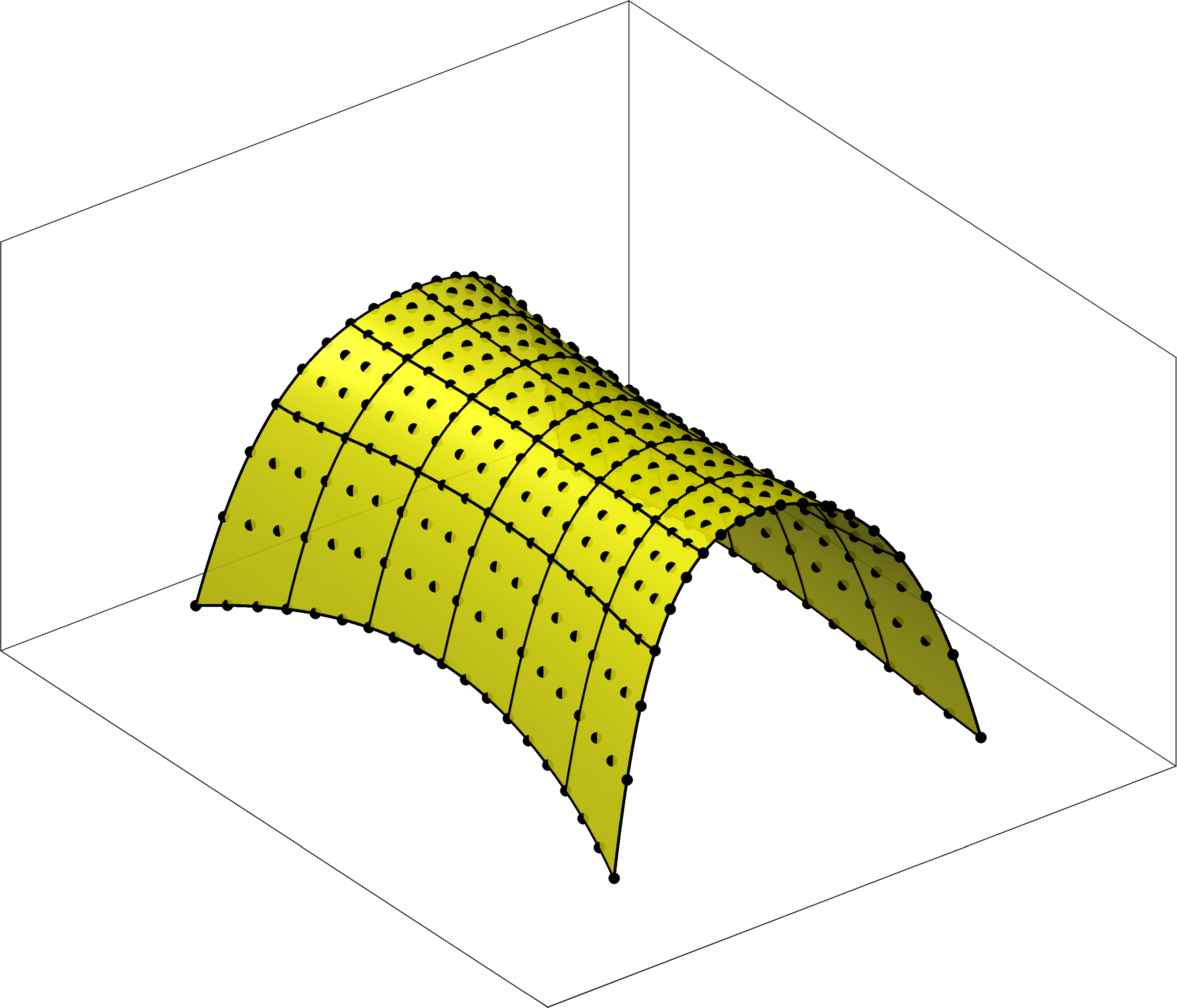}\label{fig:introtraceFEM5}}
	\hfil
	\subfloat[integration points on the surface mesh]{\includegraphics[width=\mywidth\textwidth]{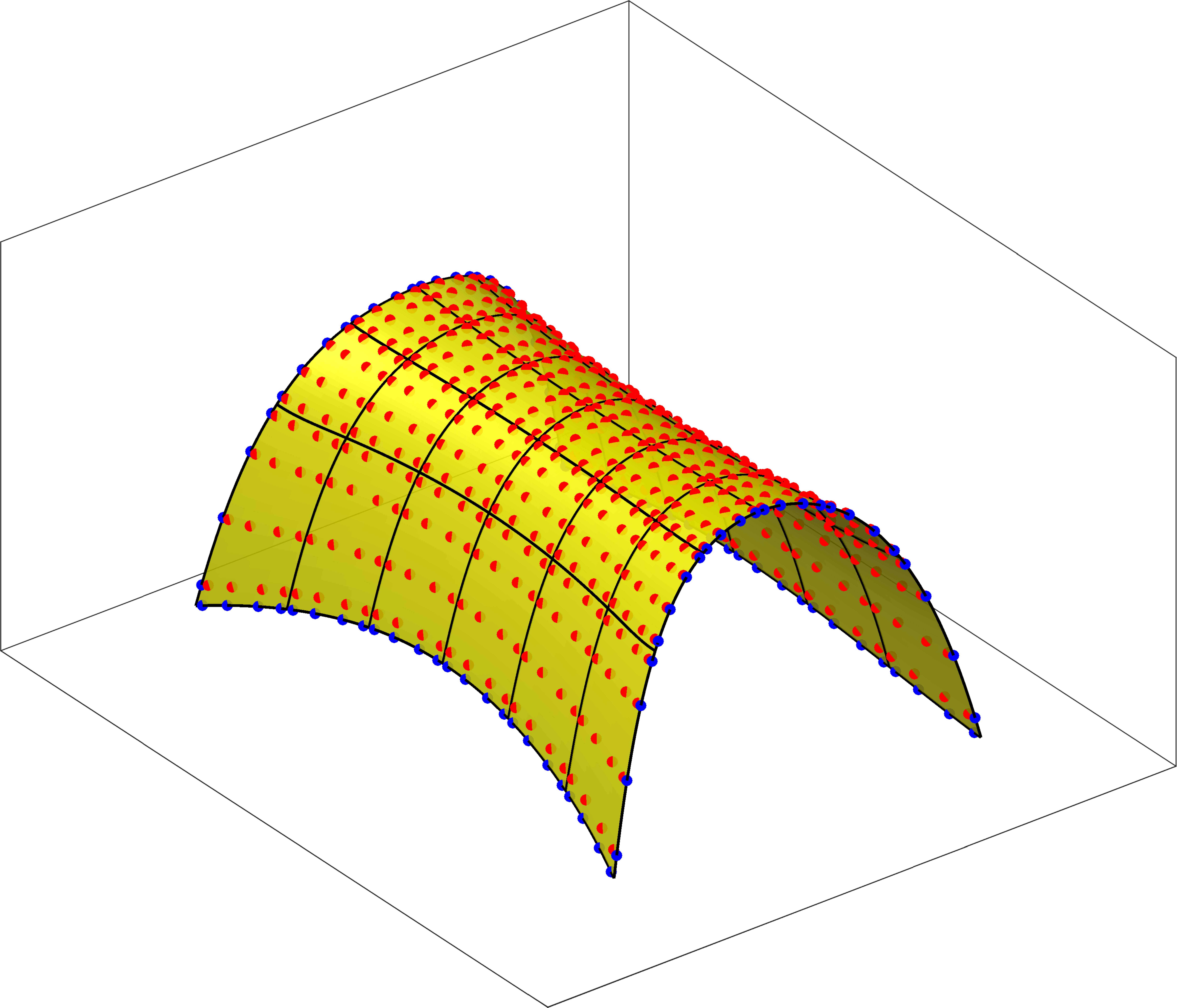}\label{fig:introtraceFEM6}}	
	\caption{Overview of the Surface FEM: (a) Explicitly defined shell obtained by an atlas of element-wise local mappings, (b) conforming surface mesh consisting of cubic 2D Lagrange elements, (c) integration points for the Surface FEM obtained by standard Gau\ss\ integration rules.\label{fig:introtraceFEMb}}
\end{figure}

The Trace FEM is a fictitious domain method for solving PDEs on manifolds, see, e.g., \cite{Reusken_2014a,Olshanskii_2009b,Olshanskii_2017a,Gross_2018a,Grande_2018a,Schoellhammer_2019c,Bonito_2019a}. This may also be called Cut FEM which, in the last years, became a popular FDM enabling higher-order accuracy \cite{Burman_2015a,Burman_2015b,Burman_2018a}. When using the Cut FEM for the solution of PDEs on manifolds as done herein, the method becomes analogous to the Trace FEM \cite{Burman_2016b,Cenanovic_2016a}. This recent finite element technique is fundamentally different compared to standard Surface FEM approaches, see, e.g., \cite{Demlow_2009a,Dziuk_2013a,Fries_2018b,Fries_2017b}. In the following, the major differences in, (1) geometry definition, (2) mesh and degrees of freedom (DOFs) and, (3) generation of integration points in these two approaches are outlined and visualized for the Trace FEM in \autoref{fig:introtraceFEM} and for the Surface FEM in \autoref{fig:introtraceFEMb}, respectively.
\begin{enumerate}
\item Firstly, the differences in the geometry definition are emphasized. In \autoref{fig:introtraceFEM1}, an implicitly defined shell by means of level-set functions is presented. In particular, the yellow surface indicates the midsurface of the shell and is defined by the zero-isosurface of a \emph{master} level-set function. The boundary of the shell is defined by additional  \emph{slave} level-set functions, i.e., the light green, purple and grey surfaces in \autoref{fig:introtraceFEM1}. In contrast to the implicit definition, an explicit representation of the shell midsurface for this example is shown in \autoref{fig:introtraceFEM4}. Then, the surface is given through an atlas of element-wise local mappings, as usual in the context of Surface FEM, implying local curvilinear coordinates. Note that a parametrization is \emph{only} available in the explicit case and is, in general, not available in implicitly defined geometries.\par 
\item Secondly, the mesh generation and the location of DOFs are considered. In the Trace FEM the domain of interest, i.e., the shell midsurface, is embedded in a three-dimensional background mesh. The 3D background mesh may consist of higher-order Lagrange elements with the only requirement that the domain of interest is completely immersed. Neither the shell surface nor the shell boundary have to conform to the background mesh. There is one master level-set function whose zero-isosurface implies the shell surface and additional slave level-set functions imply the boundaries. The set of cut elements is labelled \emph{active} mesh and is visualized in \autoref{fig:introtraceFEM2} where cubic tetrahedral elements are used as an example. For the numerical simulation, the DOFs are located at the nodes of the active mesh, which are clearly \emph{not} on the midsurface of the shell. The corresponding shape functions are those of the active mesh and are restricted to the trace. This means that for the integration of the weak form, the 3D shape functions are only evaluated on the zero-isosurface of the master level-set function. In contrast, in the Surface FEM a boundary conforming surface mesh, see \autoref{fig:introtraceFEM5}, is defined through an atlas of element-wise local mappings and the DOFs are located at nodes of the surface mesh which are on the discrete midsurface of the shell. The corresponding shape functions are the 2D shape functions living only on the surface mesh. The location of the DOFs and the different dimensionality of the shape functions are the most important differences between the two finite element techniques.\par
\item Lastly, integration points need to be placed on the shell midsurface for the integration of the weak form. In the case of the Trace FEM, this is not a trivial task especially when a higher-order accurate integration scheme is desired \cite{Fries_2015a,Fries_2017a,Fries_2017b,Fries_2019a,Mueller_2013a,Lehrenfeld_2016a}, further details are given below. Regarding the Surface FEM standard Gau\ss\ integration rules are applicable and mapped from the reference to the surface elements in the usual manner. In \autoref{fig:introtraceFEM3} and \autoref{fig:introtraceFEM6}, integration points in the domain (red) and on the boundary (blue) are visualized for the Trace FEM and Surface FEM, respectively.\par
\end{enumerate}

As mentioned above, the Trace FEM approach is a fictitious domain method and the following three well-known  implementational aspects require special attention: (i) integration of the weak form, (ii) stabilization and, (iii) enforcement of essential boundary conditions. Regarding the integration of the weak form, suitable integration points with higher-order accuracy for multiple level-set functions have to be provided. Herein, the approach, which naturally extends to multiple level-set functions as outlined by the authors in \cite{Fries_2015a,Fries_2017a,Fries_2017b,Fries_2019a} is employed. Other higher-order integration schemes for implicitly defined surfaces with \emph{one} level-set function are presented, e.g., in \cite{Mueller_2013a,Lehrenfeld_2016a}. A stabilization of the stiffness matrix is necessary due to small supports caused by unfavourable cut scenarios and the restriction of the shape functions to the trace. An overview and analysis of the different stabilization techniques in the Trace FEM is presented in \cite{Olshanskii_2017a}. Herein, the ``normal derivative volume stabilization'' first introduced for scalar-valued problems in \cite{Grande_2016a,Burman_2016b} and for vector-valued problems in \cite{Gross_2018a,Olshanskii_2018a} is used. The advantage of this particular stabilization technique is that it is suitable for higher-order accuracy; a straight forward implementation and a rather flexible choice of the stabilization parameter is possible. The essential boundary conditions need to be enforced in a weak manner due to the fact that a strong enforcement by prescribing nodal values does not apply (because the nodes of the background mesh are not on the shell boundary). Therefore, a similar approach as presented by the authors in \cite{Fries_2019a} is employed. In particular, the non-symmetric version of Nitsche's method is used, see, e.g., \cite{Burman_2012a,Schillinger_2016a,Guo_2019a,Fries_2019a}.\par

The Trace FEM approach is applied in a wide range of applications. In particular, transport problems are presented, e.g., in  \cite{Dziuk_1988a,Burman_2016b,Bonito_2019a,Olshanskii_2017a},  flow problems are considered, e.g., in \cite{Lehrenfeld_2016a,Gross_2018a,Brandner_2019a,Olshanskii_2018a,Jankuhn_2019a,Jankuhn_2020a}, moving and evolving manifolds are detailed in \cite{Olshanskii_2017a,Olshanskii_2017b,Lehrenfeld_2018a}. The first application of the Trace FEM to membranes has just recently been achieved for the linear membrane in \cite{Cenanovic_2016a} and for large deformation membranes by the authors in \cite{Fries_2019a}. This is considerably more challenging because in structural mechanics, the governing equations of membranes and shells have been, until recently, only formulated based on curvilinear coordinates. This, however, does not directly apply for an implicit shell definition and the use of the Trace FEM. Therefore, the reformulation of classical (curvilinear) models for shells and membranes during the last years was a crucial preliminary step for the use of the Trace FEM. Most importantly, the use of the Tangential Differential Calculus (TDC) was found highly useful and generalizes the mechanical models in the sense that they become valid for explicit \emph{and} implicit geometry definitions \cite{Hansbo_2014a,Hansbo_2015a,Schoellhammer_2018a,Schoellhammer_2019a,Schoellhammer_2019c,Fries_2019a}. In contrast, in flow and transport applications on curved surfaces, the general coordinate-free definition of the boundary value problems is a standard for a long time \cite{Dziuk_1988a,Demlow_2009a,Dziuk_2013a,Jankuhn_2017a,Fries_2018b}, thus enabling the application of the Trace FEM earlier than in structural mechanics as proposed herein. Herein, to the best knowledge of the authors, this is the first time, that a Trace FEM approach is applied to curved Reissner--Mindlin shells. Furthermore, the proposed approach is also higher-order accurate, thus enabling optimal higher-order convergence rates if the involved fields are sufficiently smooth. Other Trace FEM applications with particular emphasis on higher-order accuracy are rather scarce. If the geometry is defined by \emph{multiple} level-set functions, we refer to \cite{Fries_2019a} and for \emph{one} level-set function, we refer to \cite{Lehrenfeld_2016a,Jankuhn_2019a,Jankuhn_2020a}.\par

The Reissner--Mindlin shell model is suitable to model thin and moderately thick shells and as mentioned above the employed shell model need to be applicable on implicitly defined surfaces, where a parametrization of the midsurface is not existent. Therefore, a formulation of the shell equations in the frame of the TDC is employed herein. The recast of the linear shell equations in the frame of the TDC is presented by the authors in \cite{Schoellhammer_2019a}. The shell formulation is based on a standard difference vector approach. The tangentiality constraint on the difference vector is accomplished by a projection of a full 3D vector onto the tangent space combined with a consistent stabilization in normal direction.\par

The paper is organized as follows: In \autoref{sec:pre}, the implicit definition of shell geometries is described. Furthermore, the employed differential surface operators and other important quantities in the frame of the TDC are briefly introduced. In \autoref{sec:method}, the TDC-based Reissner--Mindlin shell model is outlined following \cite{Schoellhammer_2019a,Schoellhammer_2019c}. The resulting BVP is formulated in strong and weak form including boundary conditions. In \autoref{sec:tracefem}, the higher-order Trace FEM is introduced in detail. Furthermore, the implementational aspects of the Trace FEM, i.e., (i) integration of the weak form, (ii) stabilization and, (iii) enforcement of essential boundary conditions are elaborated. The discrete weak form of the equilibrium is introduced and the discretization of the difference vector is considered. In \autoref{sec:numres}, numerical results of the proposed approach are presented with a set of benchmark examples. The results confirm that optimal higher-order convergence rates are  achieved when the solution is sufficiently smooth. The paper ends in \autoref{sec:conc} with a summary and conclusions.
\section{Preliminaries}
\label{sec:pre}

\begin{figure}[ht]
	\centering\def\mywidth{.32}
	\subfloat[isosurfaces of $\phi$]{\includegraphics[width=\mywidth\textwidth]{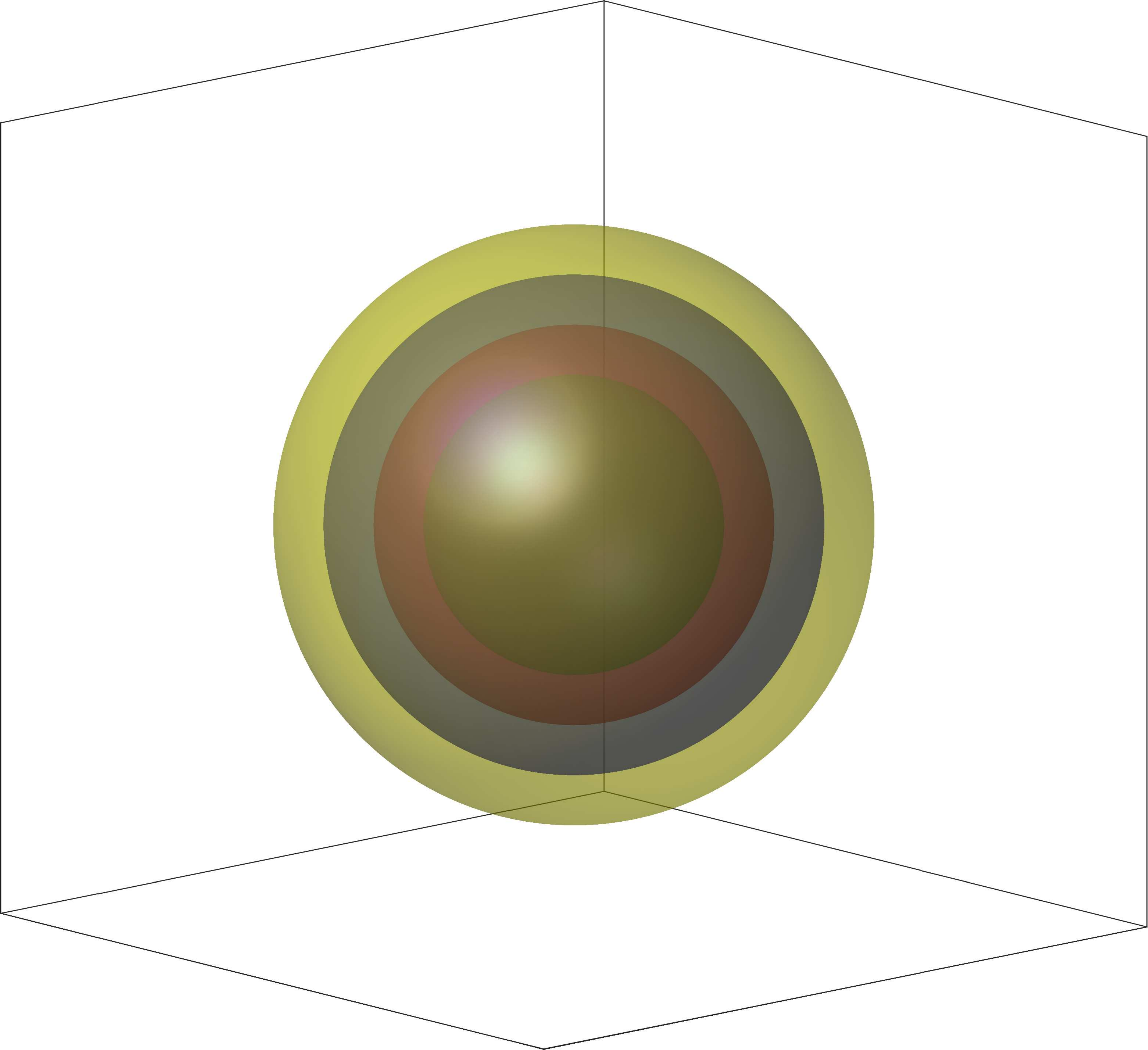}\label{fig:impgeom1}}
	\hfil
	\subfloat[implicit shell midsurface]{\includegraphics[width=\mywidth\textwidth]{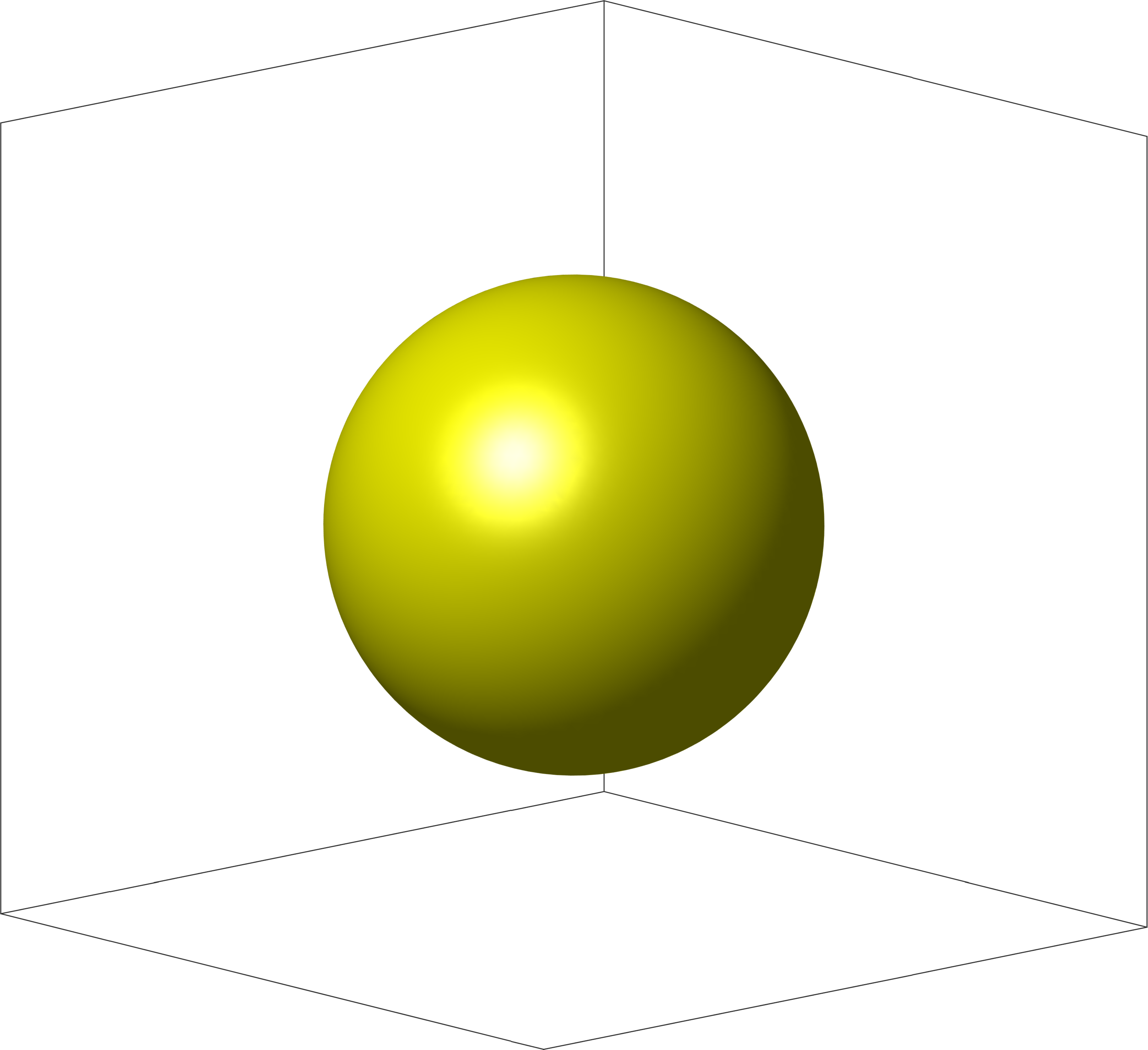}\label{fig:impgeom2}}\\
	\subfloat[some bounded shell midsurface]{\includegraphics[width=\mywidth\textwidth]{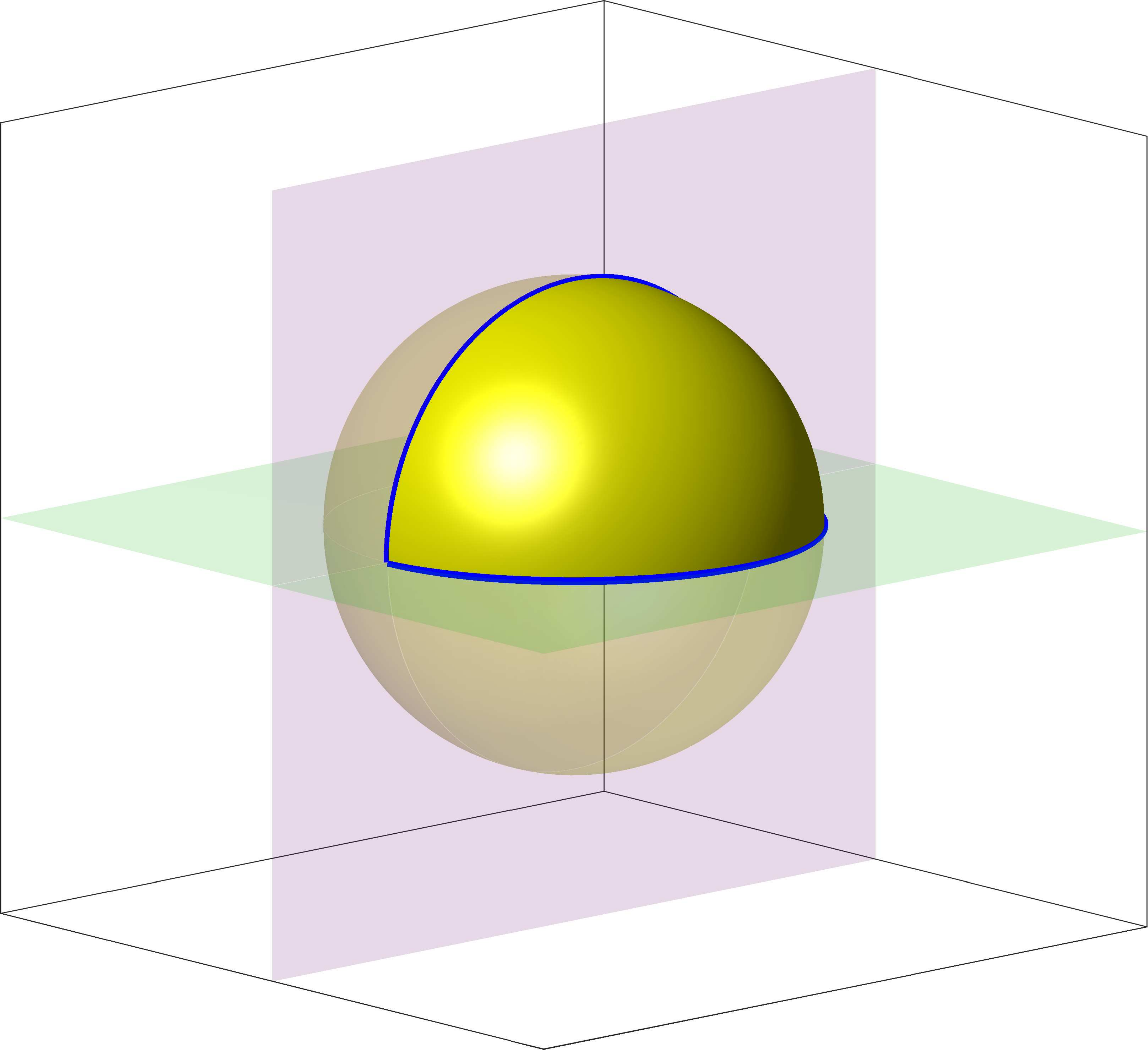}\label{fig:impgeom3}}
	\hfil
	\subfloat[normal vector and local triads]{\begin{overpic}[width=\mywidth\textwidth]{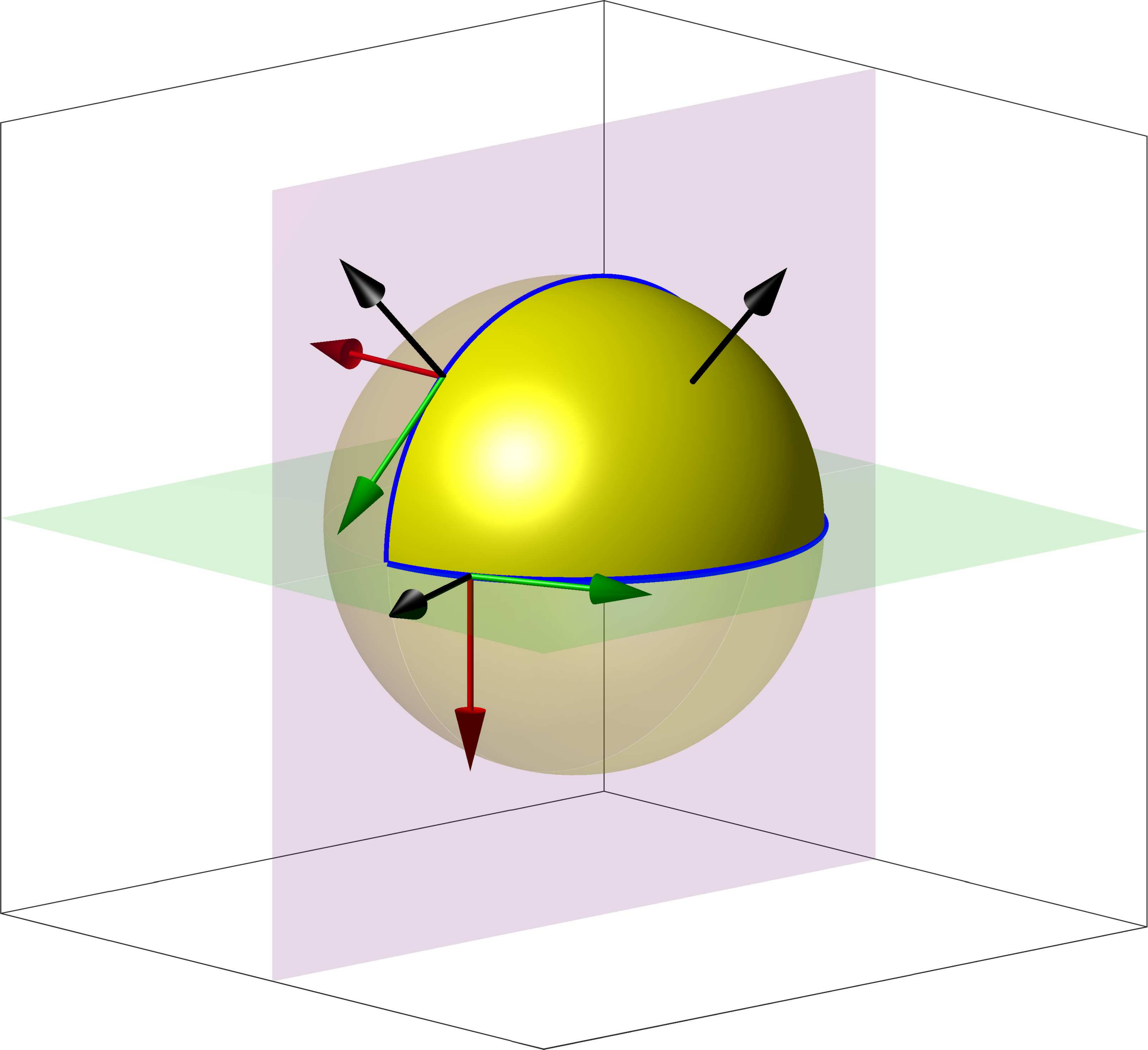}
			\put(69,63){\contour{white}{$\nG$}}
			\put(30,70){\contour{white}{$\nG$}}
			\put(16,58){\contour{white}{\color{myred}$\vek{n}_{\p\Gamma,1}$}}
			\put(17,47){\contour{white}{\color{mygreen}$\vek{t}_{\p\Gamma,1}$}}
			\put(30,32){\contour{white}{$\nG$}}
			\put(43,22){\contour{white}{\color{myred}$\vek{n}_{\p\Gamma,2}$}}
			\put(50,33){\contour{white}{\color{mygreen}$\vek{t}_{\p\Gamma,2}$}}
		\end{overpic}\label{fig:impgeom4}}
	\caption{Implicit definition of a (bounded) spherical shell: $\phi(\vek{x}) = \Vert\vek{x}\Vert - r, \psi_1(\vek{x}) = x, \psi_2(\vek{x}) = z$. (a) The colors are different isosurfaces of $\phi$, (b) implicit shell midsurface defined by the zero-isosurface of $\phi$, (c) definition of boundaries with additional slave level-set functions $\psi_1$ (purple) and $\psi_2$ (green), (d) normal (black), co-normal (red) and tangent (green) vectors.\label{fig:impgeom}}
\end{figure}
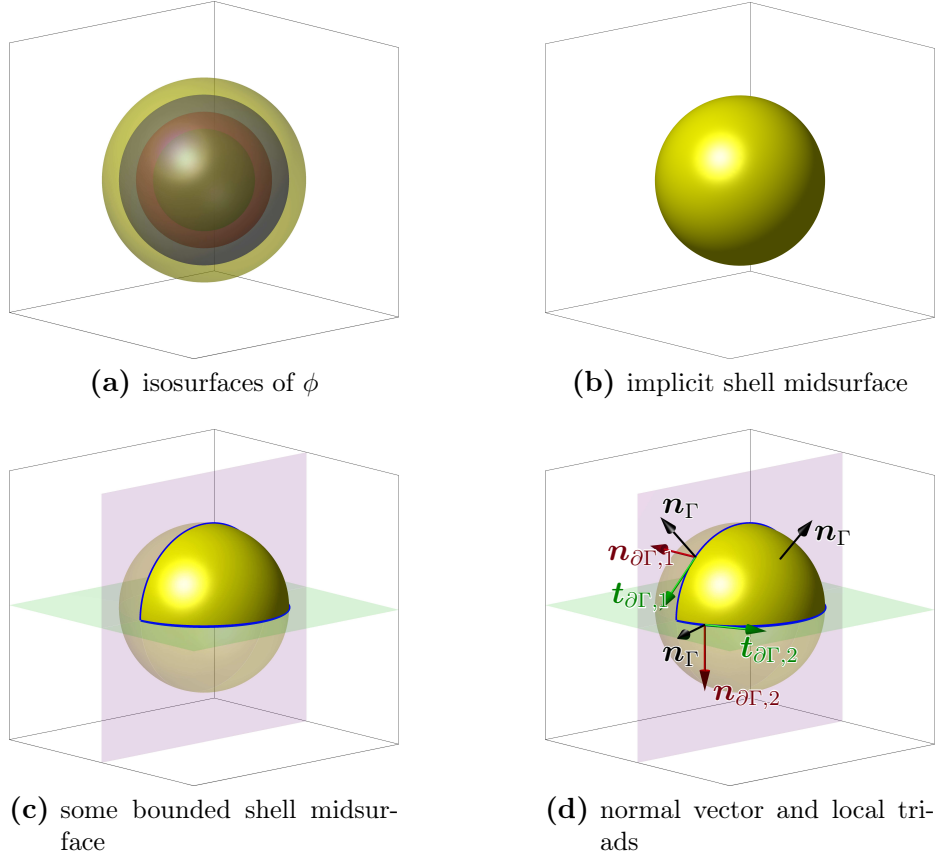

Shells are a thin-walled, possibly curved structures with thickness $t$. For the modelling, the 3D shell body $\Omega$ may be reduced to its midsurface $\Gamma$ embedded in the physical space $\mathbb{R}^3$. In general, the midsurface of a shell can be defined \emph{explicitly}, see, e.g., \cite{Bischoff_2017a,Basar_1985a}, or \emph{implicitly}, see, e.g., \cite{Schoellhammer_2018a,Schoellhammer_2019a,Fries_2017b}. Herein, the shell geometry is implicitly defined by means of (multiple) level-set functions. Let there be a master level-set function $\phi(\vek{x}):\mathbb{R}^3 \to \mathbb{R}$ whose zero-isosurface defines the (unbounded) midsurface of the shell in $\mathbb{R}^3$, see \autoref{fig:impgeom2}. The boundaries of the shell $\p\Gamma_i$ are defined by of additional \emph{slave} level-set functions $\psi_i$ with $i = 1,\ldots,n_{\t{Slaves}}$, see \autoref{fig:impgeom3}, where the blue lines indicate the boundary of the shell $\p\Gamma$. For the sake of simplicity, all slave level-set functions shall feature the same orientation, i.e., positive inside the domain and negative outside. The bounded midsurface $\Gamma$ and the boundaries $\p\Gamma_i$ of the shell are then defined by
\begin{align}
\Gamma &:= \left\lbrace \phi(\vek{x}) = 0\ \cap\ \psi_i(\vek{x}) > 0\ \forall\ i : \ \forall\  \vek{x} \in \mathbb{R}^3 \right\rbrace \ ,\\
\p\Gamma_i &:=  \left\lbrace \phi(\vek{x}) = 0\ \cap\ \psi_i(\vek{x}) = 0\ \forall\  \vek{x} \in \mathbb{R}^3 \right\rbrace \ ,
\end{align}
where the union of all boundaries define $\p\Gamma := \underset{i}{\cup}\,\p\Gamma_i$. Slave level-set functions are not necessarily needed when the master-level set function is restricted to some bounded domain of definition $\Omega^\ast \subset \mathbb{R}^3$ rather than $\mathbb{R}^3$. Both variants for the implicit definition of the bounded shell geometry work equally well in the method proposed herein.\par

The normal vector of the shell is given through the normalized gradient of the master level-set function $\phi$
\begin{align}
\nG(\vek{x}) = \dfrac{\nabla\phi(\vek{x})}{\Vert \nabla\phi(\vek{x}) \Vert} \ .
\end{align}
Along the boundaries $\p\Gamma_i$, there is an associated tangent vector $\vek{t}_{\p\Gamma,i}$ which is defined by the normalized cross product of the corresponding slave level-set function $\psi_i$ and the master level-set function $\phi$
\begin{align}
\vek{t}_{\p\Gamma,i}(\vek{x}) = \dfrac{\psi_i(\vek{x}) \times \phi(\vek{x})}{\Vert \psi_i(\vek{x}) \times \phi(\vek{x}) \Vert}\,, \quad \vek{x} \in \p\Gamma_i\ .
\end{align}
The orientation of the tangent vector $\vek{t}_{\p\Gamma,i}$ is defined with the orientation of the slave level-set function $\psi_i$. Furthermore, the co-normal vector $\vek{n}_{\p\Gamma}$ at $\p\Gamma$ pointing ``outwards'' and being perpendicular to the boundary yet in the tangent plane $T_p\Gamma$ is
\begin{align}
\vek{n}_{\p\Gamma,i}(\vek{x}) = \vek{t}_{\p\Gamma,i}(\vek{x}) \times \nG(\vek{x})\,, \quad \vek{x} \in \p\Gamma_i\ ,
\end{align}
see \autoref{fig:impgeom4}, where the normal vector $\nG$ and the local triad $(\nG,\vek{n}_{\p\Gamma,i},\vek{t}_{\p\Gamma,i})$ at the boundaries are visualized.

\subsection{Tangential differential calculus (TDC)}
\label{sec:tdc}

The tangential differential calculus (TDC) provides a framework to define surface operators independently of the concrete surface definition. This is an important generalization compared to classical models based on curvilinear coordinates which rely on parametrizations, i.e., explicit geometry definitions. In the following, {\revStart the employed geometrical and differential operators in the frame of the TDC are briefly introduced.\revEnd} For a more detailed information and derivation we refer to, e.g., \cite{Fries_2019a,Schoellhammer_2019a,Schoellhammer_2019b,Delfour_2011a}.\par

{\revStart On the manifold $\Gamma$, there are two projection operators $\mat{Q}(\vek x) = \nG(\vek x)\otimes\nG(\vek x)$ and $\mat P(\vek x) =\mathbb{I} - \mat{Q}$, where $\mat{Q}$ projects an arbitrary vector $\vek{v}(\Gamma) \in \mathbb{R}^3$ in normal direction of $\Gamma$ and $\mat{P}$ projects an arbitrary vector onto the tangent space $T_P\Gamma$ of $\Gamma$. \revEnd}\par


The tangential gradient operator $\nabla_{\Gamma}$ of a differentiable
\emph{scalar}-valued function $u:\Gamma\to\mathbb{R}$ on the surface is given by
\begin{align}
\gradG{u(\vek x)} & = \mat P\left(\vek x\right)\cdot\nabla\tilde{u}\left(\vek x\right)\ , \quad \vek{x} \in \Gamma\label{eq:TangGradImplicit}
\end{align}
where $\nabla$ is the standard gradient operator, and $\tilde{u}$ is an arbitrarily, but sufficiently smooth extension of $u$ in a neighborhood $\mathcal{U}$ of the manifold $\Gamma$. In the context of  the Trace FEM, the function $\tilde{u}$ is naturally available because all quantities are defined in the higher-dimensional space, i.e.,  $\mathbb{R}^3$ or $\Omega^\ast$. Therefore, $\tilde{u}$ is scalar-valued function in $\mathbb{R}^3$, i.e., $\tilde{u}:\mathbb{R}^3 \to \mathbb{R}$ and the function $u$ is defined by the restriction of $\tilde{u}$ to $\Gamma$,  i.e., $u := \tilde{u}_{\rvert_{\Gamma}}$ which means that $\tilde{u}$ is only evaluated on $\Gamma$.\par

{\revStart  Applying the tangential gradient operator $\nabla_{\Gamma}$ to each component of a vector-valued function $\vek u\left(\vek x\right):\Gamma\to\mathbb{R}^{3}$, gives the directional gradient of $\vek u$
\begin{align*}	\gradGD{\vek{u}\left(\vek x\right)} &= \gradGD{\begin{bmatrix}
	u(\vek x)\\
	v(\vek x)\\
	w(\vek x)
	\end{bmatrix}}=\begin{bmatrix}
(\gradG{u})^\T\\
(\gradG{v})^\T\\
(\gradG{w})^\T
\end{bmatrix}=\nabla\tilde{\vek u}\cdot\mat P\ ,
\end{align*}	
whereas the covariant gradient is defined by the projection of the directional gradient onto the tangent space, i.e., $\nabla_{\Gamma}^{\mathrm{cov}}\vek u\left(\vek x\right) = \mat P\cdot\gradGD{\vek u\left(\vek x\right)}$. The surface gradients of second-order tensor functions are defined accordingly.
\revEnd}\par

Concerning the surface divergence of vector-valued functions $\vek u\left(\vek x\right):\Gamma\to\mathbb{R}^{3}$
and tensor-valued functions $\mat A\left(\vek x\right):\Gamma\to\mathbb{R}^{3\times3}$,
there holds 
\begin{align*}
\divG{\vek u\left(\vek x\right)} & = \divG{(u,v,w)} =  \mathrm{tr}\left(\gradGD{
	\vek u}\right)=\mathrm{tr}\left(\gradGC{\vek u}\right) ,\\
\divG{\mat A\left(\vek x\right)} & =  \begin{bmatrix}
\divG{(A_{11},A_{12},A_{13})}\\
\divG{(A_{21},A_{22},A_{23})}\\
\divG{(A_{31},A_{32},A_{33})}
\end{bmatrix}\ .
\end{align*}
{\revStart Lastly, the surface gradient of the normal vector $\nG$ yields the Weingarten map $\mat{H} = \gradGD{\nG} = \gradGC{\nG}$ \cite{Delfour_1995a,Delfour_1996a,Jankuhn_2017a,Schoellhammer_2019a}. The Weingarten map is a symmetric in-plane tensor and its two non-zero eigenvalues are the principal curvatures $\kappa_{1,2} = - \t{eig}(\mat{H})$.  \revEnd}

\section{Governing equations}
\label{sec:method}

\subsection{Reissner--Mindlin shells}
\label{sec:rmshells}

The shell is implicitly defined by means of level-set functions. Therefore, it is a crucial requirement that the shell theory extends to this situation and the obtained boundary value problem is well-defined also on implicitly defined geometries, where a parametrization of the midsurface does not exist. Herein, the linear Reissner--Mindlin shell theory {\revStart formulated \revEnd} in the frame of the TDC, as presented in \cite{Schoellhammer_2019a,Schoellhammer_2019c}, {\revStart is employed \revEnd}. The main advantage is that the used surface operators are independent of the concrete surface definition. Therefore, the usage of this shell theory in the frame of the TDC is an appealing choice. In the following, the shell equations from \cite{Schoellhammer_2019a} are briefly recalled.\par

\begin{figure}[h]\centering
	
			\ifdefined\bUseTikzExternalize
				\def\tkzscale{0.275}
				\centering
				\tikzsetnextfilename{Fig_4}  
				\input{tikz/Fig_4}
			\else
				\includegraphics[width=.8\textwidth]{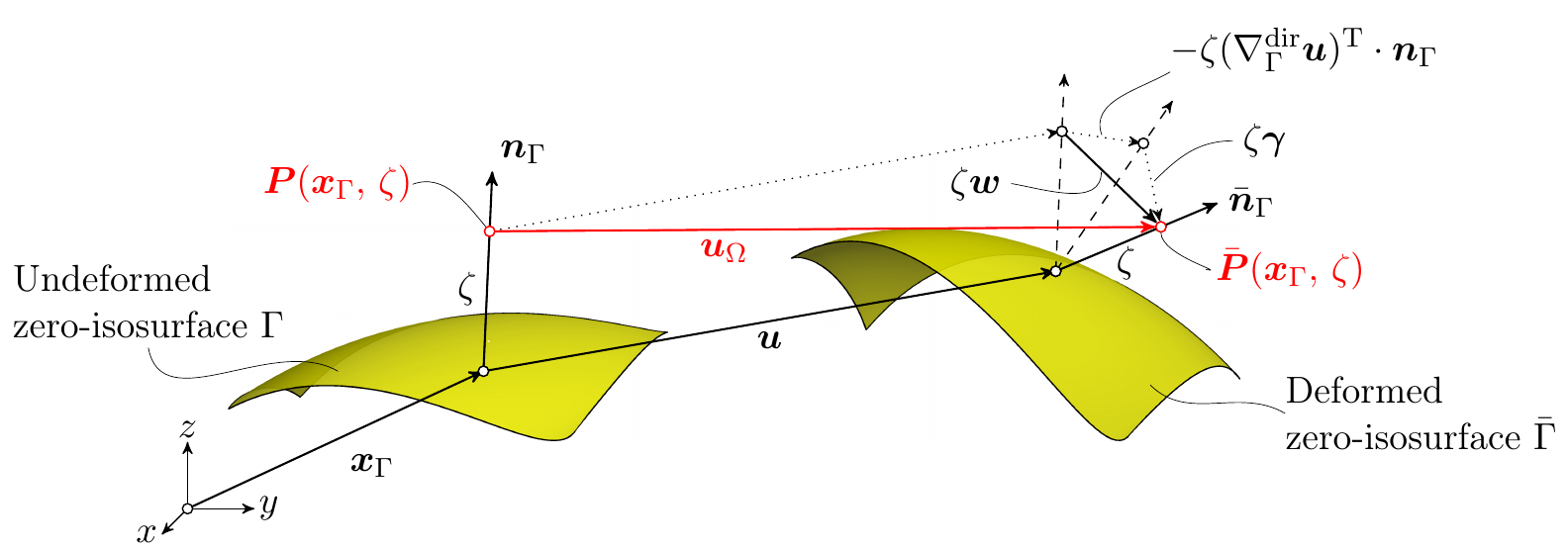}
			\fi			

	\caption{Displacement field $\vek{u}_\Omega$ of the Reissner-Mindlin shell.}
	\label{fig:rmshelldisp}
\end{figure}

In \cite{Schoellhammer_2019a}, the displacement field $\vek{u}_{\Omega}(\vek{x})$ is decomposed into the displacement of the midsurface $\vek{u}$ and the rotation of the normal vector is modelled with a difference vector approach $\vek{w}$, see \autoref{fig:rmshelldisp}. The total rotation of the normal vector is split into two parts: (1) due to bending of the midsurface, (2) and due to transverse shear deformations $\vek{\gamma}$. The coordinate in thickness direction is labelled $\zeta$ and provided that the master level-set function is a signed-distance function, $\zeta = \phi(\vek{x})$. Note that the difference vector is a tangential vector.

Based on the displacement field $\vek{u}_{\Omega}$, one may obtain the strain tensor $\ten{\varepsilon}_\Gamma$ by computing the symmetric part of the surface gradient $\gradGD{\vek{u}_{\Omega}}$. The strain tensor $\ten{\varepsilon}_\Gamma$ is split into in-plane strains $\ten{\varepsilon}_\Gamma^\t{P}$ and transverse shear strains $\ten{\varepsilon}_\Gamma^\t{S}$ using the projectors from \autoref{sec:tdc}
\begin{align}
\ten{\varepsilon}_\Gamma(\vek{x}) &= \ten{\varepsilon}_\Gamma^\t{P}(\vek{x}) + \ten{\varepsilon}_\Gamma^\t{S}(\vek{x})\ ,
\intertext{with}
\ten{\varepsilon}_\Gamma^\t{P}(\vek{x}) &= \mat{P} \cdot \ten{\varepsilon}_\Gamma \cdot \mat{P} = \ten{\varepsilon}_{\Gamma,\,\t{Mem}}^\t{P} + \zeta\, \ten{\varepsilon}_{\Gamma,\,\t{Bend}}^\t{P}\ ,\\
\ten{\varepsilon}_\Gamma^\t{S}(\vek{x}) &= \mat{Q} \cdot \ten{\varepsilon}_\Gamma + \ten{\varepsilon}_\Gamma \cdot \mat{Q}\ .
\end{align}
The in-plane strains are further decomposed into membrane strains $\ten{\varepsilon}_{\Gamma,\,\t{Mem}}^\t{P}$ and bending strains $\ten{\varepsilon}_{\Gamma,\,\t{Bend}}^\t{P}$. {\revStart In detail, the strain components are\revEnd}
\begin{align}
\ten{\varepsilon}_{\Gamma,\,\t{Mem}}^\t{P}(\vek{u}) &= \dfrac{1}{2} \left[\gradGC{\vek{u}} + (\gradGC{\vek{u}})^\T \right]\ ,\\
\ten{\varepsilon}_{\Gamma,\,\t{Bend}}^\t{P}(\vek{u}, \vek{w}) &= \dfrac{1}{2} \left[ \mat{H}\cdot\gradGD{\vek{u}} + (\gradGD{\vek{u}})^\T\cdot \mat{H} + \gradGC{\vek{w}} + (\gradGC{\vek{w}})^\T\right]\ , \\
\ten{\varepsilon}_\Gamma^\t{S}(\vek{u},\,\vek{w}) &= \dfrac{1}{2} \left[ \mat{Q} \cdot \gradGD{\vek{u}} + (\gradGD{\vek{u}})^\T \cdot \mat{Q} + \nG \otimes \vek{w} + \vek{w} \otimes \nG \right]\ .
\end{align}
For simplicity, we shall use a linear elastic material governed by Hooke's law with the modified Lam\'e constants $\mu = \frac{E}{2(1+\nu)}$, $\lambda = \frac{E\nu}{1-\nu^2}$ in order to eliminate the normal stress in thickness direction. Based on that, the linear stress tensor yields
\begin{align}
\ten{\sigma}_\Gamma(\vek{x}) = 2\mu\ten{\varepsilon}_\Gamma(\vek{x}) + \lambda\t{tr}[\ten{\varepsilon}_\Gamma(\vek{x})]\mathbb{I}\ .
\end{align}
A decomposition of the stress tensor in a similar manner than the strain tensor gives the membrane, bending and transverse shear stresses. With the assumption of a constant shifter in thickness direction, an analytical pre-integration w.r.t.~the thickness is possible and the stress resultants, such as moment tensor $\mat{m}_\Gamma$, effective normal force tensor $\tilde{\mat{n}}_\Gamma$ and transverse shear force tensor $\mat{q}_\Gamma$ are identified as{\revStart
\begin{align}
	\begin{split}
\mat{m}_\Gamma &= \int_{-\sfrac{t}{2}}^{\sfrac{t}{2}} \zeta\, \mat{P} \cdot \ten{\sigma}_\Gamma \cdot \mat{P}\ \d\zeta = \dfrac{t^3}{12} \ten{\sigma}_\Gamma^\t{P}(\ten{\varepsilon}^\t{P}_{\Gamma,\t{Bend}}) \ ,\\ 
\tilde{\mat{n}}_\Gamma &=\int_{-\sfrac{t}{2}}^{\sfrac{t}{2}}  \mat{P} \cdot \ten{\sigma}_\Gamma \cdot \mat{P}\ \d\zeta = t \ten{\sigma}_\Gamma^\t{P}(\ten{\varepsilon}^\t{P}_{\Gamma,\t{Mem}}) \ ,\\ 
\mat{q}_\Gamma &=\int_{-\sfrac{t}{2}}^{\sfrac{t}{2}} \mat{Q} \cdot \ten{\sigma}_\Gamma + \ten{\sigma}_\Gamma \cdot \mat{Q} \ \d\zeta = t \ten{\sigma}_\Gamma^\t{S}(\ten{\varepsilon}^\t{S}_{\Gamma}) \ .
\end{split}
\end{align}
\revEnd} The moment and effective normal force tensors are symmetric, in-plane tensors and their two non-zero eigenvalues are the principal moments or forces, respectively. Note that in the case of curved shells, the physical normal force tensor is $\mat{n}^{\t{real}}_\Gamma = \tilde{\mat{n}}_\Gamma + \mat{H}\cdot\mat{m}_\Gamma$ and is, in general, not symmetric, but features one zero eigenvalue just as $\tilde{\mat{n}}_\Gamma$.\par

Based on the stress resultants, the force and moment equilibrium for an implicitly defined Reissner--Mindlin shell in strong form becomes
\begin{align}
\divG{\mat{n}^{\t{real}}_\Gamma} + \mat{Q}\cdot\divG{\mat{q}_\Gamma} + \mat{H}\cdot(\mat{q}_\Gamma\cdot\nG) &= - \vek{f}\ , \label{eq:sff}\\
\mat{P}\cdot\divG{\mat{m}_ {\Gamma}} - \mat{q}_\Gamma \cdot\nG &= -\vek{c}\ , \label{eq:sfm}
\end{align} 
where $\vek{f} \in \mathbb{R}^3$ is the load vector per area and $\vek{c} \in T_P\Gamma$ is a distributed moment vector on the zero-isosurface $\Gamma$.
The equilibriums in strong form are a set of second-order surface PDEs and with suitable boundary conditions, the complete boundary value problem (BVP) of an \emph{implicitly} defined shell is defined.\par

For each field $\vek{u}$ and $\vek{w}$ there exist two non-overlapping parts at the boundary of the shell $\p\Gamma$. In particular, the Dirichlet boundary $\p\Gamma_\t{D,i}$ and the Neumann boundary $\p\Gamma_\t{N,i}$, with $i = \lbrace\vek{u},\,\vek{w}\rbrace$. The corresponding boundary conditions are
\begin{align}
\begin{alignedat}{2}\label{eq:pdebcs}
\vek{u} &= \hat{\vek{g}}_{\vek{u}} &&\t{ on } \p\Gamma_{\t{D},\vek{u}}\ , \\
\mat{n}^{\t{real}}_\Gamma \cdot \nCo + (\nG\cdot\mat{q}_\Gamma\cdot\nCo)\cdot\nG &= \hat{\vek{p}} &&\t{ on } \p\Gamma_{\t{N},\vek{u}}\ ,\\
\vek{w} &= \hat{\vek{g}}_{\vek{w}} &&\t{ on } \p\Gamma_{\t{D},\vek{w}}\ , \\
\mat{m}_\Gamma \cdot \nCo &= \hat{\vek{m}}_{\p\Gamma} &&\t{ on } \p\Gamma_{\t{N},\vek{w}}\ ,
\end{alignedat}
\end{align}
where $\hat{\vek{g}}_{\vek{u}}$ are the displacements, $\hat{\vek{g}}_{\vek{w}}$ are the rotations of the normal vector, $\hat{\vek{p}}$ are the forces and $\hat{\vek{m}}_{\p\Gamma}$ are the bending moments at their corresponding part of the boundary. 
{\revStart For a more detailed discussion regarding the natural and essential boundary conditions, we refer to \cite{Schoellhammer_2019a}. \revEnd}

\subsection{Equilibrium in weak form}
\label{sec:cwf}
In order to convert the equilibrium in strong form to the weak form, we introduce the following function spaces
\begin{align} \label{eq:su}
\mathcal{S}_{\vek{u}} &= \left\lbrace \vek{v} \in \left[\mathcal{H}^1(\Gamma)\right]^3 : \vek{v} = \hat{\vek{g}}_{\vek{u}} \t{ on } \p\Gamma_{\t{D},\vek{u}}\right\rbrace\ ,\\
\mathcal{V}_{\vek{u}} &= \left\lbrace \vek{v} \in \left[\mathcal{H}^1(\Gamma)\right]^3 : \vek{v} = \vek{0} \t{ on } \p\Gamma_{\t{D},\vek{u}}\right\rbrace\ ,\\
\mathcal{S}_{\vek{w}} &= \left\lbrace \vek{v} \in \left[\mathcal{H}^1(\Gamma)\right]^3 : \vek{v} \cdot \nG = 0\ ;\ \vek{v} = \hat{\vek{g}}_{\vek{w}} \t{ on } \p\Gamma_{\t{D},\vek{w}}\right\rbrace\ ,\\
\mathcal{V}_{\vek{w}} &= \left\lbrace \vek{v} \in \left[\mathcal{H}^1(\Gamma)\right]^3 : \vek{v} \cdot \nG = 0\ ;\ \vek{v} = 0 \t{ on } \p\Gamma_{\t{D},\vek{w}}\right\rbrace\ , \label{eq:vw}
\end{align}
where $\mathcal{H}^1$ is the space of functions with square integrable first derivatives. Note that the functions are 3D functions on the midsurface, i.e, $\vek{v}(\vek{x}) : \Gamma \to \mathbb{R}^3,  \vek{x}\in\Gamma \subset \mathbb{R}^3$. {\revStart The non-tangential spaces $(\mathcal{S}_{\vek{u}}, \mathcal{V}_{\vek{u}})$ are employed for the trial and test functions of the midsurface displacement, whereas the tangential function spaces $(\mathcal{S}_{\vek{w}}, \mathcal{V}_{\vek{w}})$ are used for the trial and test functions of the difference vector, which need to be tangential according to the Reissner--Mindlin kinematics.\revEnd} Later on for the discrete problem in the frame of the Trace FEM, these functions are defined in the physical space $\mathbb{R}^3$ and then restricted to $\Gamma$, with the additional condition that the functions need to be in $[\mathcal{H}^1(\Gamma)]^3$. It is emphasized that the definition of the functions in the higher-dimensional space and then restricting them to the trace (midsurface) is one of the major differences compared to classical Surface FEM.\par

{\revStart With the above defined function spaces, see \autoref{eq:su} -  \autoref{eq:vw}, the  weak form of the equilibrium reads as follows: Given material parameters $(E,\nu) \in \mathbb{R}^{+}$, body forces $\vek{f} \in \mathbb{R}^3$ on $\Gamma$, tractions $\hat{\vek{p}}$ on $\p\Gamma_{\t{N},\vek{u}}$,\revEnd} find $\vek{u} \in \mathcal{S}_{\vek{u}}$ and $\vek{w} \in \mathcal{S}_{\vek{w}}$ such that for all $\vek{v}_{\vek{u}} \in \mathcal{V}_{\vek{u}}$, there holds	
\begin{align}
\begin{split}\label{eq:wff}
\int_{\Gamma} \gradGD{\vu} : \tilde{\mat{n}}_\Gamma + (\mat{H}\cdot\gradGD{\vu}):\mat{m}_\Gamma + (\mat{Q}\cdot\gradGD{\vu}):\mat{q}_\Gamma\ \d A &= \int_{\Gamma} \vu \cdot \vek{f}\ \d A + \int_{\p\Gamma_{\t{N},\vek{u}}} \vu \cdot \hat{\vek{p}}\ \d s\ .
\end{split}
\end{align}
The weak form of the moment equilibrium {\revStart reads as follows: Given material parameters $(E,\nu) \in \mathbb{R}^{+}$, distributed moments $\vek{c} \in T_P\Gamma$ on $\Gamma$, bending moments $\hat{\vek{m}}_{\p\Gamma}$ on $\p\Gamma_{\t{N},\vek{w}}$,\revEnd} find $\vek{u} \in \mathcal{S}_{\vek{u}}$ and $\vek{w} \in \mathcal{V}_{\vek{w}}$ such that for all $\vek{v}_{\vek{w}} \in \mathcal{V}_{\vek{w}}$, there holds
\begin{align}
\begin{split}\label{eq:wfm}
\int_\Gamma \gradGD{\vw}:\mat{m}_\Gamma + \vw \cdot (\mat{q}_\Gamma\cdot\nG)\ \d A &=  \int_{\Gamma} \vw\cdot\vek{c}\ \d A + \int_{\p\Gamma_{\t{N},\vek{w}}} \vw \cdot \hat{\vek{m}}_{\p\Gamma}\ \d s\ .
\end{split}
\end{align}
For further details regarding the derivation of the shell equations in the frame of the TDC and the used identities in order to obtain the equilibrium in weak form, we refer the interested reader to \cite{Schoellhammer_2019a}.
\section{Trace Finite Element Method}
\label{sec:tracefem}

The obtained weak form of the equilibrium is discretized with a trace finite element approach (Trace FEM), see, e.g., \cite{Olshanskii_2009a,Olshanskii_2009b,Reusken_2014a,Olshanskii_2017a,Gross_2018a,Grande_2018a,Schoellhammer_2019c}. This approach is a fictitious domain method for surface PDEs, where the implicitly defined domain of interest $\Gamma$ is completely immersed in a background domain $\Omega_{\t{B}} \in \mathbb{R}^3$ and a parametrization of $\Gamma$ is neither available nor needed. Let us first introduce fundamental terms and quantities, which are required in order to discretize a surface PDE with the Trace FEM. The main ingredients for the Trace FEM are the definition of (1) the background mesh, (2) the discrete domain of interest $(\Gamma^h,\p\Gamma^h)$, and (3) the Trace FEM function space.\par

Firstly, a background mesh $\Omega_{\t{B}}$, which completely immerses the domain of interest, i.e., the midsurface of the shell $\Gamma \in \Omega_{\t{B}}$, needs to be defined. Without loss of generality, the mesh can be defined by a set $\tau_{\t{B}}$ of 3D elements and is not restricted to a certain element type. Herein, the background mesh consists of tetrahedral elements $T$ of complete order $k \ge 1$. The background mesh is then defined by $\Omega_{\t{B}} := \underset{T \in \tau_h}{\cup} T \in C^0$. Associated to the reference element $\bar{T}$, there is a fixed set of basis functions $\lbrace N_i^k(\vek{r}) \rbrace$, with $i = 1 , \ldots, n_{\text{nodes}}$ being the number of nodes per element. The shape functions are Lagrange basis functions, i.e., $N_i^k(\vek{r_j}) = \delta_{ij}$ and $N_i^k(\vek{r}) \in \mathbb{P}^k(\bar{T})$, where $\mathbb{P}^k(\bar{T})$ is the polynomial basis for 3D tetrahedral Lagrange elements of complete order $k$. With an atlas of local, element-wise mappings from the 3D reference element to the 3D physical elements $\vek{x}(\vek{r}) : \bar{T} \to T$, with $\vek{x}(\vek{r}) = N^k_i(\vek{r}) \vek{x}_i$, $\vek{r} \in \bar{T}$, $\vek{x}_i \in T$, where $\vek{x}_i$ are the nodal coordinates, one may define shape functions $N_i^k(\vek{x})$ be means of the isoparametric concept. The union of all elements forms a global set of $C^0$-continuous basis functions $\lbrace M_l^k(\vek{x})\rbrace$ in $\Omega_{\t{B}}$ with $l = 1 ,\ldots, n$ and $n$ being the total number of nodes of the mesh. A general finite element space is then defined as
\begin{align}\label{eq:femspace}
\mathcal{Q}^k_{\Omega_{\t{B}},h} := \left\lbrace v_h \in C^0(\Omega_{\t{B}})\ \vert\ v_h = \sum_{l=1}^{n} M_l^k(\vek{x}) \hat{v}_l\ \vert\ \hat{v}_l \in \mathbb{R} \right\rbrace \subset \mathcal{H}^1(\Omega_{\t{B}})\ .
\end{align}
Secondly, the implicitly defined geometry of the shell is defined with a set of level-set functions, see \autoref{sec:pre}. The continuous level-set functions $(\phi,\psi_i)$ are interpolated with the shape functions $\lbrace M_l^k(\vek{X})\rbrace$ of the background mesh based on their nodal values, i.e., $\hat{\phi}_l = \phi(\vek{x}_l),\hat{\psi}_{j,l} = \psi_j(\vek{x}_l)$. This means that the level-set data is only needed at the nodes of the background mesh $\Omega_{\t{B}}$. {\revStart The discrete shell midsurface $\Gamma^h$ is then implied by $\phi^h$ and the discrete boundaries of the shell $\p\Gamma^h$ may be defined either through the discrete slave level-set functions $\psi_i^h$ or the boundary of the background mesh.\revEnd} In the following, the discrete boundary is only defined with additional slave level-set functions, otherwise the overall approach would be limited to a boundary conforming background mesh. {\revStart The discrete normal vector $\nG^h$ and the discrete local triad at the boundaries $(\nG^h, \nCo^h, \tB^h)$ are evaluated on the discrete zero-isosurface of $\phi^h$. For the computation of the normal vector and local triads, one may employ the exact gradients of the level-set functions $(\phi,\psi_i)$ or, alternatively, the interpolated level-set functions $(\phi^h,\psi_i^h)$. Herein, the latter approach based on the interpolated level-set data is used, resulting in
\begin{align*}
& \nG^h(\vek{x}) = \dfrac{\nabla\phi^h(\vek{x})}{\Vert \nabla\phi^h(\vek{x}) \Vert} \,,
&&\vek{t}_{\p\Gamma,i}^h(\vek{x}) = \dfrac{\psi_i^h(\vek{x}) \times \phi^h(\vek{x})}{\Vert \psi_i^h(\vek{x}) \times \phi^h(\vek{x}) \Vert}\,,
&&&\vek{n}_{\p\Gamma,i}^h = \vek{t}^h_{\p\Gamma,i}(\vek{x}) \times \nG^h(\vek{x})\,,
\end{align*}
where $\vek{x} \in \Gamma^h \cup \p\Gamma^h$. The projectors $\mat{P}$ and $\mat{Q}$ are then computed by means of the discrete normal vector $\nG^h$. It is clear that the gradients of $(\phi^h,\psi_i^h)$ are not exact and an additional source of error is added. However, the level-set functions are interpolated using quasi-uniform, higher-order background elements and it is shown in the numerical results in \autoref{sec:numres} that optimal convergence rates are achieved. Thus, the additional error caused by employing interpolated level-set data is often acceptable.\revEnd}\par

The set of elements with a non-empty intersection with $\Gamma^h$ is denoted by $\tau_{\Omega,h}^\Gamma$ and defines the active mesh $\Omega^\Gamma_h := \underset{T\in\tau_{\Omega,h}^\Gamma}{\cup} T$. The definition of the active mesh is a crucial task, because the nodes of the \emph{active} mesh imply the degrees of freedom in the numerical simulation.\par

Lastly, the Trace FEM function space $\mathcal{T}_{h}$ is established by the restriction of a ``higher-dimensional finite element space'' the \emph{active} mesh $\Omega^\Gamma_h$. The finite element space of the active mesh $\mathcal{Q}^k_{\Omega^\Gamma_h}$ is defined in a similar manner as $\mathcal{Q}^k_{\Omega_{\t{B}},h}$, but contains only cut background elements. A general Trace FEM function space $\mathcal{T}_{h}$ is then defined by
\begin{align}\label{eq:tracefemspace}
\mathcal{T}_{h} = \left\lbrace v \in \mathcal{Q}^k_{\Omega^\Gamma_h}\ : \ v\rvert_{\Gamma^h} \in \mathcal{H}^1(\Gamma^h) \right\rbrace\ \subset \mathcal{H}^1(\Omega_{\t{B}})\ . 
\end{align}

\subsection{Implementational aspects of the Trace FEM}
\label{sec:fdmcha}
As mentioned above, the Trace FEM is a fictitious domain method and, compared to standard Surface finite element approaches, three well-known challenges arise which are further outlined below: (i) integration of the weak form on the discrete zero-isosurface and its boundaries, (ii) stabilization of the stiffness matrix due to the restriction of the shape functions to the trace and small supports due to unfavourable cut scenarios and (iii) the enforcement of essential boundary conditions. In the following, these challenges are addressed in detail enabling a higher-order Trace FEM approach.\par

\subsubsection{Higher-order accurate integration of the discrete zero-isosurface}
\label{sec:int}
The numerical integration of the domain of interest, i.e., on the zero-isosurface of $\phi$, is a non-trivial task, in particular with higher-order accuracy. One approach which is suitable for higher-order is presented in \cite{Lehrenfeld_2016a}. The numerical integration is based on a higher-order accurate lift of a linear reconstruction of the zero-isosurface. However, the extension to multiple level-set functions, where the zero-isosurface of the master level-set function is restricted with additional slave-level-set functions has not been addressed so far.\par

\begin{figure}[h]\centering\def\mywidth{.32}
	\subfloat[active background elements]{\includegraphics[width=\mywidth\textwidth]{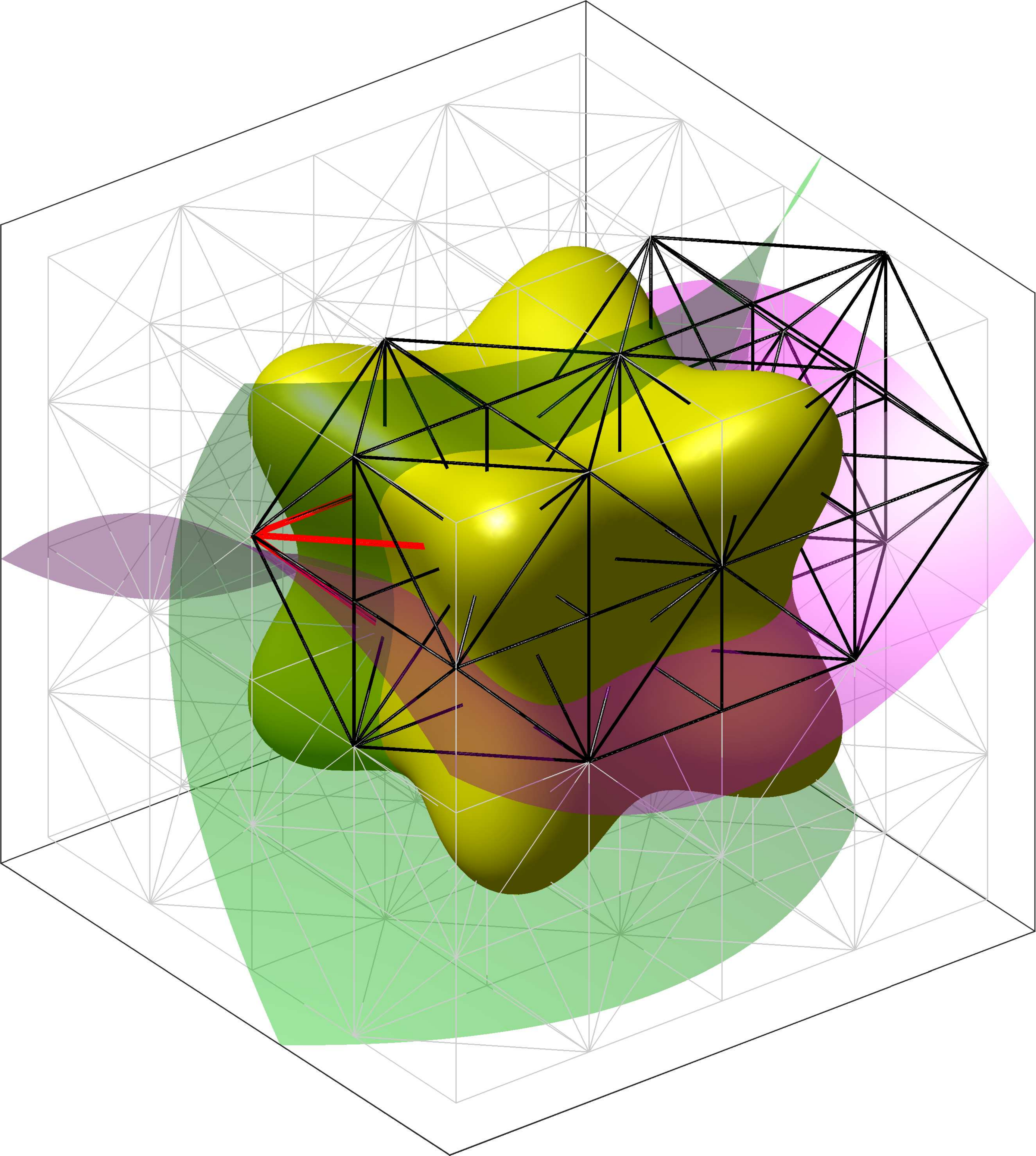}\label{fig:visint1}}
	\hfil
	\subfloat[integration points in reference element]{\includegraphics[width=\mywidth\textwidth]{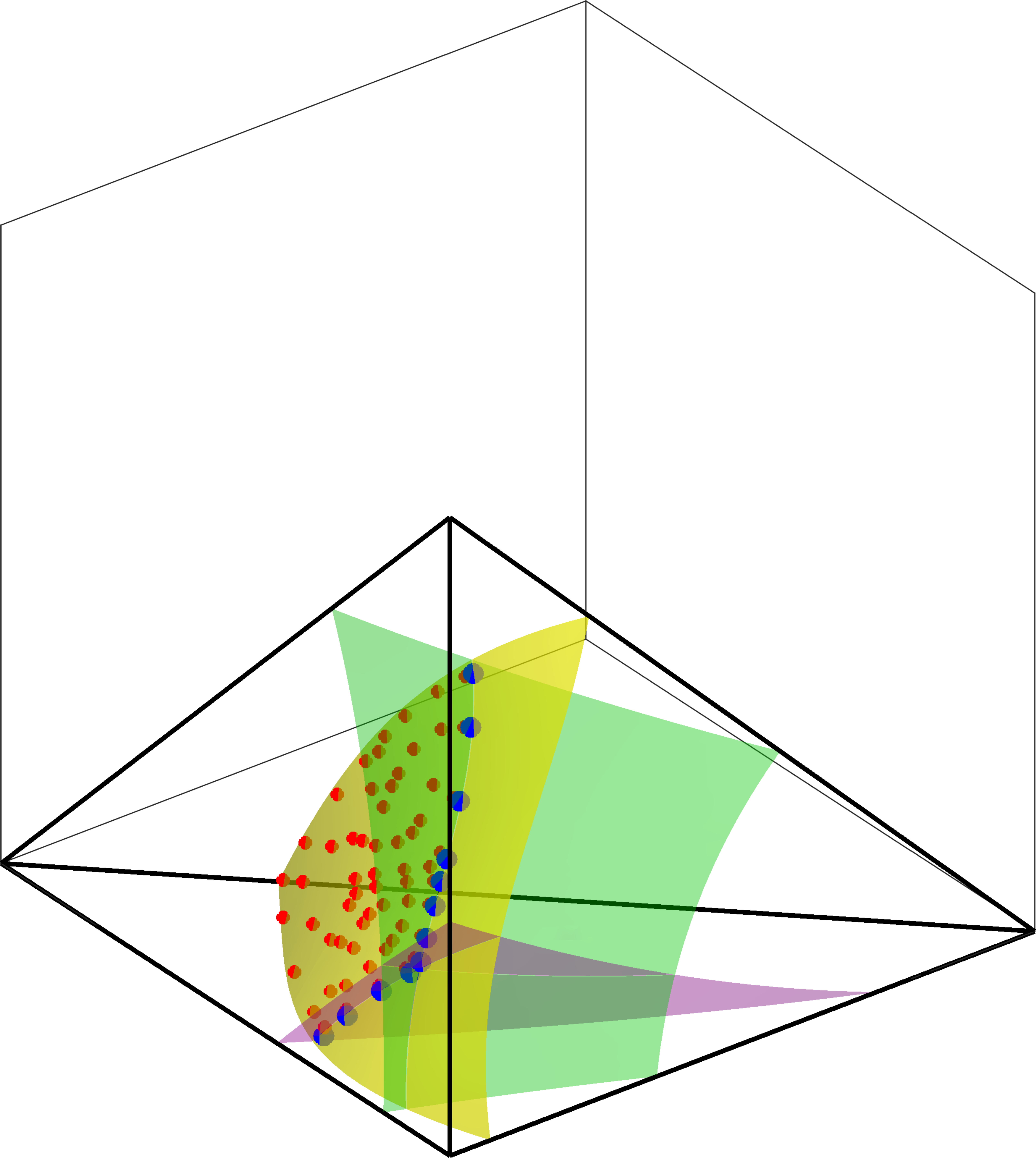}\label{fig:visint2}}
	\hfil
	\subfloat[integration points in physical elements]{\includegraphics[width=\mywidth\textwidth]{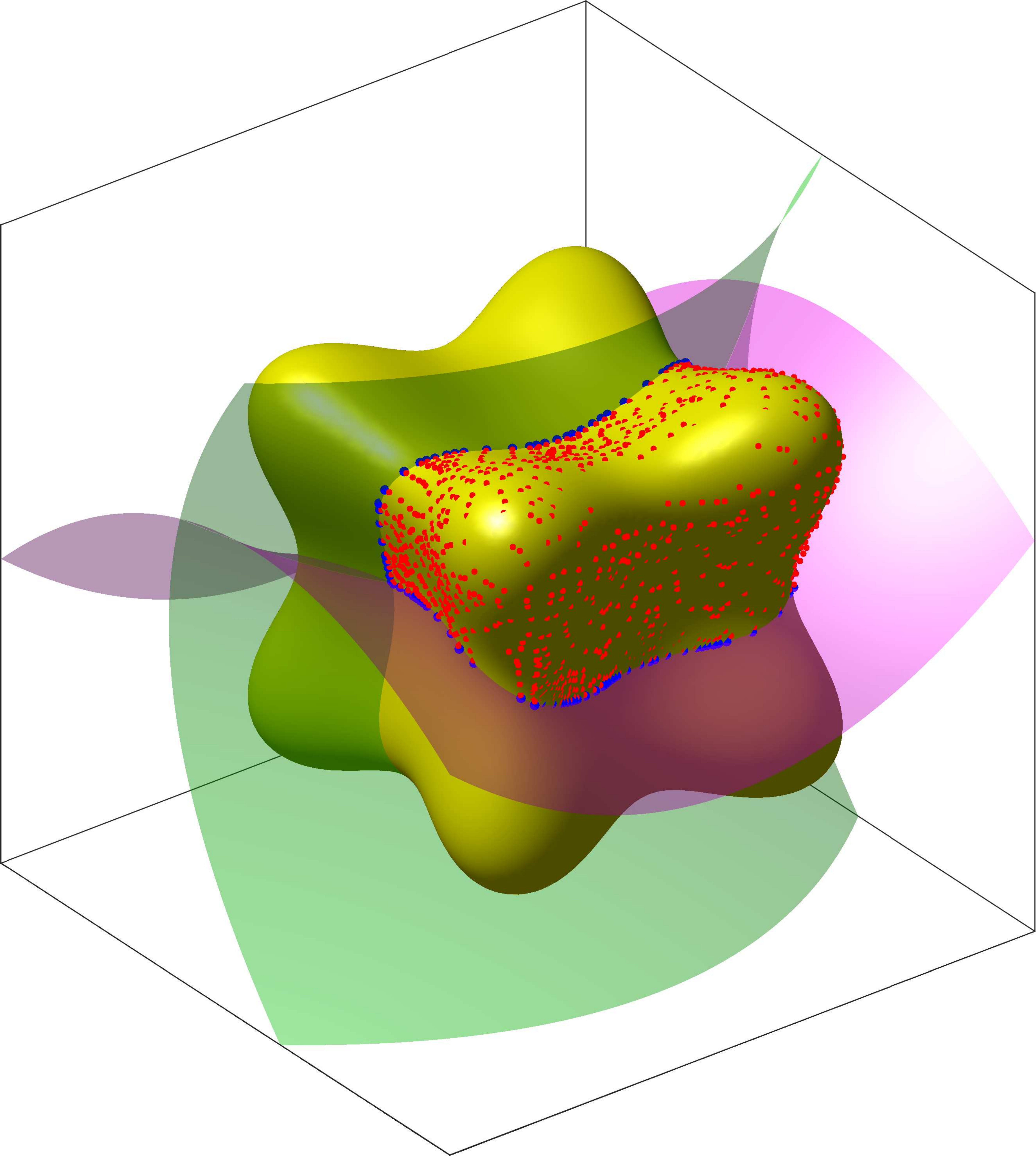}\label{fig:visint3}}
	\caption{ (a) Active, (black) elements in a background mesh are intersected by the shell midsurface, (b) integration points in one 3D reference element based on a (recursive) decomposition w.r.t.~the master level-set function $\phi^h$ and further restriction to the slave level-set functions $\psi_i^h$, (c) integration points in the physical domain (red points) and integration points on the shell boundary (blue points).}\label{fig:integration}
\end{figure}

Herein, we employ the integration strategy outlined by the authors in \cite{Fries_2015a,Fries_2017a,Fries_2017b,Fries_2019a}. The advantages of this particular approach are an optimal higher-order accurate integration and a natural extension to multiple level-set functions. In this approach, the placement of the integration points on the discrete zero-isosurface is based on a higher-order, recursive reconstruction in the cut \emph{reference} element. That is, the shell midsurface is reconstructed by higher-order surface elements. It is important that the reconstructed surface element is \emph{only} used for the generation of integration points in the 3D reference element and may be interpreted as an integration cell. In \autoref{fig:integration}, an overview of the procedure is illustrated. The yellow surface is the implicitly defined zero-isosurface of $\phi^h$ and is restricted by additional slave level-set functions $\psi_i^h$ (green and purple surfaces). In \autoref{fig:visint2}, the cut scenario and integration points in the reference space of the red marked element from \autoref{fig:visint1} are visualized. In \autoref{fig:visint3}, the integration points in the physical domain (red) and on the boundaries (blue) are shown. For further information and details, we refer to \cite{Fries_2015a,Fries_2017a}.

\subsubsection{Stabilization}
\label{sec:stab}
\begin{figure}[h]\centering\def\mywidth{.32}
	\subfloat[overview]{\includegraphics[width=\mywidth\textwidth]{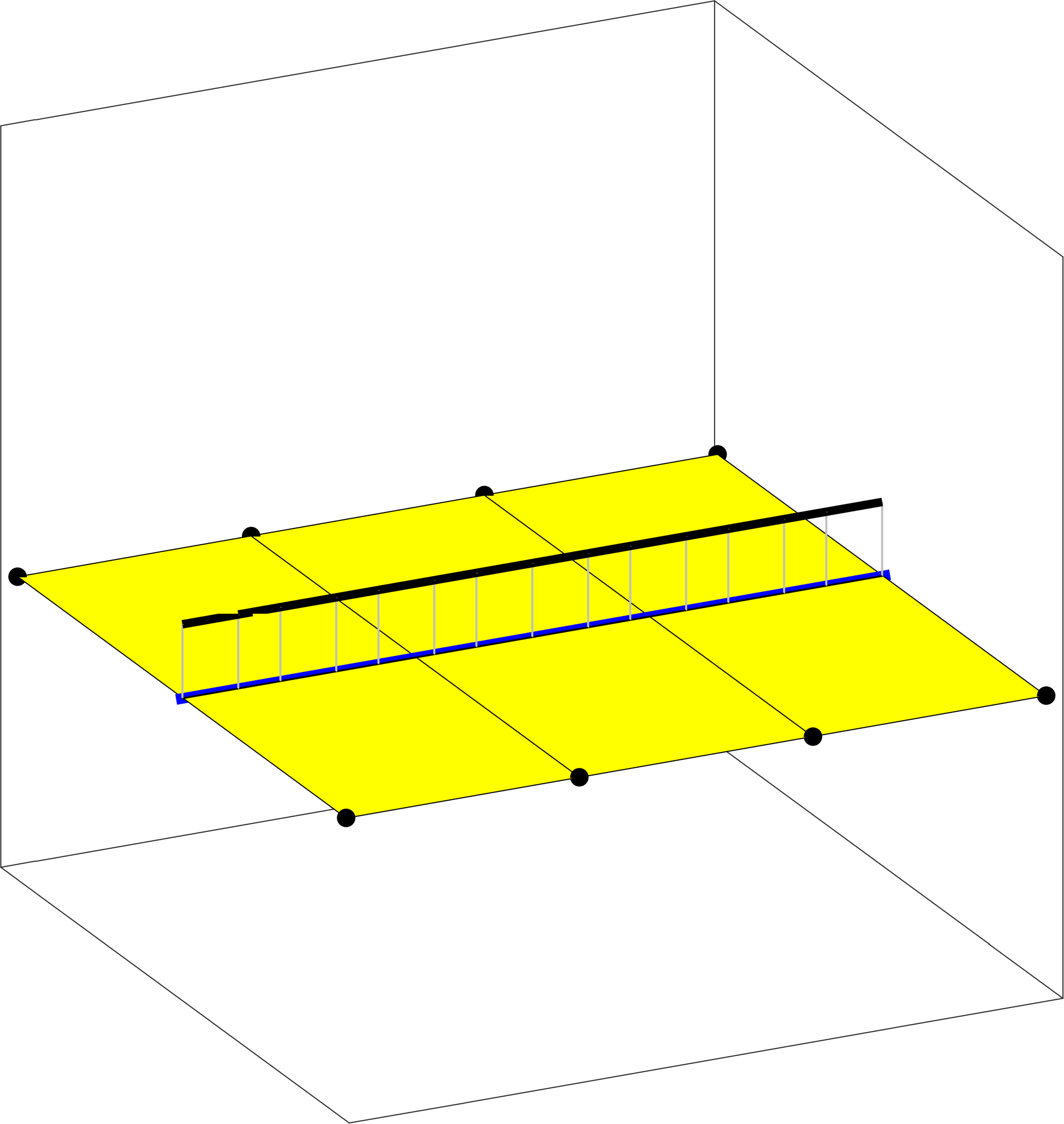}\label{fig:tracefemstaba}}
	\hfil
	\subfloat[interpolation]{\includegraphics[width=\mywidth\textwidth]{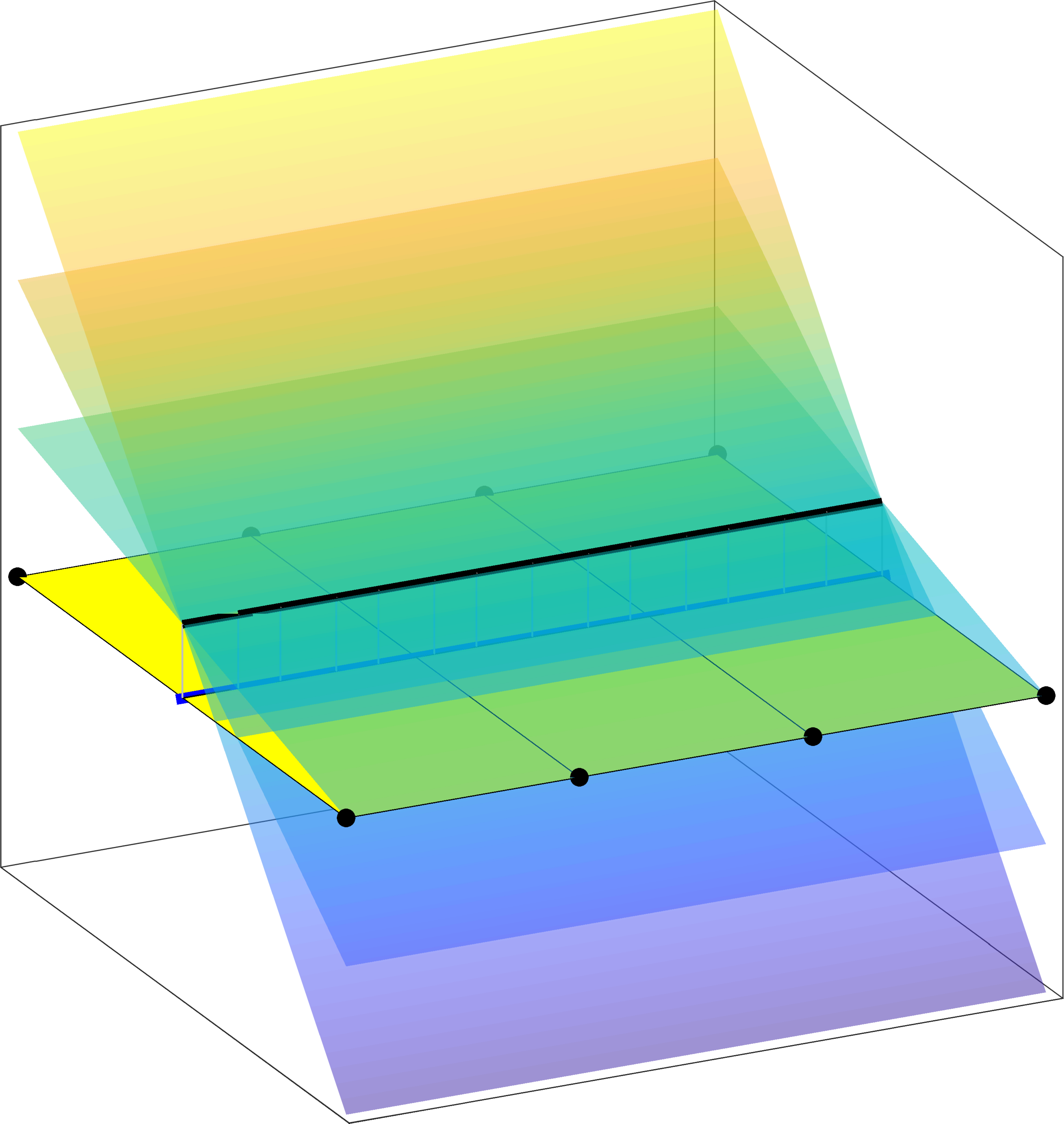}\label{fig:tracefemstabb}}
	\hfil
	\subfloat[stabilization]{\includegraphics[width=\mywidth\textwidth]{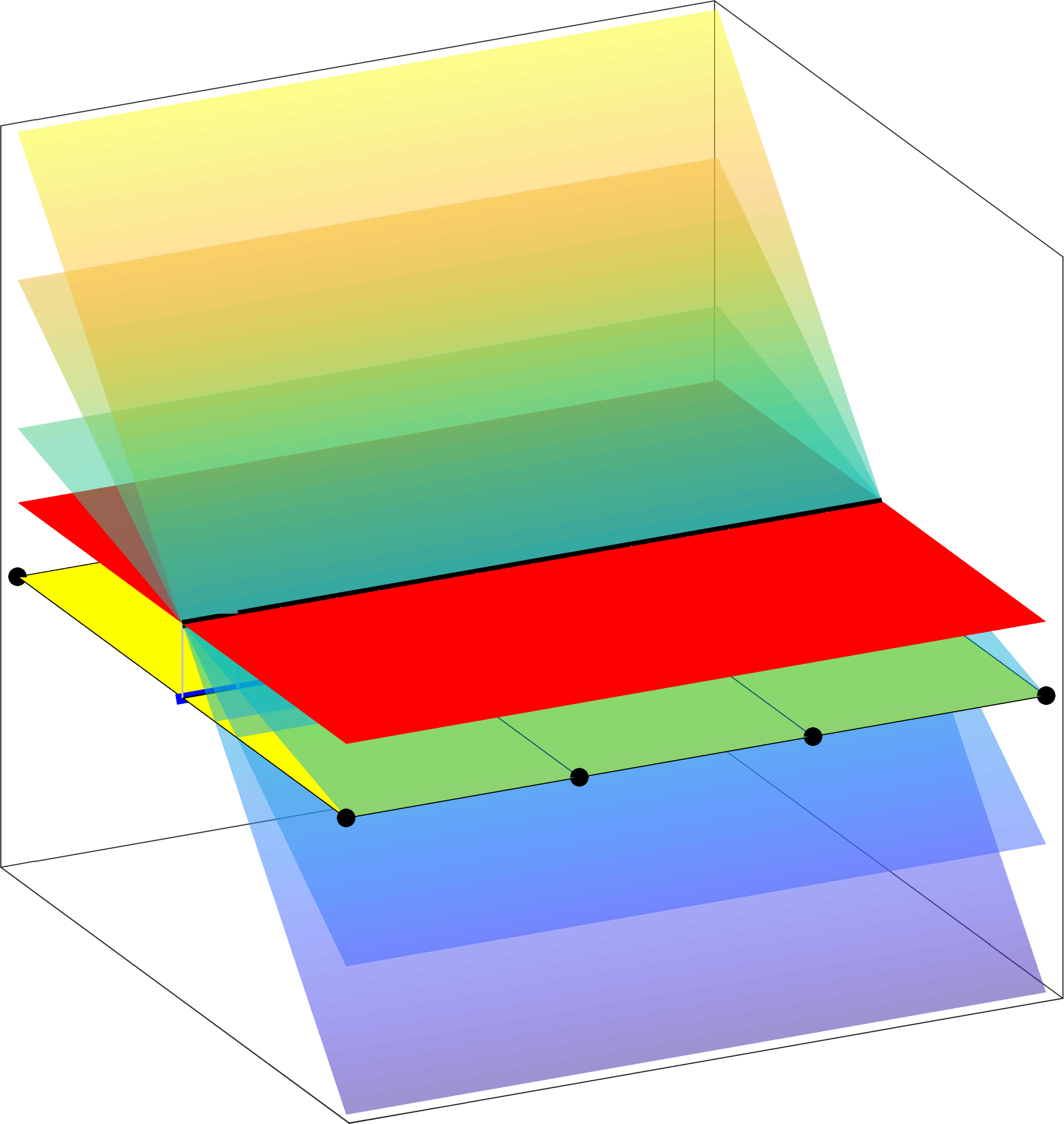}\label{fig:tracefemstabc}}
	\caption{Background meshes do not uniquely define a (constant) function on the zero-level set: (a) The black line is a constant, scalar-valued function on the manifold (blue line), which is embedded in three, bi-linear quadrilateral background elements, (b) different possibilities for the interpolation of the function on the manifold (trace), (c) the stabilization adds a constraint in normal direction of the manifold resulting in a unique interpolation (red surface).}\label{fig:tracefemstab}
\end{figure}

Due to the restriction of the shape functions to the trace, see \autoref{eq:tracefemspace}, the shape functions on the manifold only form a \emph{frame}, which is, in general, not a \emph{basis} \cite{Reusken_2014a,Olshanskii_2017a}. In \autoref{fig:tracefemstab}, the consequences of the restriction to the trace of the manifold is visualized for a simple example.\par

Let us consider a 1D manifold (blue line) embedded in three bi-linear, quadrilateral 2D elements, see \autoref{fig:tracefemstaba}. Furthermore, a constant function on the manifold (black line) shall be interpolated based on the nodal values of the active background mesh. As shown in \autoref{fig:tracefemstabb}, the choice of the nodal values in order to interpolate the function on the manifold is not unique. In particular, three different configurations, which share the same values on the manifold are visualized.
Therefore, a suitable stabilization term need to be added to the discrete weak form, otherwise the obtained linear system of equations does not have a unique solution w.r.t.~the nodal values. In addition to the restriction, depending on unfavourable cut scenarios of cut background elements, unbounded small contributions to the stiffness matrix may occur, which causes an ill-conditioned system of linear equations.\par

The used stabilization technique addresses both issues and is introduced for scalar-valued problems in \cite{Grande_2016a,Burman_2016b}. This stabilization technique is called ``normal derivative volume stabilization'' and in \cite{Gross_2018a,Olshanskii_2018a}, the stabilization is applied to vector-valued problems. The stabilization term added to each unknown field in the discrete weak form is
\begin{align} \label{eq:ndvstab}
s_h(\vek{u}^h,\vek{v}^h) := \rho \int_{\Omega^\Gamma_h} \left(\nabla\vek{u}^h\cdot\vek{n}_\Gamma^{e,h} \right) \cdot \left(\nabla\vek{v}^h\cdot\vek{n}_\Gamma^{e,h} \right)\ \d V\ ,
\end{align}
where $\vek{n}_\Gamma^{e,h}(\vek{x}) = \frac{\nabla\phi^h(\vek{x})}{\Vert \nabla\phi^h(\vek{x}) \Vert}, \vek{x} \in \Omega^\Gamma_h$ is a sufficiently smooth extension of the normal vector $\nG^h$ at the zero-isosurface of $\phi^h$. It is noteworthy, that the integral is performed over the whole active background mesh and not restricted to the trace. However, the integrand is sufficiently smooth and, therefore, a standard numerical integration scheme w.r.t.~the \emph{active} elements is applicable, i.e., a standard 3D Gauss rule. By adding this constraint to the linear system of equations, the resulting system of equations features a unique solution, see \autoref{fig:tracefemstabc}. It is recommended in \cite{Grande_2016a} that the stabilization parameter can be chosen within the following range
\begin{align}
h \lesssim \rho \lesssim h^{-1}\ ,
\end{align}
where $h$ is the element size of the elements from the active mesh. This stabilization technique is suitable for higher-order shape functions, does not change the sparsity pattern of the stiffness matrix, and only first-order derivatives are needed. In addition, the implementation is straightforward and the choice of the stabilization parameter is rather flexible. Other stabilization techniques are presented in \cite{Olshanskii_2017a,Burman_2016b}. {\revStart A recent approach where two stabilizations techniques, i.e., face stabilization of the cut elements and the normal derivative stabilization on the zero-isosurface, are combined is presented in \cite{Larson_2020a}.\revEnd}\par

\subsubsection{Essential boundary conditions}
\label{sec:ebc}
As outlined in \cite{Fries_2019a}, the enforcement of essential of boundary conditions is a challenging task in FDMs due to the fact that it is not possible to directly prescribe nodal values of the active background elements. The situation in the case of shells may be quite delicate due to complex boundary conditions, e.g., membrane support, symmetry support, clamped edges, etc., and, therefore, the treatment of boundary conditions requires special attention.\par

The essential (Dirichlet) boundary conditions may, in principle, be enforced in a weak manner using penalty methods, Lagrange multiplier methods, or Nitsche’s method. Herein, the non-symmetric version of Nitsche's method is used to enforce the Dirichlet boundary conditions \cite{Burman_2012a}. The advantage of this method in the context of FDMs is that it does not require additional stabilization terms and the discretization of auxiliary fields such as Lagrange multipliers is not needed. Furthermore, Nitsche's method is a consistent approach to enforce essential boundary conditions, which may be an advantage if higher-order convergence rates shall be achieved.  For further details about the non-symmetric version of Nitsche's method we refer to, e.g., \cite{Burman_2012a,Schillinger_2016a,Guo_2019a,Fries_2019a}. The additional terms in the discrete weak form resulting from Nitsche's method are presented in \autoref{sec:dwf}.

\subsection{Discretization of the Reissner--Mindlin shell}
\label{sec:drm}

In this section, the continuous weak form of the equilibrium, see \autoref{eq:wff} and \autoref{eq:wfm} is discretized with the Trace FEM as described above. The discrete function spaces for the trial and test functions of the midsurface displacement field are
\begin{align}
\mathcal{S}_{\vek{u}}^h &= \left\lbrace \vek{u}^h \in \left[\mathcal{T}_{h}\right]^3 \right\rbrace \ ,\\
\mathcal{V}_{\vek{u}}^h &= \left\lbrace \vu^h \in \left[\mathcal{T}_{h}\right]^3 \right\rbrace\ .
\end{align}
{\revStart For the stabilization of the discrete midsurface displacement, the above introduced normal derivative volume stabilization is employed. Regarding \revEnd}the discrete difference vector $\vek{w}^h$, the situation is more complicated due to the kinematic assumptions as the difference vector needs to be tangential. The discretization of tangent vector fields on implicitly defined manifolds is {\revStart not straightforward \revEnd}and detailed in \autoref{sec:discdiffvec}.

\subsubsection{Discrete difference vector}
\label{sec:discdiffvec}
Different approaches for the discretization of tangent vector fields are presented in, e.g., in \cite{Schoellhammer_2019a,Olshanskii_2018a,Jankuhn_2020a}. Herein, the difference vector $\vek{w}^h$ and its corresponding test function $\vw^h$ are defined in the general Trace FEM function space without the tangentiality constraint, similar to \cite{Olshanskii_2018a}. The corresponding function spaces are
\begin{align}
\mathcal{S}_{\vek{w}}^h &= \left\lbrace \vek{w}^h \in \left[\mathcal{T}_{h}\right]^3 \right\rbrace \ ,\\
\mathcal{V}_{\vek{w}}^h &= \left\lbrace \vw^h \in \left[\mathcal{T}_{h}\right]^3 \right\rbrace\ ,
\end{align}
\emph{but} in the discrete weak form, see \autoref{eq:dwf} and \autoref{eq:dwm}, only the \emph{projected} difference vector and test function are used, i.e., $\widetilde{\vek{w}}^h = \mat{P}\cdot \vek{w}^h\, ,\  \widetilde{\vek{v}}_w^h = \mat{P}\cdot{\vw}^h$. {\revStart The directional and covariant gradient of the discrete, projected difference vector $\gradGD{\widetilde{\vek{w}}^h}$ and $\gradGC{\widetilde{\vek{w}}^h}$ can be directly computed with the product rule
\begin{align}
\gradGD{\widetilde{\vek{w}}^h} &= \gradGD{\left(\mat{P}\cdot \vek{w}^h \right)} = \begin{bmatrix}
\nabla_{\Gamma\,x}^\t{dir}\mat{P}\cdot\vek{w}^h & \nabla_{\Gamma\,y}^\t{dir}\mat{P}\cdot\vek{w}^h & \nabla_{\Gamma\,z}^\t{dir}\mat{P}\cdot\vek{w}^h
\end{bmatrix} + \gradGC{\vek{w}^h} \ ,\\
\gradGC{\widetilde{\vek{w}}^h} &= \mat{P} \cdot \gradGD{\widetilde{\vek{w}}^h} = \mat{P} \cdot \begin{bmatrix}
\nabla_{\Gamma\,x}^\t{dir}\mat{P}\cdot\vek{w}^h & \nabla_{\Gamma\,y}^\t{dir}\mat{P}\cdot\vek{w}^h & \nabla_{\Gamma\,z}^\t{dir}\mat{P}\cdot\vek{w}^h
\end{bmatrix} + \gradGC{\vek{w}^h}\ .
\end{align}	\revEnd}
One may argue that in this approach the derivatives of the projector $\mat{P}$ occur, which involves surface derivatives of the normal vector $\nG^h$. This does not lead to additional computational costs, because in the case of the Reissner--Mindlin shell, the Weingarten map $\mat{H}$ directly appears in the weak form and, therefore, the surface derivatives of the normal vector $\nG^h$ are required independently of this approach.\par

As a result of this projection, only the tangential part of $\vek{w}^h$ and $\vw^h$ is considered in the discrete weak form and, therefore, the tangentiality constraint is built-in automatically. {\revStart The employed stabilization technique, i.e., normal derivative volume stabilization, see \autoref{sec:stab}, ensures unique nodal values and prevents an ill-conditioned system of equations due to small supports for general vector fields. However, due to the projection, the normal part of $\vek{w}^h$, i.e., $w^h_n = \vek{w}^h \cdot \nG^h$, does not appear in the discrete weak form and, therefore, $\vek{w}^h_n$ is not unique.
It is clear, that without any further measures this would lead to an ill-conditioned system of equations as a consequence. In order to address this issue, a simple and consistent additional stabilization term, similar to the penalty term in \cite{Olshanskii_2018a}, is introduced \revEnd}
\begin{align}
s_{w,h} := \rho_w \int_{\Gamma^h} \left(\vek{w}^h\cdot\nG^h\right) \left(\vw^h\cdot\nG^h\right)\ \d A\ ,
\end{align}
where $\rho_w$ is a suitable stabilization parameter. {\revStart In other words, for the stabilization of the projected, discrete difference vector $\widetilde{\vek{w}}^h$, a combination of the normal derivative volume stabilization and the above introduced stabilization term for the normal part of $\vek{w}^h$ is employed, see \autoref{eq:dwm}. \revEnd}\par

A series of numerical studies regarding the choice of the stabilization parameter {\revStart$\rho_w$ \revEnd} for the Reissner--Mindlin shell has been conducted on flat and curved shell geometries. In detail, the dependency on the (1) material parameter $E$, (2) thickness $t$ and (3) element size on $h$ w.r.t.~the condition number of the stiffness matrix and the influence on the results was investigated. Summarizing the outcome of the numerical studies, the stabilization parameter can be chosen independently of $h$ and for a suitable scaling of the stabilization term, the parameter is set to
\begin{align}
\rho_w = E \cdot t\ .
\end{align}
A difference of the proposed approach and the method shown in \cite{Olshanskii_2018a} is that only the projected part of the vector field, i.e., $\widetilde{\vek{w}}^h = \mat{P}\cdot \vek{w}^h$, is used in the discrete weak form which directly enforces the tangentiality constraint. Furthermore, the used stabilization parameter, which is in \cite{Olshanskii_2018a,Jankuhn_2020a} a penalty parameter, does not depend on $h$ and only a suitable scaling of the stabilization term may be required.

\subsubsection{Discrete weak form}
\label{sec:dwf}
Based on the previous definitions, the discrete weak form with the Trace FEM of the force equilibrium, see \autoref{eq:wff}, reads as follows: Given material parameters $(E,\nu) \in \mathbb{R}^{+}$, body forces $\vek{f} \in \mathbb{R}^3$ on $\Gamma^h$, tractions $\hat{\vek{p}}$ on $\p\Gamma^h_{\t{N},\vek{u}}$, stabilization parameter $\rho \in \mathbb{R}^{+}$ find the displacement fields $(\vek{u}^h,\vek{w}^h) \in \mathcal{S}_{\vek{u}}^h \times \mathcal{S}_{\vek{w}}^h$ such that for all test functions $(\vu^h,\vw^h) \in \mathcal{V}_{\vek{u}}^h\times\mathcal{V}_{\vek{w}}^h$ there holds in $\Gamma^h$
\begin{align}
\begin{split}\label{eq:dwf}
&\int_{\Gamma^h} \gradGD{\vu} : \tilde{\mat{n}}_\Gamma(\vek{u}^h) + (\mat{H}\cdot\gradGD{\vu}):\mat{m}_\Gamma(\vek{u}^h,\widetilde{\vek{w}}^h) + (\mat{Q}\cdot\gradGD{\vu}):\mat{q}_\Gamma(\vek{u}^h,\widetilde{\vek{w}}^h)\ \d A \\[.25cm] 
&-\underbrace{\int_{\p\Gamma_{\t{D},\vek{u}}^h} \vu \cdot \vek{p}(\vek{u}^h,\widetilde{\vek{w}}^h)\ \d s}_{\t{boundary term due to\ } \vu^h \neq \vek{0} \t{ on } \p\Gamma_{\t{D},\vek{u}}} + \underbrace{\int_{\p\Gamma_{\t{D},\vek{u}}^h} \vek{u}^h \cdot \vek{p}(\vu^h,\widetilde{\vek{v}}_w^h)\ \d s}_{\t{Nitsche term for displ.~on LHS}} \\[.25cm] 
&+ \underbrace{\rho \int_{\Omega^\Gamma_h} \left(\nabla\vek{u}^h\cdot\vek{n}_\Gamma^{e,h} \right) \cdot \left(\nabla\vu^h\cdot\vek{n}_\Gamma^{e,h} \right)\ \d V}_{\t{Trace FEM stabilization, see \autoref{sec:stab}}} \\[.25cm] 
&=\hspace{0cm} \int_{\Gamma^h} \vu^h \cdot \vek{f}\ \d A + \underbrace{\int_{\p\Gamma_{\t{D},\vek{u}}^h} \hat{\vek{g}}_{\vek{u}} \cdot \vek{p}(\vu^h,\widetilde{\vek{v}}_w^h)\ \d s}_{\t{Nitsche term for displ.~on RHS}} + \int_{\p\Gamma_{\t{N},\vek{u}}^h} \vu^h \cdot \hat{\vek{p}}\ \d s\ ,
\end{split}
\end{align}
where $\vek{p} = \mat{n}_\Gamma^{\t{real}}\cdot \nCo^h + \left(\nG^h\cdot\mat{q}_\Gamma \cdot \nCo^h \right)\nG^h$, see \autoref{eq:pdebcs}, are the conjugated forces at the Dirichlet boundary $\p\Gamma_{\t{D},\vek{u}}^h$.\par

The discrete weak form of the moment equilibrium, see \autoref{eq:wfm} reads as follows: Given material parameters $(E,\nu) \in \mathbb{R}^{+}$, distributed moments $\vek{c} \in T_{P}\Gamma^h$ on $\Gamma^h$, bending moments $\hat{\vek{m}}_{\p\Gamma}$ on $\p\Gamma^h_{\t{N},\vek{w}}$, stabilization parameters $(\rho,\rho_w) \in \mathbb{R}^{+}$ find the displacement fields $(\vek{u}^h,\vek{w}^h) \in \mathcal{S}_{\vek{u}}^h \times \mathcal{S}_{\vek{w}}^h$ such that for all test functions $(\vu^h,\vw^h) \in \mathcal{V}_{\vek{u}}^h\times\mathcal{V}_{\vek{w}}^h$ there holds in $\Gamma^h$
\begin{align}
\begin{split}\label{eq:dwm}
&\int_{\Gamma^h} \gradGD{\widetilde{\vek{v}}_w^h}:\mat{m}_\Gamma(\vek{u}^h,\widetilde{\vek{w}}^h) + \widetilde{\vek{v}}_w^h \cdot \left[\mat{q}_\Gamma(\vek{u}^h,\widetilde{\vek{w}}^h)\cdot\nG\right]\ \d A \\[.25cm] 
&-\underbrace{\int_{\p\Gamma_{\t{D},\vek{w}}} \widetilde{\vek{v}}_w^h \cdot \vek{m}_{\p\Gamma}(\vek{u}^h,\widetilde{\vek{w}}^h)\ \d s}_{\t{boundary term due to\ } \vw^h \neq \vek{0} \t{ on } \p\Gamma_{\t{D},\vek{w}}} + \underbrace{\int_{\p\Gamma_{\t{D},\vek{w}}} \widetilde{\vek{w}}^h \cdot \vek{m}_{\p\Gamma}(\vu^h,\widetilde{\vek{v}}_w^h)\ \d s}_{\t{Nitsche term for rot.~on LHS}} \\[.25cm] 
&+ \underbrace{\rho \int_{\Omega^\Gamma_h} \left(\nabla\vek{w}^h\cdot\vek{n}_\Gamma^{e,h} \right) \cdot \left(\nabla\vw^h\cdot\vek{n}_\Gamma^{e,h} \right)\ \d V}_{\t{Trace FEM stabilization, see \autoref{sec:stab}}} + \underbrace{\rho_w \int_{\Gamma^h} \left(\vek{w}^h\cdot\nG^h\right) \left(\vw^h\cdot\nG^h\right)\ \d A}_{\t{stabilization term for\ } \widetilde{\vek{w}}^h \t{\ see \autoref{sec:discdiffvec}}}\\[.25cm] 
&=\hspace{0cm} \int_{\Gamma^h} \vw^h \cdot \vek{c}\ \d A + \underbrace{\int_{\p\Gamma_{\t{D},\vek{w}}^h} \hat{\vek{g}}_{\vek{w}} \cdot \vek{m}_{\p\Gamma}(\vu^h,\widetilde{\vek{v}}_w^h)\ \d s}_{\t{Nitsche term for rot.~on RHS}} + \int_{\p\Gamma_{\t{N},\vek{w}}^h} \vw^h \cdot \hat{\vek{m}_{\p\Gamma}}\ \d s\ ,
\end{split}
\end{align}
where $\vek{m}_{\p\Gamma} = \mat{m}_\Gamma \cdot \nCo^h$, see \autoref{eq:pdebcs}, are the conjugated bending moments at the Dirichlet boundary $\p\Gamma_{\t{D},\vek{w}}^h$. {\revStart Note that in the implementation one may employ the identity $\mat{A} : \mat{B} = (\mat{P}\cdot \mat{A} \cdot \mat{P}) : \mat{B}^{\t{dir}}$ with $(\mat{A}, \mat{B}) \in \mathbb{R}^{3\times 3}$ and $\mat{B} = \mat{P}\cdot \mat{B}^{\t{dir}} \cdot \mat{P}$ in order to further simplify	 the obtained terms with the projected difference vector. As a result, only directional derivatives of the discrete, projected difference vector are required which simplifies the implementation significantly.\revEnd}\par

The usual element assembly w.r.t.~the active elements yields a linear system of equations in the following form
\begin{align}
\underbrace{\left(\mat{K}_{\t{Stiff}}  + \mat{K}_{\t{Nitsche}} + \mat{K}_{\t{Stab}}\right)}_{\mat{K}} \cdot \begin{bmatrix}
\underline{\hat{\vek{u}}} \\
\underline{\hat{\vek{w}}}
\end{bmatrix} &= \underbrace{\left(\vek{b}_{\t{Load}} + \vek{b}_{\t{Nitsche}}\right)}_{\vek{b}}\ ,
\end{align}
with $[\underline{\hat{\vek{u}}}, \underline{\hat{\vek{w}}}]$ being the sought displacements and rotations of the normal vector at the nodes of the active elements. The matrix $\mat{K}$ and the load vector $\vek{b}$ are split into: (1) standard terms for the stiffness matrix, (2) boundary terms, (3) stabilization terms, respectively.
\section{Numerical results}
\label{sec:numres}

In this section, the proposed numerical method for implicitly defined Reissner--Mindlin shells is tested on a set of benchmark examples, consisting of the partly clamped hyperbolic parabolid from \cite{Bathe_2000a,Chapelle_1998a}, the partly clamped gyroid from \cite{Gfrerer_2019a} and a clamped flower-shaped shell inspired by \cite{Schoellhammer_2018a,Schoellhammer_2019a}. In the case of the first two examples the shells are rather thin and locking phenomena can be expected, especially in the case of low ansatz orders. However, when increasing the order $p$, locking phenomena decrease significantly and, therefore, no further measures against locking phenomena are considered herein.\par

In the convergence studies, quasi-regular background meshes consisting of tetrahedral elements are used. For all unknown fields, i.e., $\vek{u}^h, \vek{w}^h$, and the interpolation of the level-set functions, the same order of shape functions are used. The orders are varied as $2 \le p \le 6$. The element size $h$ is proportional to the factor $n$ which is related to the number of elements and is varied between $2 \le n \le 128$.\par

In the presented examples, the stabilization parameter $\rho$ for the normal derivative volume stabilization is set to $\rho = \sfrac{1000}{h}$. In order to achieve a proper scaling of the stiffness matrix, the parameter $\rho_w$, is set to $\rho_w = Et$, as proposed in \autoref{sec:discdiffvec}.

\subsection{Hyperbolic paraboloid}
\label{sec:hyperhyper}

\begin{figure}[h]
	\begin{minipage}{.48\textwidth}\centering
		\includegraphics[width=.9\textwidth, height = .9\textwidth, keepaspectratio]{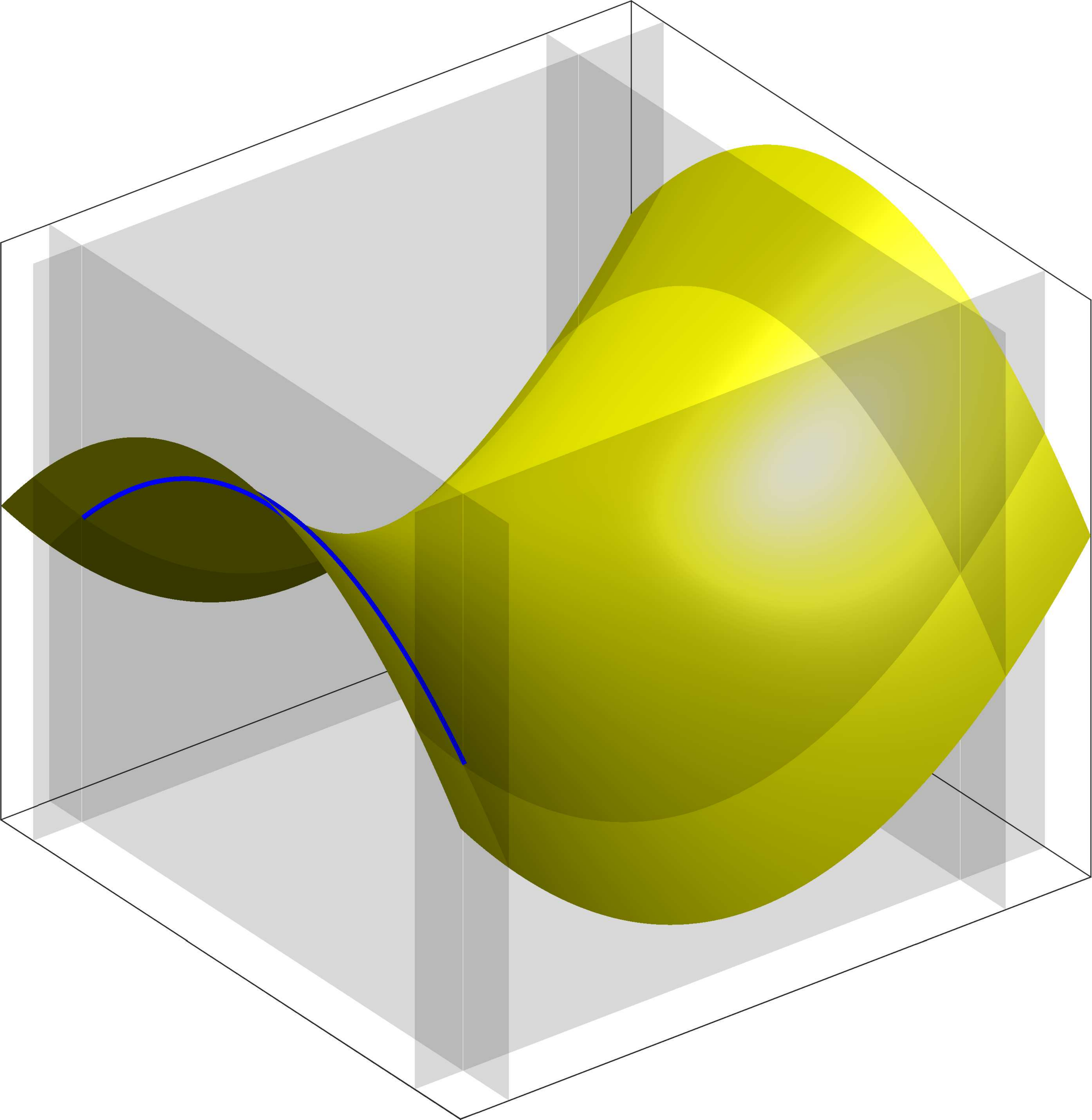}
	\end{minipage}
	\begin{minipage}{.5\textwidth}\tiny\flushleft
		\begin{tabular}[h]{ll}
			Geometry: & Hyperbolic paraboloid\\
			& $\phi(\vek{x}) = x^2 - y^2 - z$ \\[.1 cm]
			& $x \in [-0.5,0.5]$ \\
			& $y \in [-0.5,0.5]$ \\[.1 cm]	
			& $t = \SI{0.01}{}$\\[.25 cm]
			Material parameters: & $E = \SI{2.0e11}{}$ \\
			& $\nu = \SI{0.3}{}$\\
			& $\alpha_{\t{s}} = 1.0$ \\[.25 cm]
			Load: & Gravity load $\vek{f} = [0,\,0,\,-8000\cdot t]^\T $\\[.1 cm]
			& \phantom{Gravity load} $\vek{c} = \vek{0}$\\[.25 cm]
			Support: & Clamped edge at $x = -\sfrac{1}{2}$\\[.25cm]
			Ref. displacement: & $\vert u_{z,i,\t{Ref}} \vert = \SI{9.3355e-5}{}$\\
			& $\vek{x}_i = (0.5,\,0,\,0.25)^\T$
		\end{tabular}
	\end{minipage}
	\caption{Definition of the partly clamped hyperbolic paraboloid problem.}
	\label{fig:overhyper}
\end{figure}

\begin{figure}[h]
	\centering
	\subfloat[active background mesh]{\includegraphics[width=.32\textwidth, height = .35\textwidth, keepaspectratio]{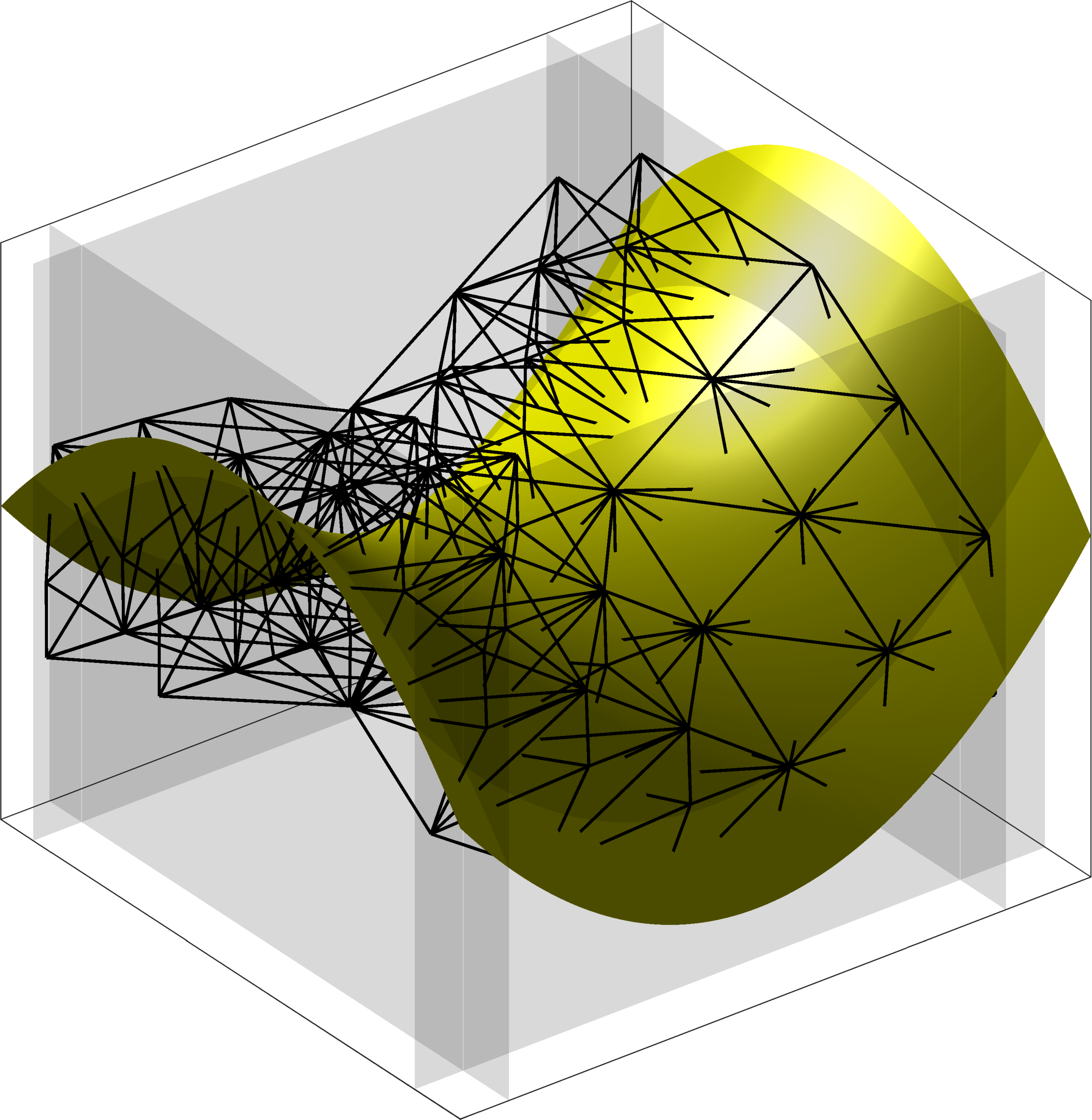}\label{fig:hypermesh}} \hfil
	\subfloat[integration points]{\includegraphics[width=.32\textwidth, height = .35\textwidth, keepaspectratio]{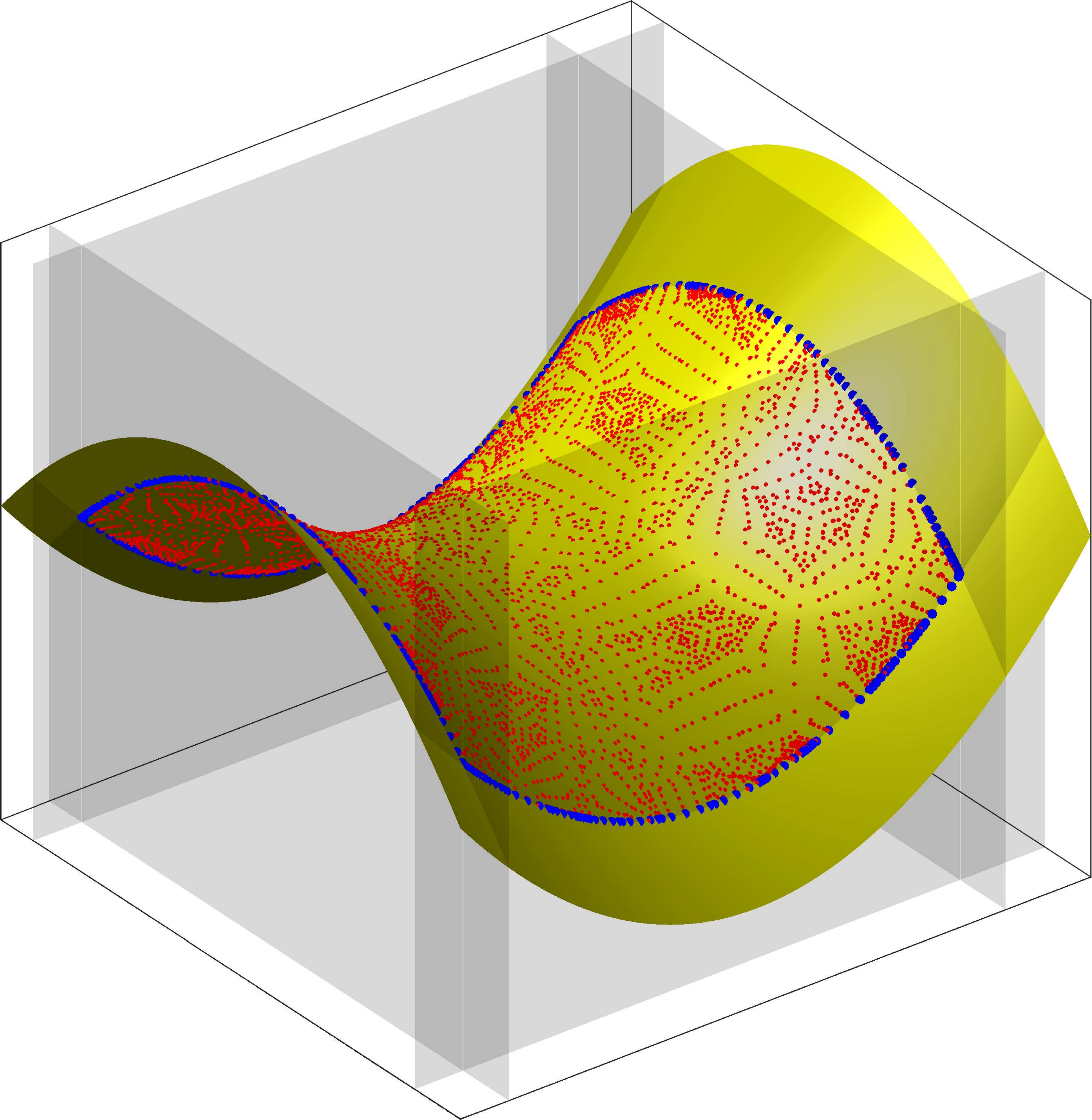}\label{fig:hyperint}} \hfil
	\subfloat[displacements]{\includegraphics[width=.32\textwidth, height = .35\textwidth, keepaspectratio]{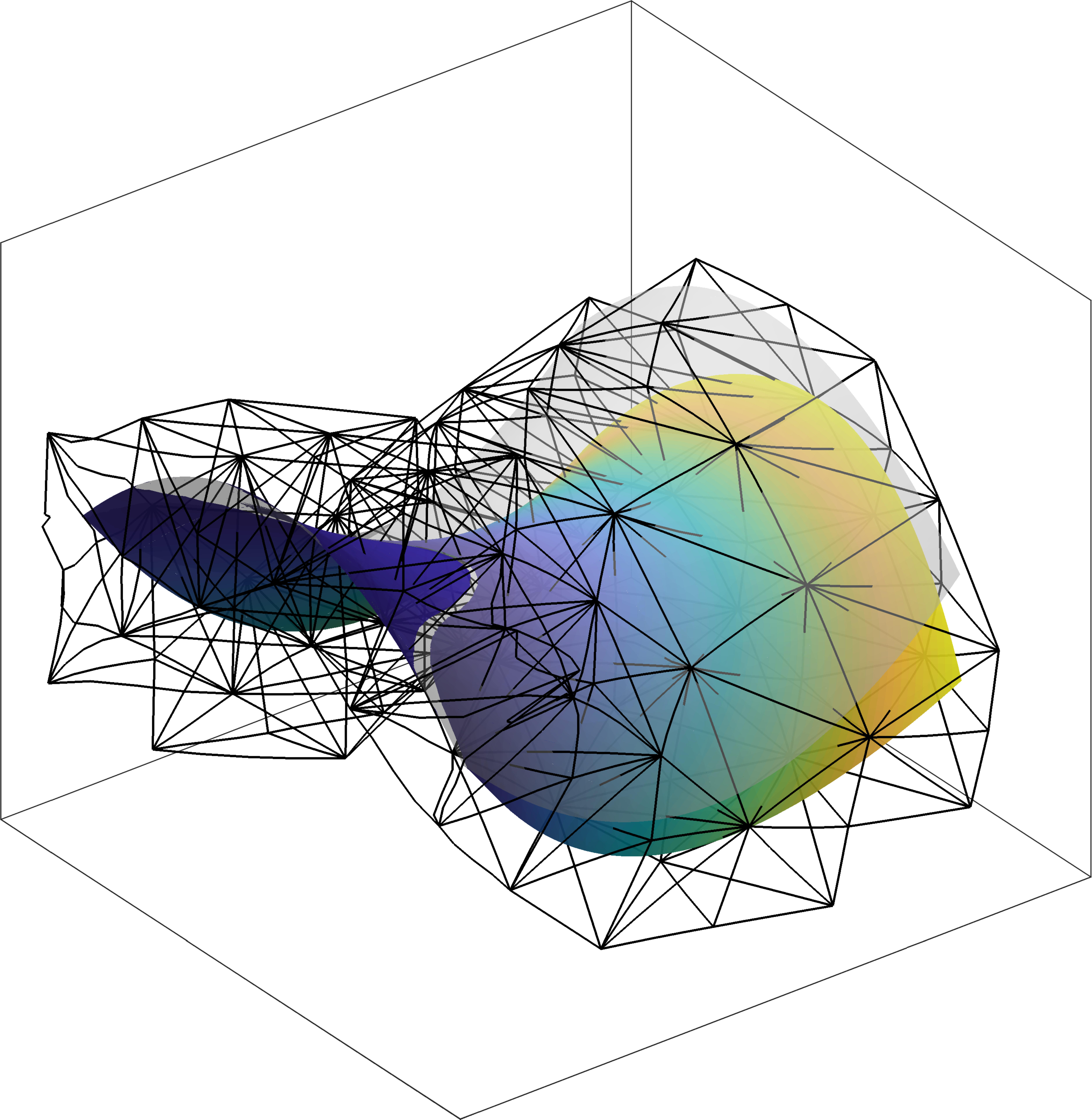}\label{fig:hyperdisp}}
	\caption{(a) Active background mesh, which consists only of cut elements, (b) automatically generated integration points in the domain (red) and on the boundaries (blue), (c) deformed zero-isosurface with scaled displacements $\vek{u}$ by a factor of $\SI{2e3}{}$.}
	\label{fig:hypermeshint}	
\end{figure}

The first example is the partly clamped hyperbolic paraboloid and is taken from \cite{Bathe_2000a,Chapelle_1998a}. The problem is defined in \autoref{fig:overhyper}. The yellow surface is the zero-isosurface of the master level-set function $\phi$ and the grey planes are the zero-isosurfaces of the slave level-set functions $\psi_j,\ j \in [1,4]$, which define the boundaries of the shell. The blue line is the clamped edge of the shell. 
In \autoref{fig:hypermesh}, the active background mesh which contains only cut elements with $p = 4$ is shown. In \autoref{fig:hyperint}, the corresponding integration points are illustrated, those in the domain are plotted in red and those on the boundaries are blue. In \autoref{fig:hyperdisp}, the numerical solution of the partly clamped hyperbolic paraboloid is presented. The grey surface is the undeformed zero-isosurface and the colors on the deformed midsurface of the shell are the Euclidean norm of the displacement field $\vek{u}^h$. The displacements are magnified by a factor of $\SI{2e3}{}$. 

\begin{figure}[h]
	\centering
	\subfloat[convergence]{\includegraphics[width=.43\textwidth, height = .43\textwidth, keepaspectratio]{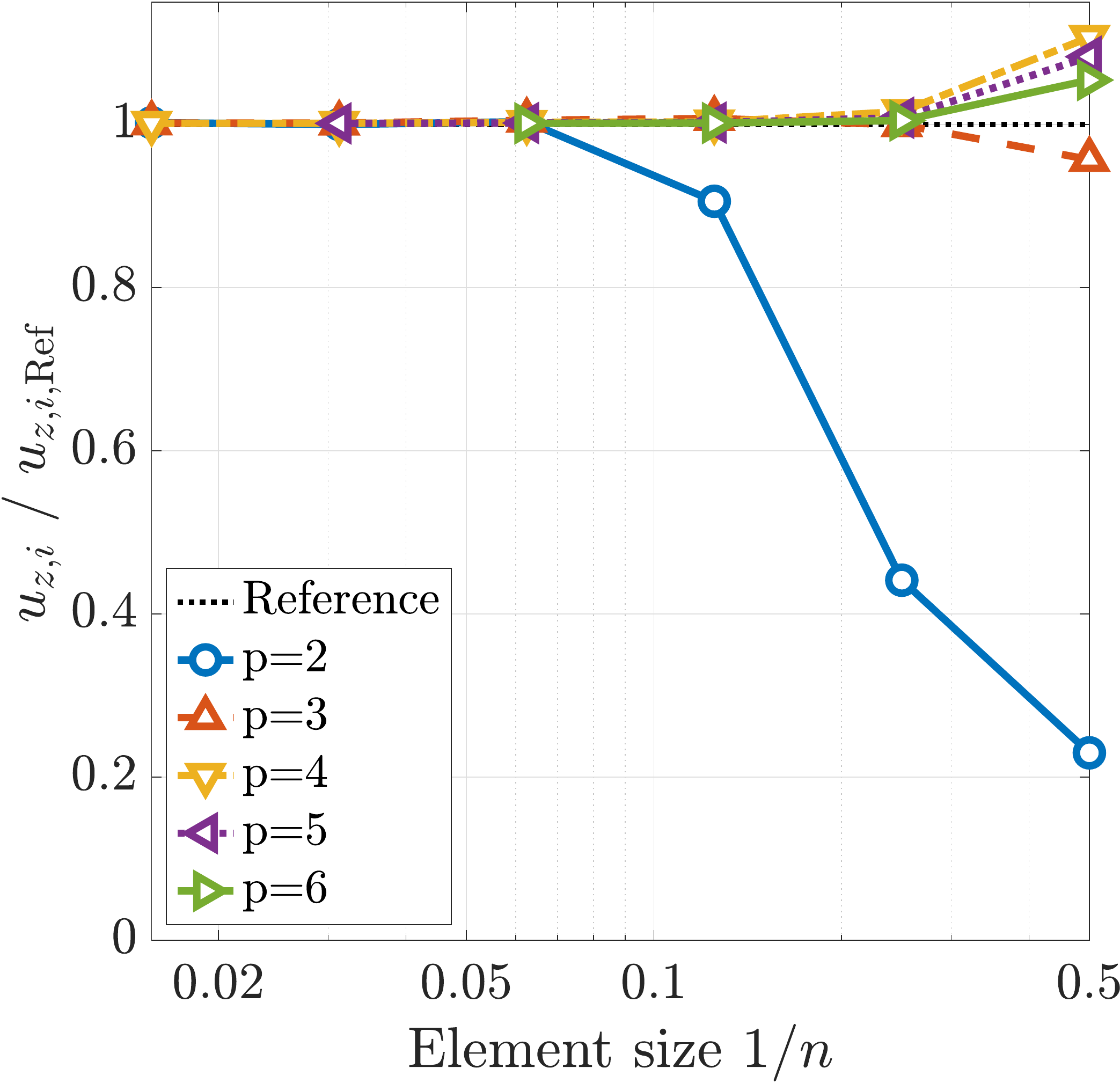}\label{fig:hyperconv}} \hfil
	\subfloat[condition number]{\includegraphics[width=.43\textwidth, height = .43\textwidth, keepaspectratio]{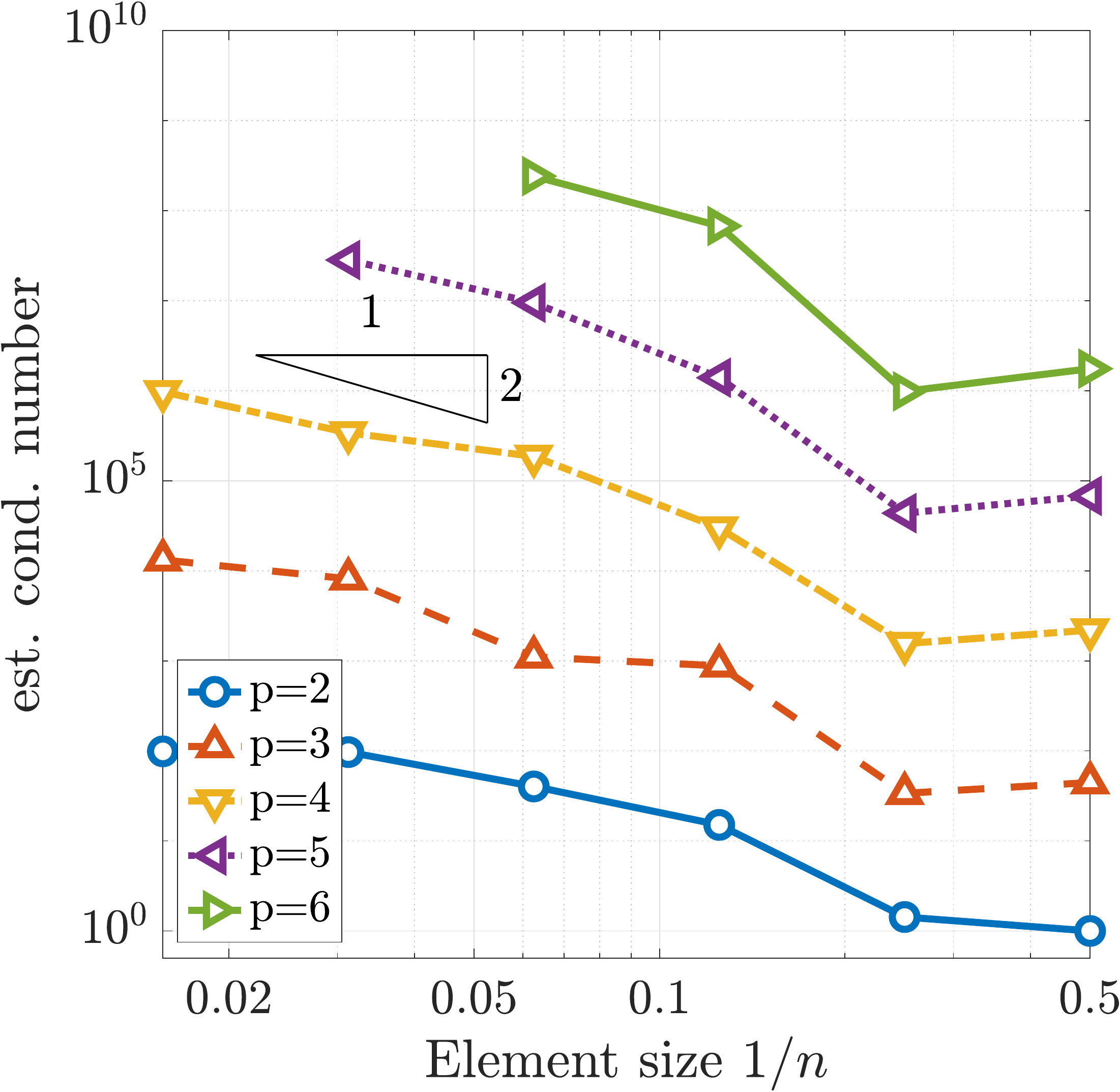}\label{fig:hypercond}}
	\caption{(a) Normalized convergence of reference displacement $u_{z,i,\text{Ref}} = \SI{-9.3355e-5}{}$ at point $\vek{x}_i = (0.5,\,0,\,0.25)^\T$, (b) normalized condition numbers, the reference value is $\SI{1.0797e+10}{}$, which is the condition number at $n = p = 2$.}
	\label{fig:hyperres}	
\end{figure}

In the convergence studies, the vertical displacement at $\vek{x}_i$ is compared with the given reference displacement. Due to the moderate complexity of the master level-set function, the numerical solution converges rather fast and the element factor $n$ is only varied between $2 \le n \le 64$. In \autoref{fig:hyperconv}, the results of the convergence study are presented. In particular, the normalized displacement $\sfrac{u_{z,i}}{u_{z,i,\text{Ref}}}$ is plotted as a function of the element size $ h \sim \sfrac{1}{n}$. The behaviour of the convergence is in agreement with the results shown, e.g., in \cite{Bathe_2000a,Kiendl_2017a,Schoellhammer_2019a}. In particular, the expected locking behaviour is more pronounced for $p=2$ and decreases significantly for higher orders. In \autoref{fig:hypercond}, the normalized, estimated condition number of the stiffness matrix is plotted as a function of the element size $h \sim \sfrac{1}{n}$. The condition numbers are obtained with the MATLAB function \emph{condest}. It can be seen that the condition numbers increase with quadratic order as expected for second-order PDEs. The jump between the element orders is well-known in the context of higher-order finite element approaches, see e.g., in \cite{Fries_2017b}.

\subsection{Gyroid}
\label{sec:gyroid}

\begin{figure}[h]
	\begin{minipage}{.48\textwidth}\centering
		\includegraphics[width=.9\textwidth, height = .9\textwidth, keepaspectratio]{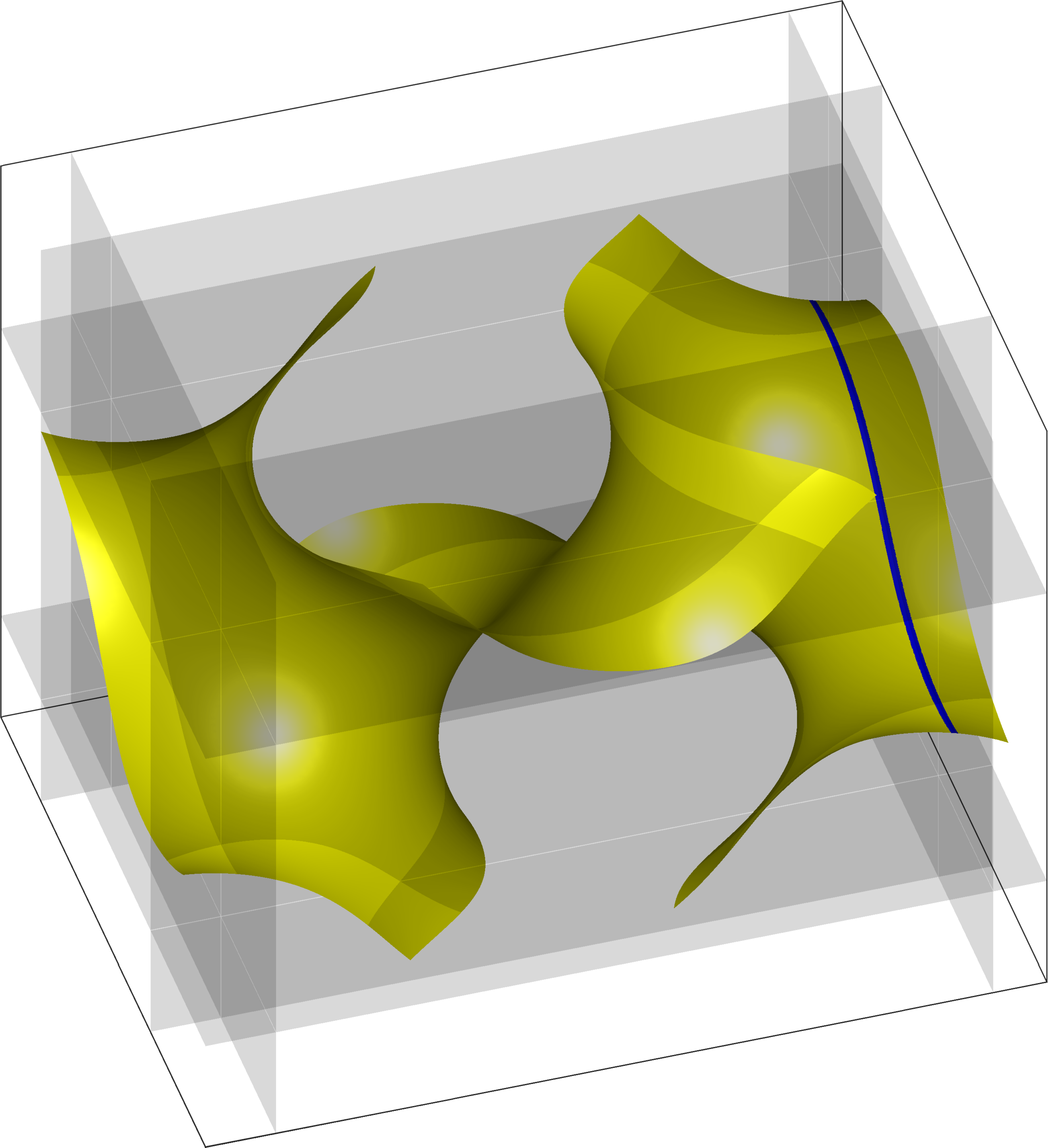}
	\end{minipage}
	\begin{minipage}{.5\textwidth}\tiny\flushleft
		\begin{tabular}[h]{ll}
			Geometry: & Gyroid\\
			& $\phi(\vek{x}) = \sin(\pi x) \cos(\pi y)\ + $\\
			& $\phantom{\phi(\vek{x}\,) =}\sin(\pi y) \cos(\pi z)\ + $\\
			& $\phantom{\phi(\vek{x}\,) =}\sin(\pi z) \cos(\pi x)$ \\[.1 cm]			
			& $x \in [0,2]$\\
			&$y \in [-0.5,0.5]$\\
			&$ z \in [-0.5,0.5]$\\[.1 cm]	
			& $t = \SI{0.03}{}$\\[.25 cm]			
			Material parameters: & $E = \SI{70e9}{}$ \\
			& $\nu = \SI{0.3}{}$\\
			& $\alpha_{\t{s}} = 1.0$ \\[.25 cm]
			Load: & Gravity load $\vek{f} = [0,\,0,\,10^7\cdot t]^\T $\\[.1 cm]
			& \phantom{Gravity load} $\vek{c} = \vek{0}$\\[.25 cm]
			Support: & Clamped edge at $x = 0$\\[.25cm]
			Ref. displacement: & $\vert u_{z,i,\t{Ref}} \vert = 0.182661$\\
			& $\vek{x}_i = (2,\,0.5,\,-0.25)^\T$
		\end{tabular}
	\end{minipage}
	\caption{Definition of the partly clamped gyroid problem.}
	\label{fig:overgyro}
\end{figure}

The next test case is a partly clamped gyroid and is taken from \cite{Gfrerer_2019a}. The problem is defined in \autoref{fig:overgyro}. Similar as above, the yellow surface is the zero-isosurface of the master level-set function $\phi$, the grey planes are the zero-isosurfaces of the slave level-set functions $\psi_j,\ j\in[1,6]$, which bound the master leve-set function. The blue curve is the clamped edge of the shell.\par

In contrast to \cite{Gfrerer_2019a}, a factor $\pi$ is inserted into the arguments of the trigonometric functions of the master level-set function, which is given in \cite[Eq.~43]{Gfrerer_2019a}. Otherwise, the obtained geometry is not in agreement with the presented geometry in \cite[Fig.~12]{Gfrerer_2019a}. In addition, the load is decreased by one order of magnitude in order to decrease the deformations, which shall be significantly smaller than the dimensions of the shell. Therefore, the given reference displacement needs to be scaled accordingly to $0.18812$. However, in \cite{Gfrerer_2019a}, a different shell model (seven-parameter shell model) is used. Herein, the classical Reissner--Mindlin shell, which is often labelled as five-parameter model, is used and, therefore, we can expect small differences in the displacements. For the Reissner--Mindlin shell model the converged reference displacement is $0.182661$, which is a relative error of $2.9\%$ compared to the seven-parameter model, which can be explained with the differences in the kinematic assumptions between the two shell models. This discrepancy could be decreased with a suitable shear correction factor.

\begin{figure}[h]
	\centering
	\subfloat[active background mesh]{\includegraphics[width=.43\textwidth, height = .43\textwidth, keepaspectratio]{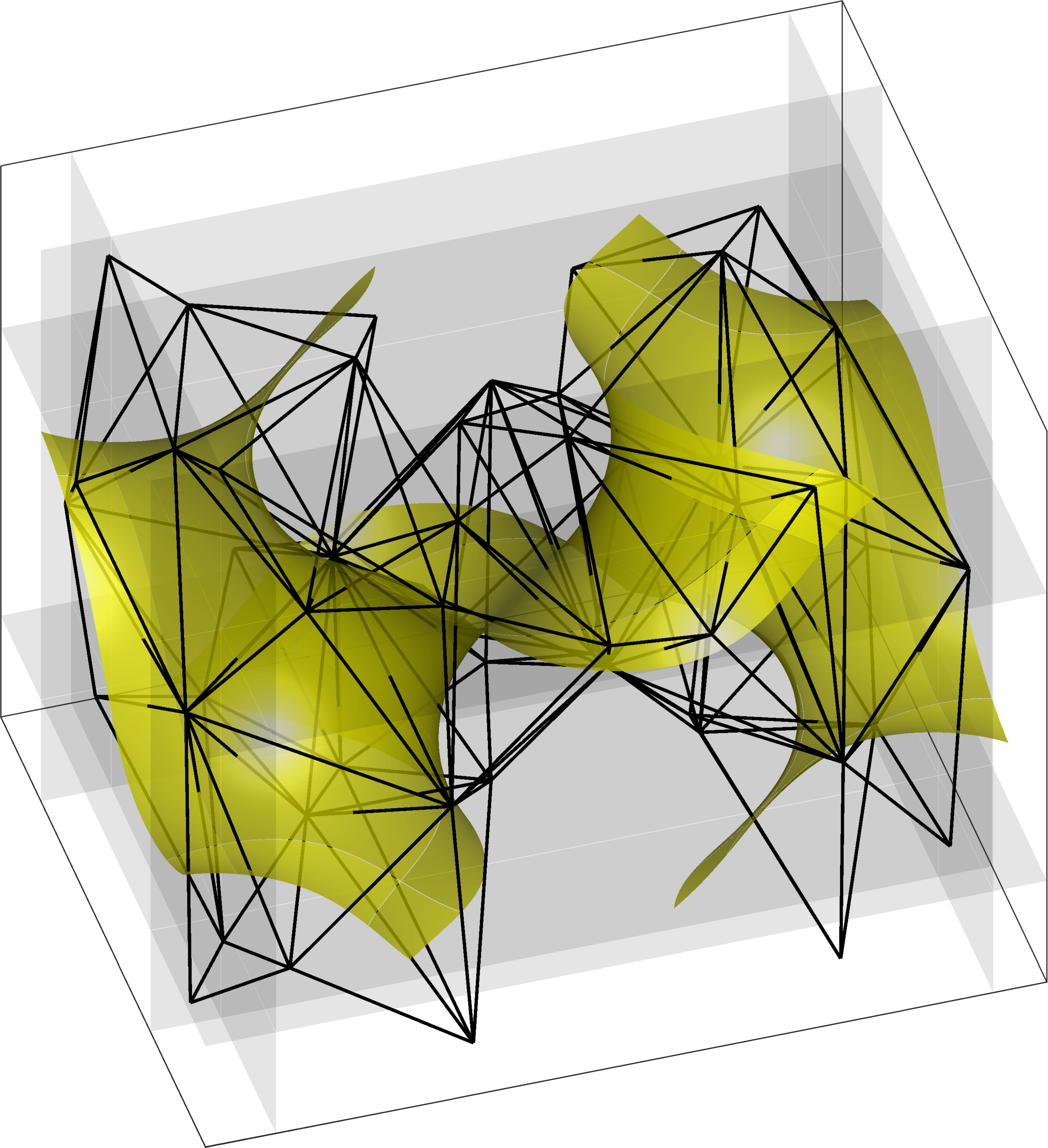}} \hfil
	\subfloat[integration points]{\includegraphics[width=.43\textwidth, height = .43\textwidth, keepaspectratio]{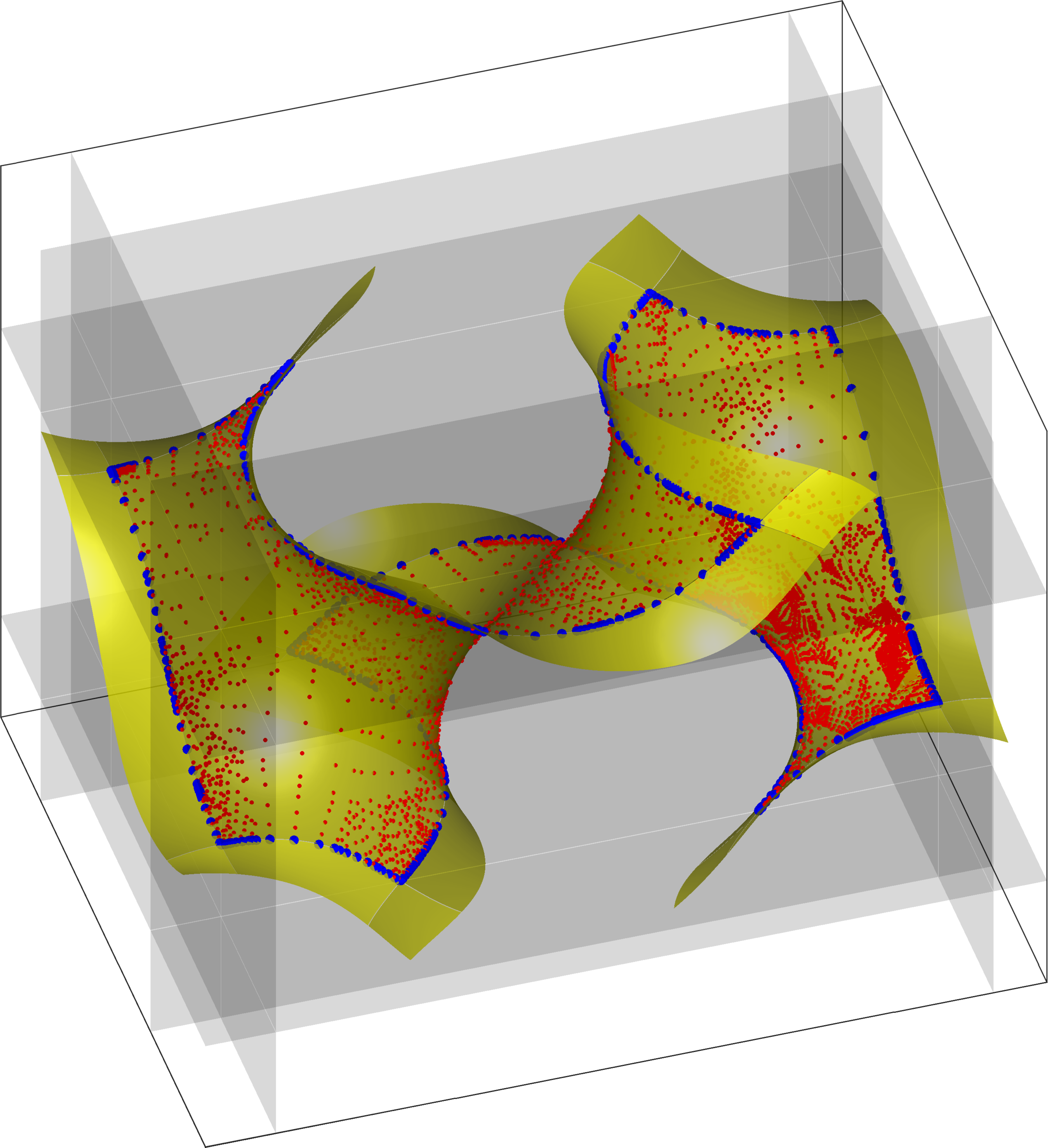}}
	\caption{(a) Active background mesh, which consists only of cut elements, (b) automatically generated integration points in the domain (red) and on the boundaries (blue).}
	\label{fig:gyromeshint}	
\end{figure}

\begin{figure}[h]
	\centering
	\subfloat[displacements]{\includegraphics[width=.43\textwidth, height = .43\textwidth, keepaspectratio]{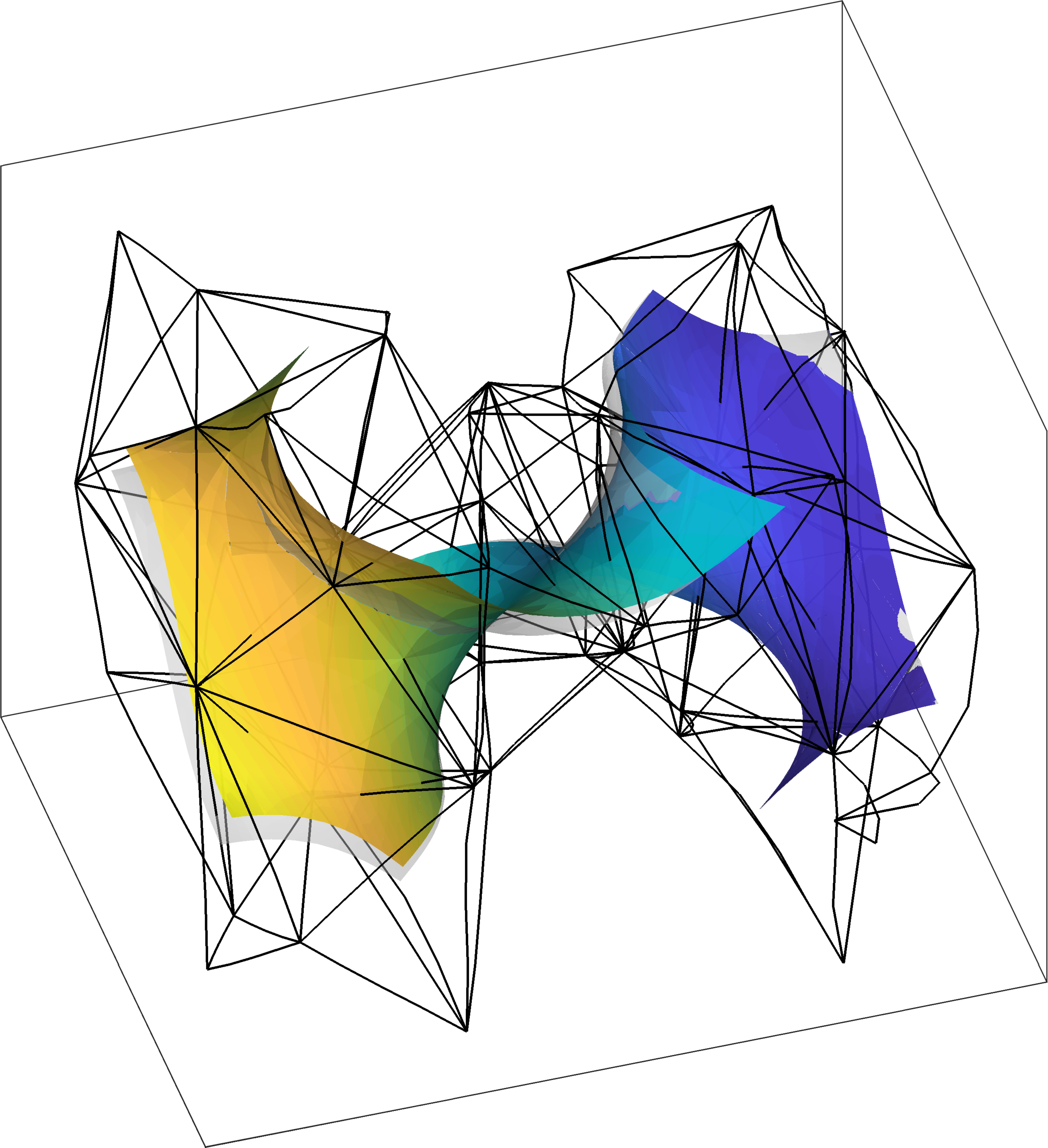}\label{fig:gyrosdisp}} \hfil
	\subfloat[convergence]{\includegraphics[width=.43\textwidth, height = .43\textwidth, keepaspectratio]{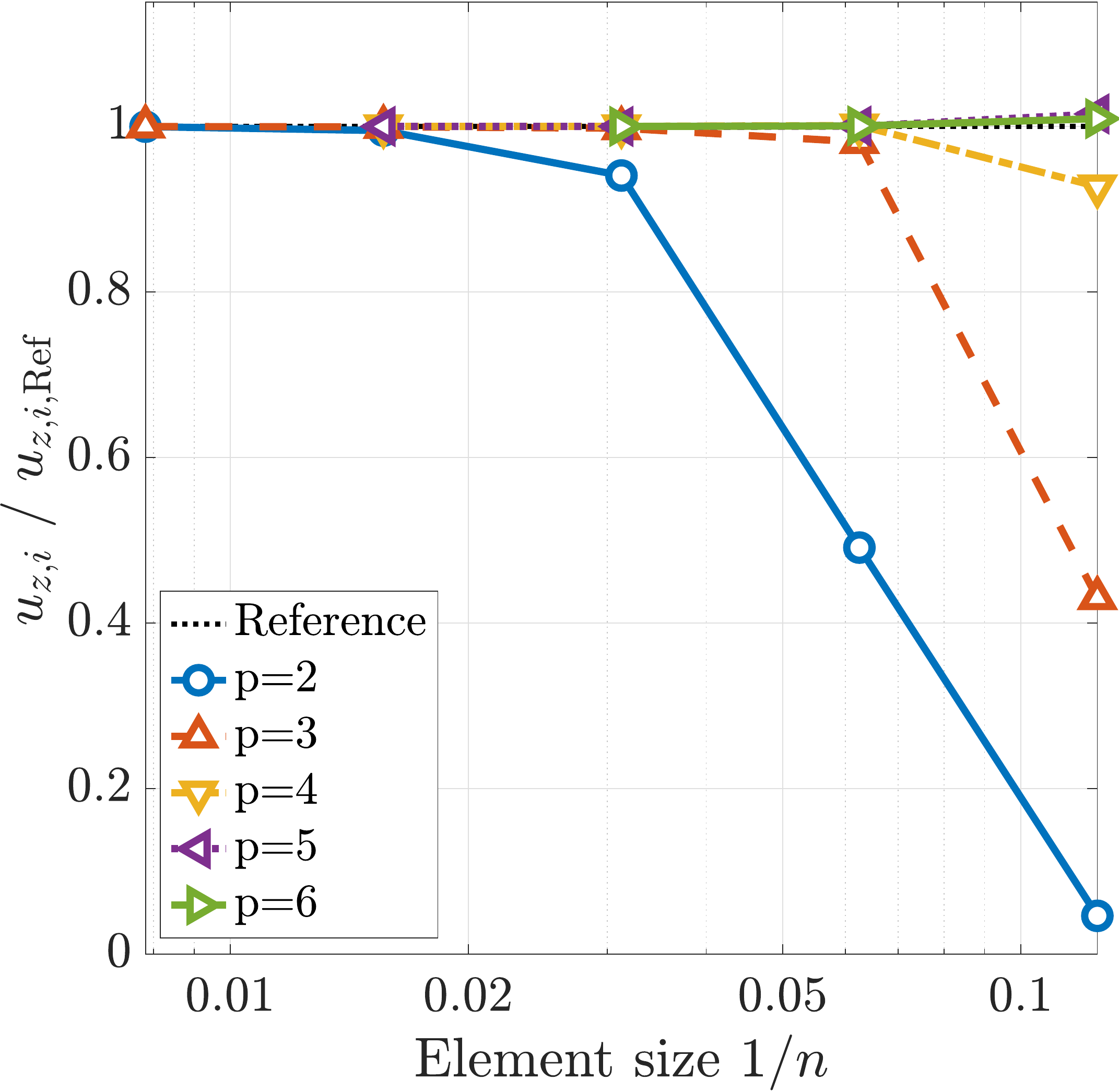}\label{fig:gyrosconv}}
	\caption{(a) Deformed zero-isosurface, (b) normalized convergence of reference displacement $u_{z,i,\text{Ref}} = 0.182661$ at point $\vek{x}_i = (2,\,0.5,\,-0.25)^\T$.}
	\label{fig:gyrosres}	
\end{figure}

Analogously to the example above, in \autoref{fig:gyromeshint}, the active background mesh for $p = 4$ and the corresponding integration points are shown, where the domain integrations points are plotted in red and the integration points on the boundaries are blue. The deformed zero-isosurface of the shell is plotted in a similar manner as in the first example in \autoref{fig:gyrosdisp}, where the colors on the surface are the Euclidean norm of the displacement field $\vek{u}^h$ and the grey surface indicates the undeformed zero-isosurface.
\par

In the convergence study, the parameter $n$ is varied between $8 \le n \le 128$. The coarser levels $n = \lbrace 2,4 \rbrace$ are skipped due to the more complex shape of the shell. In \autoref{fig:gyrosconv}, the results of the convergence analyses are presented. Similar to the first test case, the normalized displacement $\sfrac{u_{z,i}}{u_{z,i,\text{Ref}}}$ is plotted as a function of the element size $ h \sim \sfrac{1}{n}$. For the lower orders $p = \lbrace2, 3\rbrace$ the expected locking phenomena is more pronounced compared to the example before. Nevertheless, it is clearly seen that the accuracy for higher orders increases significantly and the behaviour of convergence is in agreement with the results shown, e.g., in \cite{Gfrerer_2019a}. The condition numbers for this example behave in a similar manner as in \autoref{fig:hypercond} and, therefore, the plot is omitted for the sake of brevity.

\subsection{Flower-shaped shell}
\label{sec:flowershell}

\begin{figure}[h]
	\begin{minipage}{.48\textwidth}\centering
		\includegraphics[width=.9\textwidth, height = .9\textwidth, keepaspectratio]{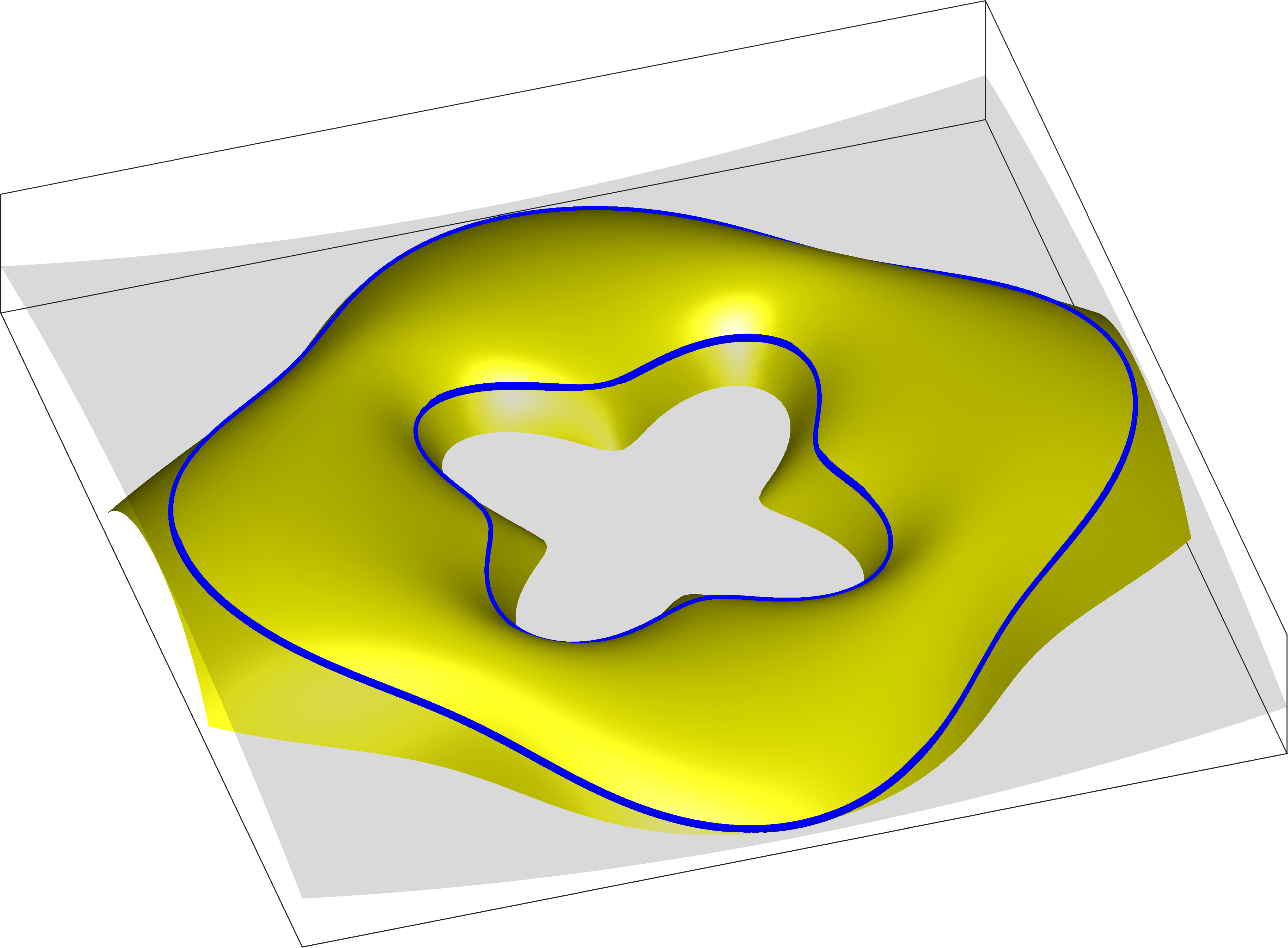}
	\end{minipage}
	\begin{minipage}{.5\textwidth}\tiny\flushleft
		\begin{tabular}[h]{ll}
			Geometry: & Flower shell\\			
			& $\phi(\vek{x}) = \sfrac{1}{3}\left(1-s^2\right) + $\\
			& $\phantom{\phi(\vek{x}\,) =}\sfrac{1}{40}\left(x^2-y^2\right) - z $\\[.1 cm]
			& $s =  \frac{r - 2.3}{0.8 + 0.3\cos(4\theta)}$ \\[.1 cm]
			& $r,\theta\ $ are polar coordinates of $x,y$\\[.1 cm]	
			& $\psi(\vek{x}) = \sfrac{1}{40}\left(x^2-y^2\right) - z$\\[.1 cm]
			& $t = \SI{0.05}{}$\\[.25 cm]			
			Material parameters: & $E = \SI{1.05e8}{}$ \\
			& $\nu = \SI{0.33}{}$\\
			& $\alpha_{\t{s}} = 1.0$ \\[.25 cm]
			Load: & $\vek{f} = -10^2[1,\,2,\,3]^\T $\\[.1 cm]
			& $\vek{c} = \vek{0}$\\[.25 cm]
			Support: & Clamped edges\\[.25cm]
			Error measurement & Residual errors
		\end{tabular}
	\end{minipage}
		\caption{Definition of shallow flower-shaped shell problem}
		\label{fig:overflower}
\end{figure}

The geometry of the last example is inspired from \cite{Schoellhammer_2019a}. The problem is defined in \autoref{fig:overflower}. Analogously as above, the yellow surface is the zero-isosurface of the master level-set function $\phi$ and the intersection with the slave level-set function $\psi$ (grey surface) defines the boundaries of the shell.\par

\begin{figure}[h]
	\centering
	\subfloat[integration points]{\includegraphics[width=.43\textwidth, height = .43\textwidth, keepaspectratio]{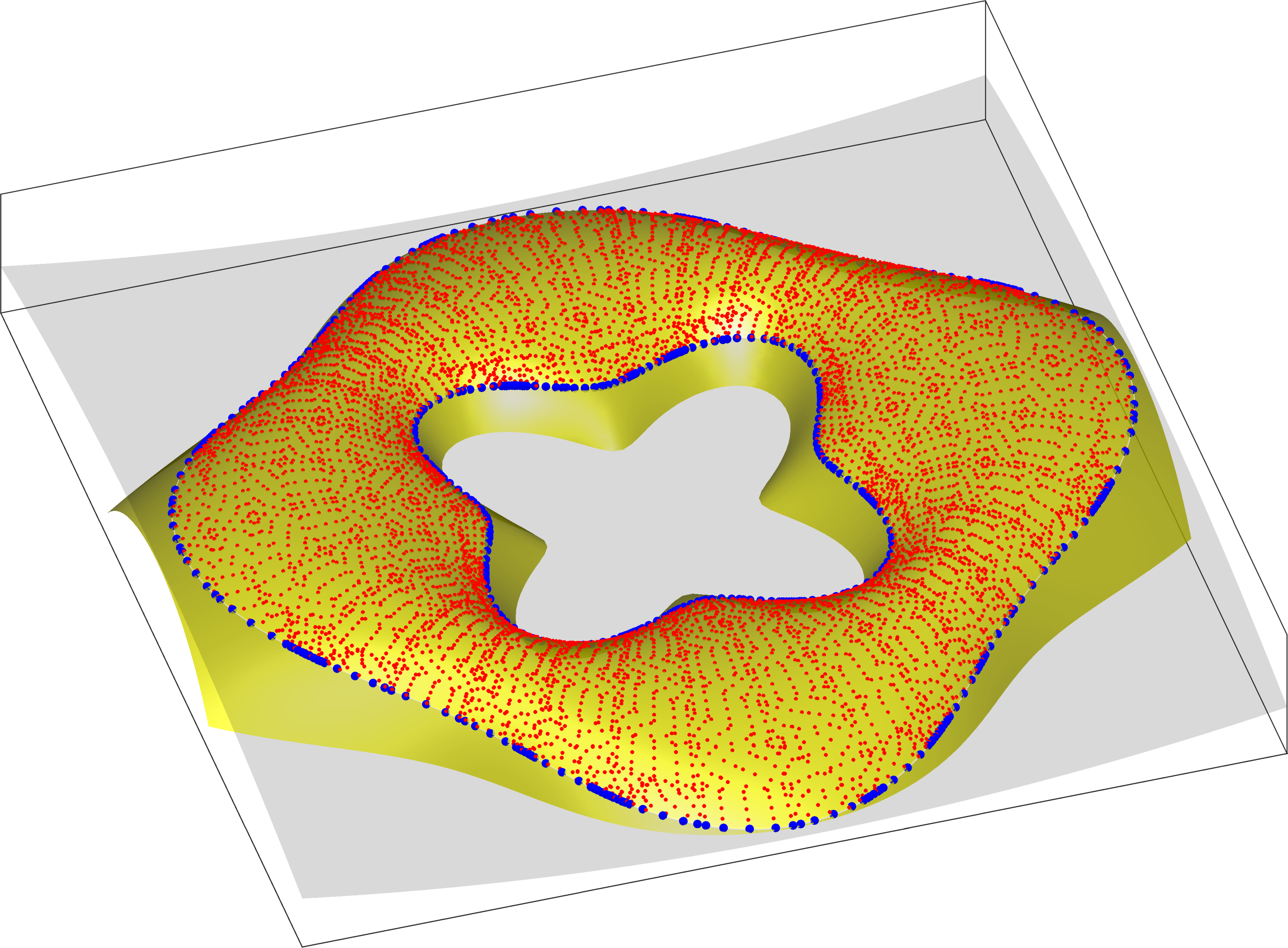}\label{fig:flowerint}} \hfil
	\subfloat[displacements]{\includegraphics[width=.43\textwidth, height = .43\textwidth, keepaspectratio]{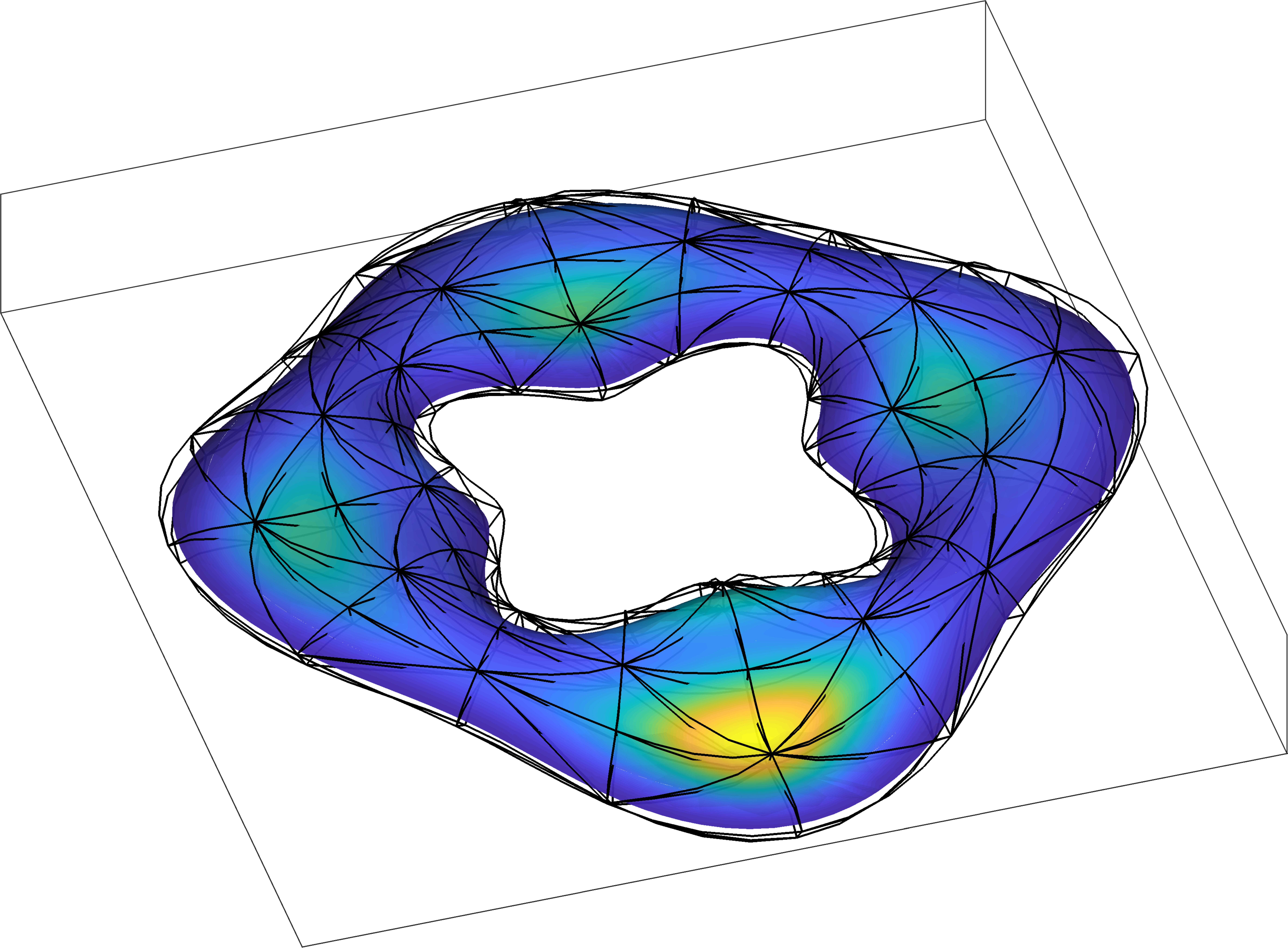}\label{fig:flowerdisp}}
	\caption{(a) Automatically generated integration points in the domain (red) and on the boundaries (blue), (b) deformed zero-isosurface with scaled displacements $\vek{u}$ by a factor of $\SI{5e2}{}$.}
	\label{fig:flowersol}	
\end{figure}

A characteristic feature of this test case is that smooth solutions in all involved fields can be expected and, therefore, optimal higher-order convergences rates are enabled. In contrast to the examples before, a reference displacement is not available for the particular example. For the error measurement we {\revStart employ \revEnd}the concept of residual errors in a similar manner as shown in \cite{Fries_2018b,Schoellhammer_2018a,Schoellhammer_2019a}. The residual errors are the evaluation of the equilibrium in strong form, see \autoref{sec:rmshells}, integrated over the domain.
In particular, the element-wise $L_2$-errors of the force and moment equilibrium are computed in the convergence analyses
\begin{align}\label{eq:resf}
	\varepsilon_{\t{rel,residual,F}}^2 &= \mathlarger{\mathlarger{\sum}}_{T=1}^{\tau_{\Omega,h}^\Gamma}\ \dfrac{\int_T \left[ \divG{\mat{n}^{\t{real}}_\Gamma} + \mat{Q}\cdot\divG{\mat{q}_\Gamma} + \mat{H}\cdot(\mat{q}_\Gamma\cdot\nG) + \vek{f} \right]^2\ \d A}{\int_T \vek{f}^2\ \d A} \ ,\\[.25 cm]
	\varepsilon_{\t{residual,M}}^2 &= \mathlarger{\mathlarger{\sum}}_{T=1}^{\tau_{\Omega,h}^\Gamma}\ \int_T \left[\mat{P}\cdot\divG{\mat{m}_ {\Gamma}} - \mat{q}_\Gamma \cdot\nG  + \vek{c}\right]^2\ \d A\ . \label{eq:resm}
\end{align}
For the computation of the residual errors, second-order surfaces derivatives are required which implies a theoretical optimal order of convergence $\mathcal{O}(p-1)$. The numerical solution of the problem is visualized in \autoref{fig:flowerdisp} for $p=4$. The displacements are scaled by a factor of $\SI{5e2}{}$. The colors on the deformed zero-isosurface are the Euclidean norm of the displacement field $\vek{u}^h$. The corresponding integration points are shown in \autoref{fig:flowerint} in the same style than before.
	
\begin{figure}[h]
	\centering
	\subfloat[force equilibrium]{\includegraphics[width=.43\textwidth, height = .43\textwidth, keepaspectratio]{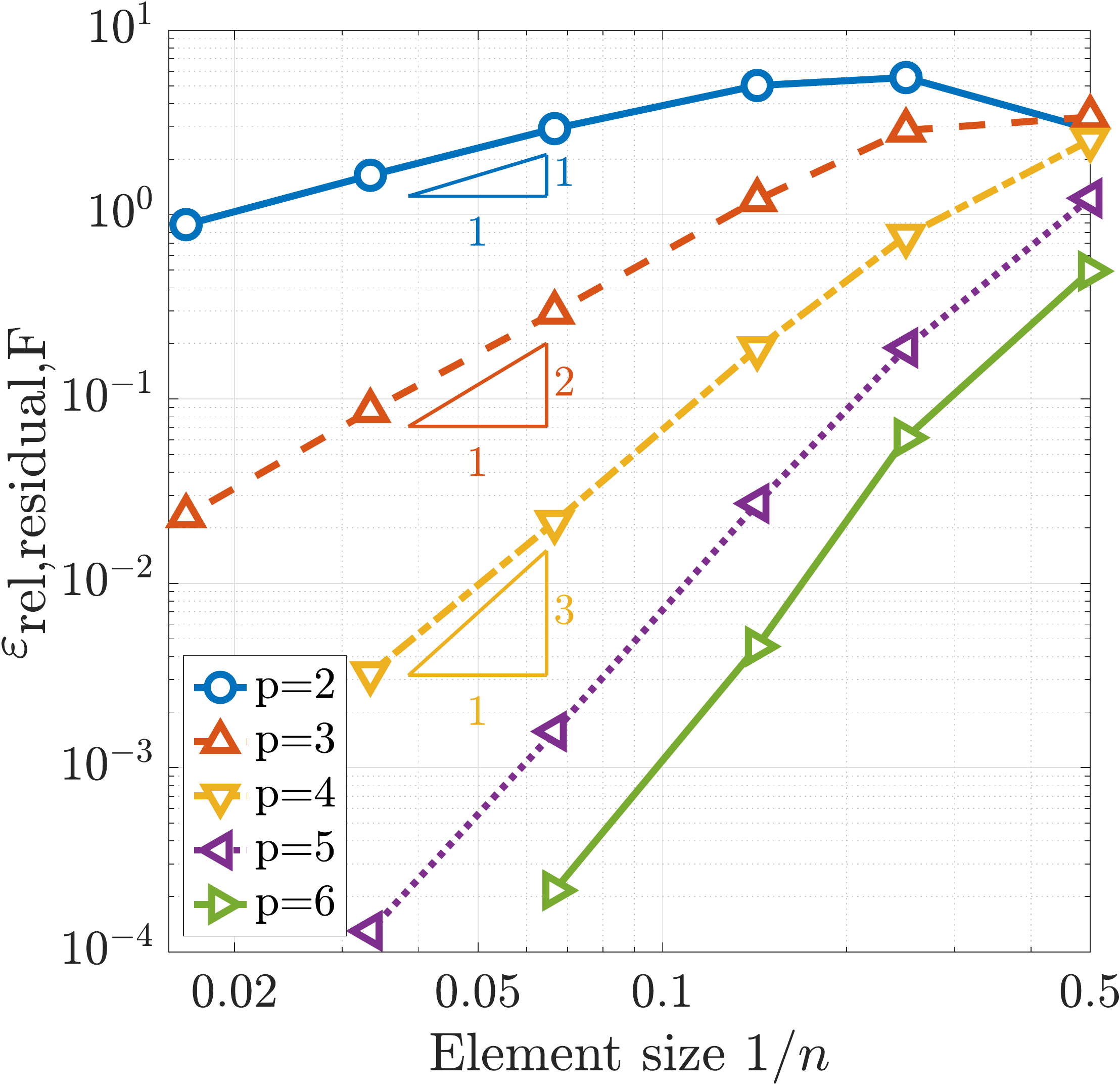}} \hfil
	\subfloat[moment equilibrium]{\includegraphics[width=.43\textwidth, height = .43\textwidth, keepaspectratio]{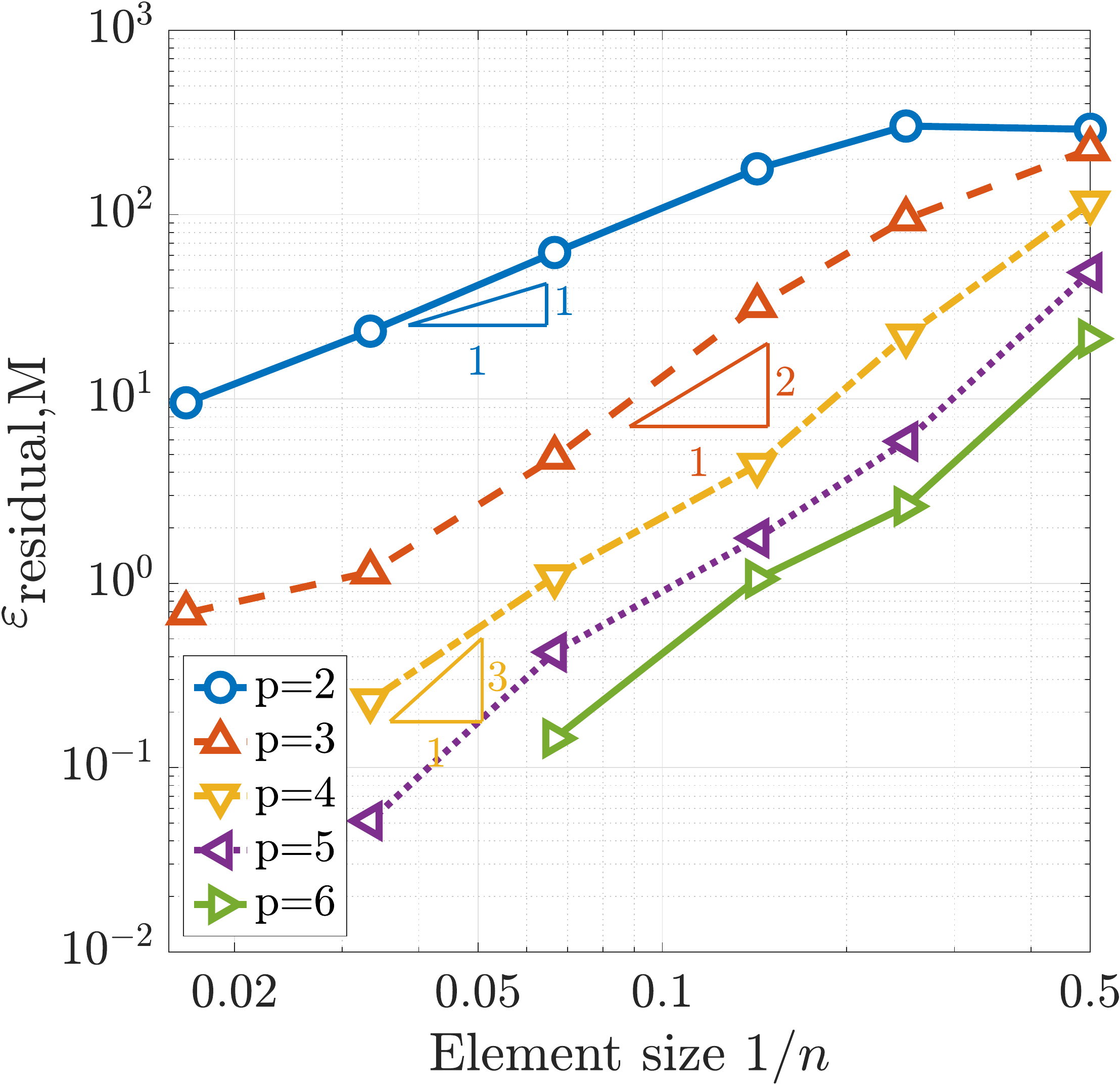}\label{fig:flowerresb}}
	\caption{(a) Residual error of the force equilibrium $\varepsilon_{\t{rel,residual,F}}$, (b) residual error of the moment equilibrium $\varepsilon_{\t{residual,M}}$.}
	\label{fig:flowerres}	
\end{figure}

In the convergence analyses the parameter $n$ is varied between $2 \le n \le 64$. The results are plotted in \autoref{fig:flowerres}. It is noticeable that the pre-asymptotic range is more pronounced for the lower ansatz orders $p=\lbrace2,3\rbrace$. Nevertheless, it is clear that higher-order convergence rates are achieved in both residual errors of Eqs.~\ref{eq:resf} and \ref{eq:resm}. In comparison to the original test case for the Reissner--Mindlin shell presented in \cite{Schoellhammer_2019a}, the behaviour of convergence in the residual errors obtained here with the Trace FEM {\revStart are \revEnd} in very good agreement to the results in \cite{Schoellhammer_2019a} obtained with isogeometric analysis. {\revStart Note that the residual error of the moment equilibrium is the absolute, element-wise $L^2$-norm, due to $\vek{c} = \vek{0}$. When comparing $\varepsilon_{\t{residual,M}}$ in \autoref{fig:flowerresb} with results obtained by the authors in \cite{Schoellhammer_2019a}, the difference in the magnitudes is traced back to the material parameters and a modified geometry.
\revEnd} 
\section{Conclusions}
\label{sec:conc}

A higher-order accurate Trace FEM approach for implicitly defined shells is presented. The shell geometry is implicitly defined by means of multiple level-set functions. Due to the implicit geometry definition, a parametrization of the midsurface is not available and the classical shell formulations of the linear Reissner--Mindlin shell are not applicable. Therefore, a more general shell formulation in the frame of the TDC, which extends also to implicitly defined shells, is employed.\par

The Trace FEM is a fictitious domain method for PDEs on manifolds enabling higher-order accuracy when the following three aspects (well-known for general FDMs) are addressed properly: (i) numerical integration, (ii) stabilization and (iii) enforcement of essential boundary conditions. Each aspect is carefully detailed herein, thus proposing an optimal higher-order accurate FDM for shells for the first time. The employed integration technique extends to multiple level-set functions and is based on a recursive reconstruction of the cut reference element into higher-order integration cells. The normal derivative volume stabilization also extends to higher-order shape functions without additional measures and the choice of the stabilization parameter is rather flexible. The essential boundary conditions are enforced with the non-symmetric version of Nitsche's method which is a consistent approach and does neither require additional stabilization terms nor the discretization of auxiliary fields. In addition, the tangentiality constraint on the difference vector is automatically built-in with an additional projection of a full 3D vector combined with a consistent stabilization term. In the numerical results, classical and new benchmark examples are presented and optimal higher-order convergence rates are achieved when the physical fields are sufficiently smooth.

\newpage

\clearpage
\newpage

\bibliographystyle{schanz}
\addcontentsline{toc}{section}{\refname}\bibliography{\pathToBibFile}

\begin{thebibliography}{10}

\bibitem{Basar_1985a}
Ba{\c{s}}ar, Y.; Kr\"{a}tzig, W.B.: \emph{Mechanik der {F}l\"{a}chentragwerke}.
\newblock Vieweg$+$Teubner Verlag, Braunschweig, 1985.

\bibitem{Bathe_2000a}
Bathe, K.J.; Iosilevich, A.; Chapelle, D.: An evaluation of the {MITC} shell
  elements.
\newblock \emph{Computers \& Structures}, \textbf{75}, 1--30, 2000.

\bibitem{Bischoff_2017a}
Bischoff, M.; Ramm, E.; Irslinger, J.: {M}odels and {F}inite {E}lements for
  {T}hin-Walled {S}tructures. In \emph{Encyclopedia of Computational Mechanics
  (Second Edition)}.
\newblock (Stein, E.; Borst, R.; Hughes, T.~J.; Hughes, T.~J., Eds.), {John
  Wiley \& Sons}, 2017.

\bibitem{Blaauwendraad_2014a}
Blaauwendraad, J.; Hoefakker, J.H.: \emph{Structural {S}hell {A}nalysis}, Vol.
  200, \emph{Solid Mechanics and Its Applications}.
\newblock {Sprin\-ger}, Berlin, 2014.

\bibitem{Bonito_2019a}
Bonito, A.; Demlow, A.; Nochetto, R.H.: Chapter 1 - Finite {E}lement {M}ethods
  for the {L}aplace-{B}eltrami {O}perator. In \emph{Geometric Partial
  Differential Equations - Part I}.
\newblock (Bonito, A.; Nochetto, R.H., Eds.), Vol.~21, \emph{Handbook of
  Numerical Analysis}, Elsevier, Amsterdam,  1--103, 2020.

\bibitem{Brandner_2019a}
Brandner, P.; Reusken, A.: Finite element error analysis of surface Stokes
  equations in stream function formulation.
\newblock \emph{arXiv e-prints}, 2019.
\newblock arXiv: 1910.09221.

\bibitem{Burman_2012a}
Burman, E.: A penalty-free nonsymmetric {N}itsche-type method for the weak
  imposition of boundary conditions.
\newblock \emph{SIAM J. Numer. Anal.}, \textbf{50}, 1959--1981, 2012.

\bibitem{Burman_2015a}
Burman, E.; Claus, S.; Hansbo, P.; Larson, M.G.; Massing, A.: Cut{FEM}:
  {D}iscretizing geometry and partial differential equations.
\newblock \emph{Internat. J. Numer. Methods Engrg.}, \textbf{104}, 472--501,
  2015.

\bibitem{Burman_2018a}
Burman, E.; Elfverson, D.; Hansbo, P.; Larson, M.G.; Larsson, K.: Shape
  optimization using the cut finite element method.
\newblock \emph{Comp. Methods Appl. Mech. Engrg.}, \textbf{328}, 242--261,
  2018.

\bibitem{Burman_2015b}
Burman, E.; Hansbo, P.; Larson, M.G.: A stabilized cut finite element method
  for partial differential equations on surfaces: The {L}aplace-{B}eltrami
  operator.
\newblock \emph{Comp. Methods Appl. Mech. Engrg.},  188--207, 2015.

\bibitem{Burman_2016b}
Burman, E.; Hansbo, P.; Larson, M.G.; Massing, A.: Cut finite element methods
  for partial differential equations on embedded manifolds of arbitrary
  codimensions.
\newblock \emph{ESAIM: Mathematical Modelling and Numerical Analysis},
  \textbf{52}, 2247--2282, 2018.

\bibitem{Calladine_1983a}
Calladine, C.~R.: \emph{Theory of {S}hell {S}tructures}.
\newblock Cambridge University Press, Cambridge, 1983.

\bibitem{Cenanovic_2016a}
Cenanovic, M.; Hansbo, P.; Larson, M.G.: Cut finite element modeling of linear
  membranes.
\newblock \emph{Comp. Methods Appl. Mech. Engrg.}, \textbf{310}, 98--111, 2016.

\bibitem{Chapelle_1998a}
Chapelle, D.; Bathe, K.J.: Fundamental considerations for the finite element
  analysis of shell structures.
\newblock \emph{Computers \& Structures}, \textbf{66}, 19--36, 1998.

\bibitem{Chapelle_2000a}
Chapelle, D.; Bathe, K.J.: The mathematical shell model underlying general
  shell elements.
\newblock \emph{Internat. J. Numer. Methods Engrg.}, \textbf{48}, 289--313,
  2000.

\bibitem{Delfour_1995a}
Delfour, M.C.; Zol\'{e}sio, J.P.: A {B}oundary {D}ifferential {E}quation for
  {T}hin {S}hells.
\newblock \emph{J. Differential Equations}, \textbf{119}, 426--449, 1995.

\bibitem{Delfour_1996a}
Delfour, M.C.; Zol\'{e}sio, J.P.: Tangential {D}ifferential {E}quations for
  {D}ynamical {T}hin {S}hallow {S}hells.
\newblock \emph{J. Differential Equations}, \textbf{128}, 125--167, 1996.

\bibitem{Delfour_2011a}
Delfour, M.C.; Zol{\'e}sio, J.P.: \emph{{S}hapes and {G}eometries: {M}etrics,
  {A}nalysis, {D}ifferential {C}alculus, and {O}ptimization}.
\newblock SIAM, Philadelphia, 2011.

\bibitem{Demlow_2009a}
Demlow, A.: Higher-order finite element methods and pointwise error estimates
  for elliptic problems on surfaces.
\newblock \emph{SIAM J. Numer. Anal.}, \textbf{47}, 805--827, 2009.

\bibitem{Dziuk_1988a}
Dziuk, G.: \emph{Finite Elements for the {B}eltrami operator on arbitrary
  surfaces},  142--155.
\newblock {Sprin\-ger}, Berlin, 1988.

\bibitem{Dziuk_2013a}
Dziuk, G.; Elliott, C.M.: Finite element methods for surface {PDE}s.
\newblock \emph{Acta Numerica}, \textbf{22}, 289--396, 2013.

\bibitem{Farshad_1992a}
Farshad, M.: \emph{Design and {A}nalysis of {S}hell {S}tructures}.
\newblock {Sprin\-ger}, Berlin, 1992.

\bibitem{Fries_2018b}
Fries, T.P.: Higher-order surface {FEM} for incompressible {N}avier-Stokes
  flows on manifolds.
\newblock \emph{Int. J. Numer. Methods Fluids}, \textbf{88}, 55--78, 2018.

\bibitem{Fries_2015a}
Fries, T.P.; Omerovi{\'c}, S.: Higher-order accurate integration of implicit
  geometries.
\newblock \emph{Internat. J. Numer. Methods Engrg.}, \textbf{106}, 323--371,
  2016.

\bibitem{Fries_2017a}
Fries, T.P.; Omerovi\'{c}, S.; Sch\"{o}llhammer, D.; Steidl, J.: Higher-order
  meshing of implicit geometries - {P}art {I}: {I}ntegration and interpolation
  in cut elements.
\newblock \emph{Comp. Methods Appl. Mech. Engrg.}, \textbf{313}, 759--784,
  2017.

\bibitem{Fries_2017b}
Fries, T.P.; Sch{\"o}llhammer, D.: Higher-order meshing of implicit geometries
  - {P}art {II}: {A}pproximations on manifolds.
\newblock \emph{Comp. Methods Appl. Mech. Engrg.}, \textbf{326}, 270--297,
  2017.

\bibitem{Fries_2019a}
Fries, T.P.; Sch{\"o}llhammer, D.: A unified finite strain theory for membranes
  and ropes.
\newblock \emph{Comp. Methods Appl. Mech. Engrg.}, \textbf{365}, 113031, 2020.

\bibitem{Gfrerer_2019a}
Gfrerer, M.H.; Schanz, M.: High order exact geometry finite elements for
  seven-parameter shells with parametric and implicit reference surfaces.
\newblock \emph{Comput. Mech.}, \textbf{64}, 133--145, 2019.

\bibitem{Grande_2018a}
Grande, J.; Lehrenfeld, C.; Reusken, A.: Analysis of a high-order trace finite
  element method for {PDE}s on level set surfaces.
\newblock \emph{SIAM J. Numer. Anal.}, \textbf{56}, 228--255, 2018.

\bibitem{Grande_2016a}
Grande, J.; Reusken, A.: A higher order finite element method for partial
  differential equations on surfaces.
\newblock \emph{SIAM}, \textbf{54}, 388--414, 2016.

\bibitem{Gross_2018a}
Gross, S.; Jankuhn, T.; Olshanskii, M.A.; Reusken, A.: A trace finite element
  method for vector-laplacians on surfaces.
\newblock \emph{SIAM J. Numer. Anal.}, \textbf{56}, 2406--2429, 2018.

\bibitem{Guo_2019a}
Guo, Y.; Do, H.; Ruess, M.: Isogeometric stability analysis of thin shells:
  {F}rom simple geometries to engineering models.
\newblock \emph{Internat. J. Numer. Methods Engrg.}, \textbf{118}, 433--458,
  2019.

\bibitem{Hansbo_2014a}
Hansbo, P.; Larson, M.G.: Finite element modeling of a linear membrane shell
  problem using tangential differential calculus.
\newblock \emph{Comp. Methods Appl. Mech. Engrg.}, \textbf{270}, 1--14, 2014.

\bibitem{Hansbo_2015a}
Hansbo, P.; Larson, M.G.; Larsson, F.: Tangential differential calculus and the
  finite element modeling of a large deformation elastic membrane problem.
\newblock \emph{Comput. Mech.}, \textbf{56}, 87--95, 2015.

\bibitem{Jankuhn_2019a}
Jankuhn, T.; ; Reusken, A.: {H}igher {O}rder {T}race {F}inite {E}lement
  {M}ethods for the {S}urface {S}tokes {E}quations.
\newblock \emph{arXiv e-prints}, 2019.
\newblock arXiv: 1909.08327.

\bibitem{Jankuhn_2017a}
Jankuhn, T.; Olshanskii, M.A.; Reusken, A.: Incompressible {F}luid {P}roblems
  on {E}mbedded {S}urfaces {M}odeling and {V}ariational and {F}ormulations.
\newblock \emph{Interfaces Free Bound.}, \textbf{20}, 353--377, 2018.

\bibitem{Jankuhn_2020a}
Jankuhn, T.; Olshanskii, M.A.; Reusken, A.; Zhiliako, A.: Error {A}nalysis of
  {H}igher {O}rder {T}race {F}inite {E}lement {M}ethods for the {S}urface
  {S}tokes {E}quations.
\newblock \emph{arXiv e-prints}, 2020.
\newblock arXiv: 2003.06972.

\bibitem{Kiendl_2017a}
Kiendl, J.; Marino, E.; {De Lorenzis}, L.: Isogeometric collocation for the
  {R}eissner-{M}indlin shell problem.
\newblock \emph{Comp. Methods Appl. Mech. Engrg.}, \textbf{325}, 645--665,
  2017.

\bibitem{Larson_2020a}
Larson, M.G.; Zahedi, S.: Stabilization of high order cut finite element
  methods on surfaces.
\newblock \emph{IMA J. Numer. Anal.}, \textbf{40}, 1702--1745, 2020.

\bibitem{Lehrenfeld_2016a}
Lehrenfeld, C.: High order unfitted finite element methods on level set domains
  using isoparametric mappings.
\newblock \emph{Comp. Methods Appl. Mech. Engrg.}, \textbf{300}, 716--733,
  2016.

\bibitem{Lehrenfeld_2018a}
Lehrenfeld, C.; Olshanskii, M.A.; Xu, X.: A stabilized trace finite element
  method for partial differential equations on evolving surfaces.
\newblock \emph{SIAM}, \textbf{56}, 1643--1672, 2018.

\bibitem{Mueller_2013a}
M\"uller, B.; Kummer, F.; Oberlack, M.: Highly accurate surface and volume
  integration on implicit domains by means of moment-fitting.
\newblock \emph{Internat. J. Numer. Methods Engrg.}, \textbf{96}, 512--528,
  2013.

\bibitem{Olshanskii_2018a}
Olshanskii, M.A.; Quaini, A.; Reusken, A.; Yushutin, V.: A finite element
  method for the surface Stokes problem.
\newblock \emph{SIAM J. Sci. Comput.}, \textbf{40}, A2492--A2518, 2018.

\bibitem{Olshanskii_2009b}
Olshanskii, M.A.; Reusken, A.: A finite element method for surface {PDE}s:
  {M}atrix properties.
\newblock \emph{Numer. Math.}, \textbf{114}, 491--520, 2009.

\bibitem{Olshanskii_2017a}
Olshanskii, M.A.; Reusken, A.: Trace finite element methods for {PDE}s on
  surfaces.
\newblock \emph{Lecture Notes in Computational Science and Engineering},
  \textbf{121}, 211--258, 2017.

\bibitem{Olshanskii_2009a}
Olshanskii, M.A.; Reusken, A.; Grande, J.: A finite element method for elliptic
  equations on surfaces.
\newblock \emph{SIAM}, \textbf{47}, 3339--3358, 2009.

\bibitem{Olshanskii_2017b}
Olshanskii, M.A.; Xu, X.: A trace finite element method for {PDE}s on evolving
  surfaces.
\newblock \emph{SIAM}, \textbf{39}, A1301--A1319, 2017.

\bibitem{Osher_2003a}
Osher, S.; Fedkiw, R.P.: \emph{Level {S}et {M}ethods and {D}ynamic {I}mplicit
  {S}urfaces}.
\newblock {Sprin\-ger}, Berlin, 2003.

\bibitem{Reusken_2014a}
Reusken, A.: Analysis of trace finite element methods for surface partial
  differential equations.
\newblock \emph{IMA J. Numer. Anal.}, \textbf{35}, 1568--1590, 2014.

\bibitem{Schillinger_2016a}
Schillinger, D.; Harari, I.; Hsu, M.C.; Kamensky, D.; Stoter, S.K.F.; Yu, Y.;
  Zhao, Y.: The non-symmetric {N}itsche method for the parameter-free
  imposition of weak boundary and coupling conditions in immersed finite
  elements.
\newblock \emph{Comp. Methods Appl. Mech. Engrg.}, \textbf{309}, 625--652,
  2016.

\bibitem{Schoellhammer_2018a}
Sch{\"o}llhammer, D.; Fries, T.P.: Kirchhoff-{L}ove shell theory based on
  tangential differential calculus.
\newblock \emph{Comput. Mech.}, \textbf{64}, 113--131, 2019.

\bibitem{Schoellhammer_2019a}
Sch{\"o}llhammer, D.; Fries, T.P.: Reissner--{M}indlin shell theory based on
  tangential differential calculus.
\newblock \emph{Comp. Methods Appl. Mech. Engrg.}, \textbf{352}, 172--188,
  2019.

\bibitem{Schoellhammer_2019b}
Sch{\"o}llhammer, D.; Fries, T.P.: {R}eissner--{M}indlin shell theory based on
  {T}angential {D}ifferential {C}alculus.
\newblock {John Wiley \& Sons}, Chichester, 2019, PAMM.

\bibitem{Schoellhammer_2019c}
Sch{\"o}llhammer, D.; Fries, T.P.: A unified approach for shell analysis on
  explicitly and implicitly defined surfaces.
\newblock  \emph{Form and {F}orce: {P}roceedings of the {IASS} {S}ymposium 2019
  - {S}tructural {M}embranes 2019, {C}. {L}{\'a}zaro, {K.U.} {B}letzinger, {E}.
  {O}{\~n}ate (eds.)}, Barcelona,  750--757, 2019, Form and {F}orce:
  {P}roceedings of the {IASS} {S}ymposium 2019 - {S}tructural {M}embranes 2019,
  {C}. {L}{\'a}zaro, {K.U.} {B}letzinger, {E}. {O}{\~n}ate (eds.).

\bibitem{Sethian_1999b}
Sethian, J.A.: \emph{Level {S}et {M}ethods and {F}ast {M}arching Methods}.
\newblock Cambridge University Press, Cambridge, 2nd edition, 1999.

\bibitem{Simo_1989a}
Simo, J.C.; Fox, D.D.: On a stress resultant geometrically exact shell model.
  {P}art I: {F}ormulation and optimal parametrization.
\newblock \emph{Comp. Methods Appl. Mech. Engrg.}, \textbf{72}, 267--304, 1989.

\bibitem{Simo_1989b}
Simo, J.C.; Fox, D.D.; Rifai, M.S.: On a stress resultant geometrically exact
  shell model. {P}art II: {T}he linear theory; {C}omputational aspects.
\newblock \emph{Comp. Methods Appl. Mech. Engrg.}, \textbf{73}, 53--92, 1989.

\bibitem{Wempner_2002a}
Wempner, G.; Talaslidis, D.: \emph{{M}echanics of {S}olids and {S}hells:
  {T}heories and {A}pproximations}.
\newblock CRC Press LLC, Florida, 2002.

\bibitem{Zingoni_2018a}
Zingoni, A.: \emph{Shell {S}tructures in {C}ivil and {M}echanical
  {E}ngineering: {T}heory and analysis}.
\newblock ICE Publishing, London, 2nd edition, 2018.

\end{thebibliography}
 
\end{document}